Université de Montréal

# Does Chance hide Necessity ?

## A reevaluation of the debate 'determinism - indeterminism' in the light of quantum mechanics and probability theory

par

Louis Vervoort

Département de Philosophie

Faculté des Arts et des Sciences

Thèse présentée à la Faculté des Arts et des Sciences
en vue de l'obtention du grade de Docteur (PhD) en Philosophie

Avril 2013

# Résumé

Dans cette thèse l'ancienne question philosophique "tout événement a-t-il une cause ?" sera examinée à la lumière de la mécanique quantique et de la théorie des probabilités. Aussi bien en physique qu'en philosophie des sciences la position orthodoxe maintient que le monde physique est *indéterministe*. Au niveau fondamental de la réalité physique – au niveau quantique – les événements se passeraient sans causes, mais par chance, par hasard 'irréductible'. Le théorème physique le plus précis qui mène à cette conclusion est le théorème de Bell. Ici les prémisses de ce théorème seront réexaminées. Il sera rappelé que d'autres solutions au théorème que l'indéterminisme sont envisageables, dont certaines sont connues mais négligées, comme le 'superdéterminisme'. Mais il sera argué que d'autres solutions compatibles avec le déterminisme existent, notamment en étudiant des systèmes physiques modèles. Une des conclusions générales de cette thèse est que l'interprétation du théorème de Bell et de la mécanique quantique dépend crucialement des prémisses philosophiques desquelles on part. Par exemple, au sein de la vision d'un Spinoza, le monde quantique peut bien être compris comme étant déterministe. Mais il est argué qu'aussi un déterminisme nettement moins radical que celui de Spinoza n'est pas éliminé par les expériences physiques. Si cela est vrai, le débat 'déterminisme – indéterminisme' n'est pas décidé au laboratoire : il reste philosophique et ouvert – contrairement à ce que l'on pense souvent.

Dans la deuxième partie de cette thèse un modèle pour l'interprétation de la probabilité sera proposé. Une étude conceptuelle de la notion de probabilité indique que l'hypothèse du déterminisme aide à mieux comprendre ce que c'est qu'un 'système probabiliste'. Il semble que le déterminisme peut répondre à certaines questions pour lesquelles l'indéterminisme n'a pas de réponses. Pour cette raison nous conclurons que la conjecture de Laplace – à savoir que la théorie des probabilités présuppose une réalité déterministe sous-jacente – garde toute sa légitimité. Dans cette thèse aussi bien les méthodes de la philosophie que de la physique seront utilisées. Il apparaît que les deux domaines sont ici solidement reliés, et qu'ils offrent un vaste potentiel de fertilisation croisée – donc bidirectionnelle.

# Abstract

In this thesis the ancient philosophical question whether ‘everything has a cause’ will be examined in the light of quantum mechanics and probability theory. In the physics and philosophy of science communities the orthodox position states that the physical world is *indeterministic*. On the deepest level of physical reality – the quantum level – things or events would have no causes but happen by chance, by irreducible hazard. Arguably the clearest and most convincing theorem that led to this conclusion is Bell’s theorem. Here the premises of this theorem will be re-evaluated, notably by investigating physical model systems. It will be recalled that other solutions to the theorem than indeterminism exist, some of which are known but neglected, such as ‘superdeterminism’. But it will be argued that also other solutions compatible with determinism exist. One general conclusion will be that the interpretation of Bell’s theorem and quantum mechanics hinges on the philosophical premises from which one starts. For instance, within a worldview à la Spinoza the quantum world may well be seen as deterministic. But it is argued that also much ‘softer’ determinism than Spinoza’s is not excluded by the existing experiments. If that is true the ‘determinism – indeterminism’ is not decided in the laboratory: it remains philosophical and open-ended – contrary to what is often believed.

In the second part of the thesis a model for the interpretation of probability will be proposed. A conceptual study of the notion of probability indicates that the hypothesis of determinism is instrumental for understanding what ‘probabilistic systems’ are. It seems that determinism answers certain questions that cannot be answered by indeterminism. Therefore we believe there is room for the conjecture that probability theory cannot not do without a deterministic reality underneath probability – as Laplace claimed. Throughout the thesis the methods of philosophy and physics will be used. Both fields appear to be solidly intertwined here, and to offer a large potential for cross-fertilization – in both directions.

# Contents

# Figures

# Abbreviations

BI = Bell Inequality

OI = Outcome Independence

PI = Parameter Independence

MI = Measurement Independence

OD = Outcome Dependence

PD = Parameter Dependence

MD = Measurement Dependence

HVT = Hidden Variable Theory

HV = hidden variable

# Remerciements – Acknowledgements

I would in particular like to thank my thesis director, Jean-Pierre Marquis, and two other Montreal professors of philosophy of science, Yvon Gauthier (Université de Montréal) and Mario Bunge (McGill), for instruction, support and encouragement. Coming from a physics background, I had much to learn; they opened a new world to me.

Of course, concerning the interpretation of the here investigated problems, our preferred philosophical positions are rarely identical and sometimes even quite different. But it is only thanks to the many detailed discussions I had with them that I could elaborate, reassess and refine my own positions on the topic. Special thanks go to Jean-Pierre also for material support, and for creating opportunities.

I would further like to thank a series of experts in the foundations of physics and quantum mechanics met at conferences or elsewhere, for detailed discussion of certain topics of the thesis, notably Guido Bacciagaluppi, Gilles Brassard, Yves Gingras, Gerhard Grössing, Michael J. W. Hall, Lucien Hardy, Gabor Hofer-Szabó, Andrei Khrennikov, Marian Kupczynski, Eduardo Nahmad, Vesselin Petkov, Serge Robert, Bas van Fraassen, Steve Weinstein; and a physicist and friend from the old days in Paris, Henry E. Fischer. From the beginning of my stay in Montreal, I remember with pleasure the gentlemanlike help of Prof. Claude Piché, and the enthusiastic encouragement of Emmanuel Dissakè. I am grateful to my teachers of the philosophy department of the Université de Montréal, Frédéric Bouchard, Ryoa Chung, Louis-André Dorion, Daniel Dumouchel, Jean Grondin, Christian Leduc, Claude Piché, Iain Macdonald, Michel Seymour, Daniel Weinstock.

Finally I thank my family and friends in Belgium and Europe, and my Montreal friends, for their lasting support and encouragement.

# Foreword to the original manuscript, April 2013

The present thesis has a somewhat specific form since it is 'by articles': it bundles four articles that are accepted or submitted for publication or published on-line as of march 2013. Only a few minor changes in vocabulary were made with respect to the original articles. Besides these four articles the thesis contains an Introduction (Chapter 1) and Conclusion (Chapter 6), in which the link between the chapters is highlighted.

The question of determinism is ancient and touches on a vast spectrum of cognitive domains; it could be studied from a variety of angles. Here this topic will be investigated from the perspective of philosophy of physics and physics. A few of the finest intellects from both fields have addressed the question – Democritus, Aristotle, Spinoza, Kant, Einstein, Bohr to name a few. It remains an active research topic in contemporary philosophy, both in philosophy of science and metaphysics. So it goes without saying that the present thesis can only treat a very small part of the wide range of questions linked to determinism – even while restricting the focus to physics and philosophy of physics. Moreover in a thesis 'by articles' many topics cannot be elaborated in the greatest detail; the texts are rather condensed by nature. Also, in order to make the articles self-contained I had to repeat some elements, especially in the introductions. So there is an inevitable overlap between Chapters 2 and 3 (on Bell's theorem); and also between Chapters 4 and 5 (on probability theory). It is however hoped that the articles allow a rather synthetic approach in a reasonable number of pages.

A few definitions are in place from the start. The question of 'determinism' will be taken here to mean "Does every event, or every phenomenon, have a cause ?" 'Deterministic' is opposed to 'indeterministic'. As we will see in Chapter 1 the latter adjective is usually taken to be equivalent with 'probabilistic' – at least if one restricts the debate to physical events, as we will mostly do here. So, according to the received view, either the physical world is deterministic and every event has a cause; either it is probabilistic and all events are characterized by probabilities; either it is a mixture of both cases (some events being deterministic, others probabilistic).

**Foreword to the present on-line version of the thesis (March 2014)**:

The present text is a reworked version of the thesis manuscript. Minor changes were made in Chapters 1-2, 4-6, while Chapter 3 is essentially a new contribution; it notably corrects an error in a preceding version detected by Lucien Hardy of the Perimeter Institute. Chapter 2 is now published (L. Vervoort, “Bell's Theorem: Two Neglected Solutions”, Foundations of Physics (2013) 43: 769-791). A condensed version of Chapters 4 and 5 is also published (L. Vervoort, “The instrumentalist aspects of quantum mechanics stem from probability theory”, Am. Inst. Phys. Conf. Proc., FPP6 (Foundations of Probability and Physics 6, June 2011, Vaxjo, Sweden), Ed. M. D’Ariano et al., p. 348 (2012)).

# Chapter 1

# Introduction and Overview

*In this Chapter the theme of the thesis will – succinctly – be put in its historical context. Then a simplified introduction to Bell's theorem will be given. Finally an overview of the remaining Chapters will be presented.*

## 1. Historical context.

The question of determinism – is everything in our world determined by causes ? - is millennia old. Already Leucippus and Democritus ($5^{th}$ cent. BC), the fathers of an incredibly modern-looking atomic theory, stated that *everything happens out of necessity, not chance*. Aristotle (384 – 322 BC) was one of the first to defend *in*determinism, objecting that some events happen by chance or hazard. Since then, countless philosophers and scholars have dealt with the topic. Leibniz' (1646 – 1716) name is forever linked to his celebrated 'principle of sufficient reason', stipulating that everything must have a reason. Kant (1724 – 1804) famously elected the thesis that all events have a cause one of his 'synthetic a priori principles'. Baruch Spinoza (1632 – 1677) put a radical determinism – even our thoughts and actions are determined – at the very heart of his Ethics, his masterpiece.

From a scientific point of view the most important advances in the debate were made by the development of, first, probability theory in say the $17^{th}$-$19^{th}$ C., and second, quantum mechanics in the $20^{th}$ C. Indeed, since the days we have a mature probability theory, scientists began to see indeterministic or random events as *probabilistic* events, governed by the rules of probability theory. This classic dichotomy in modern physics between deterministic and probabilistic events directly parallels a second dichotomy, namely in the type of systems: also physical systems are either deterministic or probabilistic. Thus, according to a standard view, individual deterministic events are determined by causes (according to laws) by which they follow with necessity. Individual indeterministic events have no causes but are nevertheless governed (as an ensemble) by probabilistic laws - the rules of probability theory. Pierre-Simon de Laplace (1749-1827), the 'Newton of France', while one of the fathers of probability theory, resolutely took the side of the 'determinists': according to him every probability and every random event only looks probabilistic or random *to us* because of our ignorance. Our limited minds cannot grasp the myriad of hidden causes that in reality, deep down, cause every event. In other words, if we would have a more efficient brain, we would not need probability theory: instead of stating 'this event E has probability X', we would know *with certainty* whether E would happen or not (X could only resume the values 0 or 1). This worldview à la Spinoza or Laplace, in which probability 'emerges' so to speak out of a fully deterministic world, seems to have been the dominant position, at least till the 1930ies. It indeed was in perfect agreement with Newton's mechanics, at the time an immensely influential theory both in physics and philosophy. But this dominance was bound to vanish.

A remarkable and abrupt moment in the history of the debate arrived with the development of quantum mechanics and its interpretation between say 1927 and 1930, by physicists as Nils Bohr, Werner Heisenberg, Erwin Schrödinger, Max Born, Wolfgang Pauli, and several others. As is well-known, the orthodox or Copenhagen interpretation of quantum mechanics stipulates that quantum events / phenomena / properties are irreducibly probabilistic, and can have no causes, *not even in principle*. The ground for this strong (metaphysical) commitment lies essentially in the fact that the theory only predicts probabilities. As an example, it can predict the energy values $E_1$, $E_2$, $E_3$,… that a quantum particle can assume in a given experimental context, and the probabilities $p_1$, $p_2$, $p_3$,… with which these values occur; but it cannot say for *one* given electron in precisely which of these energy-states it will be.

It is useful to remark from the start that quantum mechanics is a physical theory, but that its interpretation is an extra-physical, i.e. philosophical theory – it is in any case not part of physics in the strict sense[1]. The development of the Copenhagen *interpretation* was for a large part Bohr's work. One of the paradigmatic texts clarifying his position was his reply to a critical article by Einstein, Podolsky and Rosen (known since as EPR) dating from 1935. As is well-known, Einstein never was in favour of the indeterminism of quantum mechanics, which he considered a feature of a provisional theory, much as Laplace would do. In the EPR paper the authors devised an ingenuous thought experiment by which they believed they could *demonstrate* that the properties of quantum particles must have determined values, in other words that these values must exist independently of whether they are measured or not - contrary to what the Copenhagen interpretation claims. (In Chapter 2 the link between 'determination' and 'existence' will be discussed.) EPR actually made a slightly stronger assertion, namely that even non-commuting properties as the position (x) and momentum ($p_x$) of a particle must exist simultaneously – in overt contradiction to Bohr's complementarity principle. (I use here the phrasings that are typically found in quantum mechanics books to describe the EPR experiment in a rather loose way. Some philosophically alert readers may by now have slipped into a frenzy, craving *to get notions defined* ! I think here of concepts as cause, probability, existence, determination etc., all subject to intensive philosophical research, sometimes since centuries. In a sense people that are critical by now are right: several of the difficulties of the interpretation of quantum mechanics are due, I believe, to superficial philosophy and imprecise definitions. It is one of the goals of the coming chapters to make this point. But let us go back to the broad lines of the story. I ask the reader to believe, for the moment, it is a coherent account that can be made precise if enough attention is paid to it.)

[1] A physical theory T can be considered (Bunge 2006) the conjunction of a strictly 'physics' part P and an interpretational part I: $T = P \cup I$. Here P is typically highly mathematized and ideally axiomatized; but this has not for all physics theories been achieved, e.g. quantum field theory. Note that I is not part of physics in the strict sense but contains hypotheses on how to apply the theory to the real world (it contains e.g. the philosophical hypothesis that the theory represents things 'out there'). In the context of quantum mechanics I is the Copenhagen interpretation – according to the orthodox view.

In short, Einstein maintained that the probabilistic predictions of quantum mechanics should still be considered as deterministic 'deep down', notwithstanding the prevailing interpretation. It is highly instructive to scrutinize Bohr's reply. The history books mention that Bohr entered in a frenzy himself when learning about the EPR paper, and that he left all his work on the side until his answer was ready. Unfortunately it is notoriously difficult, as even experts as Bell commented (Bell 1981). As we will see in some detail in Chapters 2 and 5, the essential point of Bohr was to insist that measurement involves an interaction between apparatus and system (through the 'quantum of action', symbolized by Planck's constant h). The reality of the system is influenced by the action of the measurement device; a quantum system forms an inseparable whole with the measurement device. In Bohr's words: "The procedure of measurement has an essential influence on the conditions on which the very definition of the physical quantities in question rests" (Bohr 1935). Now x and $p_x$ cannot be measured simultaneously by the same apparatus; *ergo they do not exist simultaneously*. This may seem a cryptic summary of Bohr's reply, but as I will argue in Chapters 2 and 5, I believe it is the gist of the argument.

Now this is surely an interesting and coherent account, worth consideration especially if coming from Bohr. *But it is not a proof within a physics theory*; it is a hypothesis of interpretation. Einstein was not impressed. He could, if he had wished, reply by repeating the arguments of his original paper. In sum, in 1935 the determinism – indeterminism debate essentially remained on a metaphysical level. Both camps could stick to their philosophical positions. Still, it is almost always believed that Bohr won the debate.

**2. Bell's theorem**

In a historical context, the next decisive moment in the debate on determinism was provided by the publication of an article by particle physicist John Stuart Bell, in a new and short-lived journal called Physics, in 1964 (Bell 1964). The article was entitled "On the Einstein-Podolsky-Rosen Paradox", and we will examine it in Chapter 2. It is generally said that the crucial achievement of Bell consisted in bringing the debate 'to the laboratory', i.e. in making the question of determinism *empirically decidable*. Notice that this seems a priori, especially for a philosopher, an extremely strong claim.

How did Bell go about ? Remarkably, his paper of 1964 seems at first sight a small extension of EPR's work (for people who know the paper, he just adds analyzer directions to the EPR experiment). Let us first go over Bell's experiment swiftly, before looking at it in more detail. Loosely said, Bell proposes in his article an experiment of which the outcome can be calculated by two theories. On the one hand by quantum mechanics, on the other by a generic deterministic theory, i.e. a theory that considers the stochastic quantum properties of particles as being deterministic, i.e. *caused* by yet unknown additional variables – so-called 'hidden variables'[2]. Then Bell proves by simple mathematics that the outcome predicted by quantum mechanics is numerically different from the result calculated within *any* deterministic theory, or 'hidden variable theory' (HVT). If this remarkable theorem would be correct, there would be an experimentally verifiable difference between quantum mechanics and deterministic or HV theories, or loosely said, between quantum mechanics and determinism. One of the two would necessarily be wrong. (We will see further that this picture needs only a little bit complicated: the actual contradiction Bell established is between quantum mechanics and *local* HVTs.)

[2] Notice this introduces the notion of 'cause'.

In some more detail, the experiment Bell proposes is the following (Fig. 1). Consider a pair of particles (say electrons) that leave a source S in which they interacted; one particle (1) goes to the left, the other (2) to the right. One measures the spin $\sigma$ of the particles, so $\sigma_1$ on the left, $\sigma_2$ on the right. Spin can be measured with Stern-Gerlach magnets in different directions, say a on the left, b on the right (a and b are angles in the plane perpendicular to the line of flight). So one measures $\sigma_1(a)$ and $\sigma_2(b)$. Suppose now that a large ensemble of pairs leaves the source, and that one determines the average product $M(a,b) = < \sigma_1(a).\sigma_2(b) >$ for the ensemble. Suppose now all particle pairs are in the singlet state, the quantum mechanical state corresponding to a spin-entangled pair that EPR had already considered. (Years after Bell proposed his experiment, physicists have succeeded in performing it; it is possible to prepare the particle pairs in such a way that they indeed are in the singlet state.) In that case it is possible to apply the rules of quantum mechanics and to calculate M(a,b); one finds M(a,b) = cos(a-b), so a simple cosine dependence on the left and right analyzer angles.

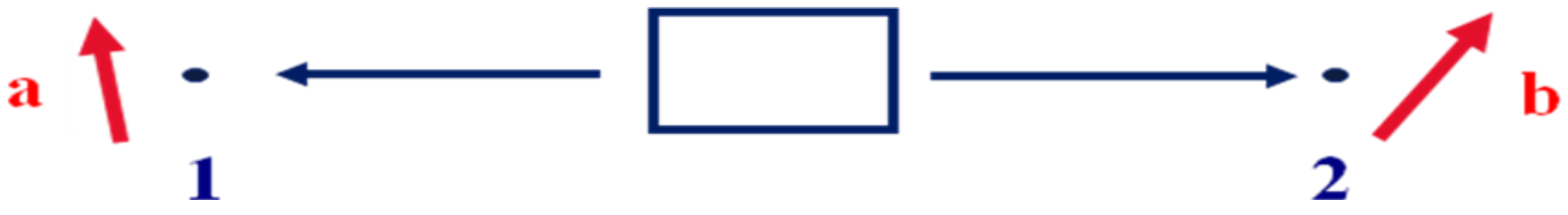

Fig. 1. Particles 1 and 2 are emitted from a source.
'a' and 'b' are the angles of the left and right magnets with which one measures the particles' spins.

In the remainder of the article all Bell does is prove that this quantum prediction for M(a,b) cannot be identical to the result calculated within any 'reasonable' deterministic theory. Here it is of course crucial to scrutinize what is reasonable for Bell; but the vast majority of researchers agree with his analysis. What is a reasonable deterministic theory ? Bell stipulates that in such a theory M(a,b) should be expressed as follows:

$$M(a,b) = < \sigma_1(a).\sigma_2(b) > = \int \sigma_1(a,\lambda).\sigma_2(b,\lambda).\rho(\lambda).d\lambda. \qquad (1)$$

A formula as (1) is the standard expression for the average value of the product of two quantities that are determined by parameters $\lambda$ having a probability distribution $\rho(\lambda)$. Indeed, (1) says first of all that the left spin ($\sigma_1$), in a deterministic theory, is in general not only a function of a, but also of yet unknown, hidden parameters $\lambda$; and similarly $\sigma_{II}$ will in general be a function of b and some other parameters $\lambda$. Here $\lambda$ can be a set of variables, they can be different for 1 and 2: Bell's theorem is extremely general as to the nature of these hidden parameters. Thus, while quantum mechanics only predicts that $\sigma_1$ and $\sigma_2$ can be equal to +1 or -1 with a probability of 50%, in a deterministic theory $\sigma_I$ and $\sigma_{II}$ are *determined* to assume one of these values by yet unknown causes, $\lambda$. If we would know these $\lambda$ and a, we would know the result of the measurement of $\sigma_1$ in advance; similarly for $\sigma_2$ and b. So a HVT for Bell's experiment 'completes' quantum mechanics by predicting the outcome of individual measurement events:

$$\sigma_1 = \sigma_1(a,\lambda) \text{ and } \sigma_2 = \sigma_2(b,\lambda). \tag{2}$$

This is the assumption of 'determinism' (as applied to Bell's experiment).

Bell acknowledged that there is one other assumption on which (1) hinges. This is the assumption of *locality*: $\sigma_1$ may depend on a but not on b, and $\sigma_2$ may depend on b but not on a. This is a crucial hypothesis, since if one relaxes it, it is easy to recover the quantum result M = cos(a-b) within a deterministic theory. Now this locality assumption is definitely reasonable, for the simple reason that physics does not know any 'nonlocal' influences, i.e. influences (forces) that a left analyzer (a) would exert on a particle (2) that is separated from it over a long distance. Indeed, the Bell experiment has been performed using photons instead of electrons while the distance between the left and right measurements was as long as 144 km (Scheidl et al. 2010) !

In sum, formula (1) gives us the average product $M(a,b) = < \sigma_1(a).\sigma_2(b) >$ as calculated within a '*local HVT*', or assuming '*local determinism*' according to the celebrated phrasings. Bell shows that this 'local-and-deterministic' prediction (the integral) cannot be equal to the cos(a-b) predicted by quantum mechanics. Ergo local HVTs are in contradiction with quantum mechanics, at least for the considered experiment. Bell's proof goes like this. He defines a quantity X = M(a,b) + M(a',b) + M(a,b') – M(a',b') where (a,b,a',b') are 4 analyzer directions, so angles in the plane perpendicular to the line of flight (a, a' are angles of the left detector, and b, b' of the right one). He then proves that if M is given by an integral as in (1), $X \leq 2$ for all choices of angles (a,b,a',b'). The inequality $X \leq 2$ is the famous 'Bell inequality' (actually one form of it), which is thus satisfied by all local HVTs. But if M is a cosine, as quantum mechanics predicts, then X can for some choices of angles be much larger than 2, up to $2\sqrt{2}$.

So indeed, to put it succinctly, local determinism is in contradiction with quantum mechanics. More precisely, *at least one of the following 3 fundamental hypotheses must necessarily be wrong*:

(i) a quantum property like spin is determined by additional variables (as in (2)),

(ii) all influences in nature are local,

(iii) quantum mechanics adequately describes Bell's experiment.

As recalled in Chapter 2, Bell's original analysis (Bell 1964) was in the decades following his discovery confirmed by several authors, deriving Bell's inequality in a somewhat different manner. Based on these works, it is generally believed, both in the physics and quantum philosophy communities, that quantum mechanics is indeed in empirical contradiction with all local HVTs. Now the remarkable point of Bell's analysis as compared to EPR's is that the experiment he proposed can be done: M(a,b) or X can be measured. In the 1980ies till very recently highly sophisticated experiments were performed, coming closer and closer to the ideal Bell experiment, by the groups of Alain Aspect, Anton Zeilinger, Nicolas Gisin and others. All these experiments clearly violated the Bell inequality and were in agreement with the quantum prediction. In other words, among the above conflicting hypotheses (i) – (iii), (iii) must be withheld. Based on these results, it is widely believed that '*local HVTs are impossible*' – to put it in a slogan. 'Local determinism is dead' is another popular slogan that resumes the consequence that is drawn from Bell's theorem in conjunction with the experimental results.

### 3. The thesis in a nutshell: Part I (Chapters 2 and 3)

The thesis basically starts here. Had Bohr already given a serious blow to determinism, Bell's theorem and the experimental confirmation of the quantum result seem to have given it its death sentence – or so it is widely believed. The object of the investigations in the next two chapters is to critically reassess this conclusion, i.e. the demise of determinism by Bell's theorem and the experiments.

If one feels something might be wrong with the general rejection of determinism (for instance, as in the author's case, starting from a philosophical intuition), the obvious starting point is to investigate whether the expression (1) is general enough to represent *all* local HVTs.

One of the first and most relevant critical analyses was provided by Shimony, Horne and Clauser (Shimony et al. 1976), about ten years after Bell's seminal work. These authors observed that (1) can be generalized in that the probability density of the HVs $\lambda$, $\rho(\lambda)$, could *at least in principle* depend on the left and right analyzer directions a and b. Shimony et al. argued this might happen if $\lambda$, a and b would have *common causes*. That would mean that $\rho(\lambda)$ should be written as a conditional probability density $\rho(\lambda|a,b)$. (I will show in some detail in Chapters 4-5 that this 'conditional' expression is indeed in agreement with a detailed interpretation of the notion of probability.) If $\lambda$ thus depends in principle on (a,b), then one may have in general that $\rho(\lambda|a,b) \neq \rho(\lambda|a',b')$ for some values of $\lambda$, a, b, a', b'. But in that case one easily shows that the Bell inequality cannot be derived anymore (Shimony et al. 1976). So Bell's reasoning hinges not only on the assumptions of 'determinism' ((i) above) and 'locality' (ii), but also on a condition now often called 'measurement independence' (MI):

$$\rho(\lambda|a,b) = \rho(\lambda|a',b') \equiv \rho(\lambda) \text{ for all relevant } \lambda, a, b, a', b' \qquad \text{(MI).} \qquad (3)$$

Note this is a condition of probabilistic independence. Shimony et al. explained that a dependence of $\lambda$ on (a,b), or more precisely violation of MI, could come about if a, b and $\lambda$ (determining the outcomes $\sigma_1$ and $\sigma_2$) would all have common causes in their past lightcones. Now they conceded, after a detailed discussion with Bell (see Shimony 1978 and the review in d'Espagnat 1984), that this seemed even to them a quite implausible solution. Indeed, if $\lambda$ would depend on (a,b) then (a,b) would depend on $\lambda$, by the standard reciprocity of stochastic dependence following from Bayes' rule. Now in Bell's experiment a and b are parameters that may be *freely* chosen by one or even two experimenters – Alice on the left, Bob on the right. So how could they depend on $\lambda$, physical parameters that would moreover determine $\sigma_1$ and $\sigma_2$ (via (2)) ? (This 'free will' argument will play an important role in the following.) To Shimony, Clauser, Horne, Bell such a dependence is in overt contradiction with any reasonable conception of free will (Shimony 1978). Since then, almost all authors have accepted this seemingly final verdict. Not accepting MI in (3) would amount to accepting hidden causes determining *both* the outcomes of a Bell experiment *and* our individual choices of analyzer directions. That has since the early days of Bell's theorem been deemed a 'superdeterministic', nay 'conspiratorial' conception of the universe.

I will interpret and contest this conclusion in essentially two ways, as explained in detail in Chapters 2 and 3. My first argument (Ch. 2) will essentially be philosophical, involve little physics, and mostly follow a known line of thought. The second argument will be backed-up by new physics results (Ch. 2-3).

In short and to start with, in Chapter 2 the 'superdeterministic' solution envisaged by Shimony et al. will simply be put in a somewhat more general historical and philosophical context than is usually done. I hope in this way to show that superdeterminism may be less exotic than is often believed. It will be recalled that the 'total' or 'superdeterminism' of the quantum philosophy community immediately (and quite obviously) derives from the ancient, historical principle of determinism; and that if every (physical) event since the Big-Bang is determined, caused, as this principle demands, there may very well be a causal link between all events – including our choices, for instance the choices of analyzer directions of a Bell experiment. Ergo MI in (2) may be violated; ergo there is not necessarily a contradiction between quantum mechanics and deterministic theories.

Such a position is in need of further explanation (Chapter 2). It seems impossible to construct a (physical) theory that would exhibit the common link between all events, even implicitly. Hence superdeterminism seems a philosophical theory, not a physical one. Also, this position seems to assume that our choices, which have a mental component, are at least partly physical events or have a physical component, such as biochemical processes in the brain. Of such physical processes one conceives more easily that they are inscribed in a causal web. Recall that Shimony et al., Bell etc. have always rejected a violation of MI (Eq. (3)) and thus justified the Bell inequality based on a 'free will' argument. But that 'free will' may be more complex than meets the eye is a favourite topic of philosophy since Antiquity ! Therefore, from a philosophical point of view superdeterminism (actually one could stick to the term determinism) seems a sane, and certainly admissible assumption. Spinoza put determinism at the basis of a cogent and comprehensive philosophical theory. In Chapter 2 I will – succinctly – argue that it corresponds to a simpler worldview than indeterminism, in that it needs fewer categories. This seems to us, in view of Occam's razor, a considerable advantage of this position.

At the same time it is important to recall that only a handful of quantum physicists and philosophers have given serious consideration to superdeterminism, among others Guido 't Hooft, a Nobel prize winner, Michael Hall, Carl Brans, Bernard d'Espagnat (references are given in Ch. 2-3).

In sum, superdeterminism, as an interpretation or 'solution' to Bell's theorem, accepts that a, b and λ may depend on common causes. It thus rejects the 'free will' argument of Shimony et al. to justify MI by interpreting 'common causes', even when related to our choices, as a harmless, necessary fact of nature. This seems however essentially a philosophical position; it may well be impossible to prove it within a physics theory.

If a brief note on the evolution of the present thesis is allowed, since there is such an overwhelming majority of experts maintaining that 'all physics is done' if one wants to stick to determinism, I was first inclined to leave things at the philosophical discussion of the topic, along above lines, which I found fascinating in itself - and which can doubtlessly be much more elaborated than I will do here. However, to my surprise, it appeared possible to go further. I believe now it is not only possible to re-interpret the 'free will' argument of Bell, Shimony et al. as we did above, but to prove it is false under certain conditions – *namely if one attributes the HVs to a background medium instead of to the particles*.

The approach proposed in Chapters 2 and 3 is to investigate physical model systems. Recall that the Bell inequality (BI) is supposed to hold for *any* deterministic and local system on which one performs a Bell-type correlation experiment. Bell's derivation of the BI is extremely general; nothing in it is restricted to the singlet state. Therefore, if one can find a classical and local system in which MI is violated if one performs a Bell-type experiment on it, then such a system could possibly serve as a model for a hidden reality explaining the real Bell experiment. If MI is violated in the system, the BI is possibly too (but this is not *necessarily* so, it must be calculated). We will investigate such systems, namely spin lattices, in Chapters 2 and 3. They are described by Ising Hamiltonians and exist in many magnetic materials (but actually the Ising Hamiltonian describes a particularly broad family of physical phenomena). These systems can be shown to be local in the usual sense, as is explained in Chapters 2-3. It will be shown that in such systems the BI can be strongly violated, and that this is due to violation of MI (Ch. 2). Now, spin-lattices appear to be a simple model system for what I will term 'background-based' theories. I will argue in Ch. 3 that basically in any theory involving a background medium in which the Bell particles move and that interacts with the Bell particles and analyzers, MI can be violated. If MI is violated for such background-based models, Bell's no-go result does trivially not apply to them. In Chapter 3 I will show how precisely such background-based theories could underlie the real Bell experiments, in particular the most sophisticated experiments that use ‘dynamic’ analyzer settings (Aspect et al. 1982, Weihs et al. 1998, Scheidl et al. 2010). I will come back to the relevance of such dynamic experiments in a moment.

But let us first ponder a moment on the significance of the above result, i.e. the violation of MI and the BI in a wide variety of spin lattices. A first essential point is that MI can be violated in these experiments (as we will unambiguously calculate) *even if there is no superdeterministic link* between (a,b) and $\lambda$. So even if there are no common causes between (a,b) and $\lambda$, causes that would determine both (a,b) and $\lambda$. How can we prove that ? Simply by showing that perfectly free-willed experimenters find violation of MI and the BI in our experiment. This calculation will thus prove that the ‘free will’ argument to justify MI is not valid. If MI is not valid, Bell’s reasoning cannot be brought to its end. There is not necessarily a contradiction between the hypotheses (i) – (iii) above.

It is important to realize in the present discussion that extremely sophisticated ‘dynamic’ experiments have been performed, all confirming the quantum prediction (Aspect et al. 1982, Weihs et al. 1998, Scheidl et al. 2010). The importance of these experiments resides in the fact that they try to exclude that some ‘trivial’ HVT may explain the quantum correlations. In particular, they try to impose locality. What is meant here is the following. We have seen that it is possible to recover the quantum correlations if $\sigma_1$ would depend on b, $\sigma_2$ on a, or $\rho(\lambda)$ on (a,b). Such a dependence could come about if an (unknown) long-range force would exist between the left and right parts of the experiment, e.g. between $\sigma_1$ and b, or between $\lambda$ and (a,b). Now the mentioned experiments try to exclude such a possibility, first of all by making the distance between the two parts very large (up to 144 km !), and especially by creating a spacelike distance between relevant events, i.e. a spacetime distance that can only be crossed at unphysical, superluminal speeds. For instance, if one sets the left analyzer to its direction ‘a’ precisely at the moment $\sigma_2$ is measured, the latter could not depend on it: there is not enough time for a physical signal to travel from left to right. Only a nonlocal, i.e. superluminal signal could do so. As we will see in some detail in Chapter 3, by this and similar draconic experimental precautions experimenters have come closer and closer to Bell’s original experiment.

The upshot is that if one wants to construct a local HVT for the Bell experiment, it should be able to explain what happens in these most sophisticated experiments. In Chapter 3 we will show that also in these dynamic experiments MI may be violated, by a mechanism very similar to the one we discovered in the (static) spin lattices, namely the interaction of the Bell particles and analyzers with a background. Our argument is thus quite general, and not linked to the specific form of the Ising Hamiltonian (of course nothing indicates that a realistic HVT should be based on an Ising Hamiltonian, which only serves as an example). Indeed, the mechanism by which MI may be violated appears, in hindsight, to be quite straightforward. It just involves local interaction of the left analyzer with 'something' (the HVs λ) in its neighbourhood, causing some change in that something, which in turn interacts with the left particle (thus determining the left outcome); and similarly on the right. A simple way to understand that 'something' is to interpret it as a background field that pervades space[3]; so the left (right) HVs are spacetime values of this 'hidden' field in the spacetime neighbourhood of the left (right) measurement. It is due to the fact that the analyzers interact with the λ (on their respective sides) and thus may change these λ, that in general MI will not be valid - much as happens in the spin lattices. In sum, MI may be violated through a classical physical process in an experiment done by perfectly free-willed experimenters. Since there is no ground to invoke 'superdeterminism' here we termed this solution to Bell's theorem 'supercorrelation' (Ch. 2). It is based on correlations that are stronger than allowed by MI.

In view of the reigning orthodoxy concerning Bell's theorem, the results and especially the conclusion of Chapters 2-3 are surprising, to some extent. Therefore further investigation of these results would be particularly welcome; in Chapter 3 several research directions are proposed (and many more are conceivable). At the moment it remains somewhat puzzling why this solution seems to have escaped from the attention of researchers. A tentative explanation is given in Chapter 3. One almost always seems to interpret ρ(λ|a,b) in (3) as the probability density of variables pertaining to the particles and/or being 'created' at the source (see e.g. the phrasing in Scheidl et al. 2010, cited in Ch. 3). In this view the HVs λ look like a property like mass or momentum 'pertaining' to the particles; it is very likely that this corresponds to Bell's initial intuition concerning the nature of the HVs. Sure, if (a,b) are set at the emission time of each pair, as in the most advanced experiment (Scheidl et al. 2010), then there can be no influence of (a,b) on the values of properties 'at the source', since they are spacelike separated. So MI holds. Or as one often reads: "the way the particles are emitted cannot depend on the simultaneous choice of distant analyzer directions". However, symbols may be interpreted quite differently, it seems. Indeed, we will attach the λ to a background medium rather than to the Bell particles[4]; and ρ(λ|a,b) can be interpreted as the probability of λ *at the moment of measurement*. And there may well be interaction *at the moment of measurement*, more precisely dependence of λ on (a,b). Such claims become fully tangible by investigating what happens in real systems, such as Ising lattices – hence their importance. Further interpretations of why this solution may have escaped till date are given in Chapter 3.

[3] Note that other authors (e.g. Grössing et al. 2012) have tried to explain quantum phenomena as double-slit interference by invoking a stochastic background field, or zero-point field, as recalled in Chapter 2 and 3. Obviously there may be a relevant connection with our findings.

Let us also emphasize that our analysis is in agreement with the conclusions reached by other authors criticizing the general validity of Bell's theorem (see Khrennikov 2008, Kupczynski 1986, Nieuwenhuizen 2009 and references therein). These authors have concluded that MI is not valid, essentially based on a detailed interpretation of the notion of probability. Our findings, while focussing on physical mechanisms to explain how violation of MI and other premises of the BI can come about, are in agreement with their conclusion. It seems there is another important link to be made with existing works, namely the HVTs that were recently developed to explain e.g. double-slit interference (Grössing et al. 2012). The link is explained in some detail in Chapters 2-3 and 6. These theories also involve a background field, and should show the same strong correlation as the spin-lattices. Whether this is the most promising road towards realistic HVTs for quantum mechanics is an open question, which can only be answered by further research.

In the following we will not only be concerned with the deterministic variant of Bell's theorem, but also with the stochastic variant. As we saw above, in a deterministic HVT the left and right spins $\sigma_1$ and $\sigma_2$ are assumed to be deterministic functions of HVs, as in (2). Stochastic HVTs are more general, less stringent: they only assume that the *probability* that $\sigma_1$ and $\sigma_2$ assume a certain value (±1) is determined *given additional variables λ*. In other words in such a HVT $\sigma_1$ and $\sigma_2$ are probabilistic variables, for which one assumes (instead of (2)) that

$$P(\sigma_1|a,\lambda),\ P(\sigma_2|b,\lambda) \text{ and } P(\sigma_1,\sigma_2|a,b,\lambda) \text{ exist.} \quad (4)$$

As is recalled in Chapter 2, the Bell Inequality can also be derived for such HVTs. To do so, one has to assume, besides MI (3), two other conditions or hypotheses which are now often termed ‘outcome independence’ (OI) and ‘parameter independence’ (PI), following a seminal analysis by Jarrett (Jarrett 1984). These assumptions are usually believed to follow from locality (see e.g. Jarrett 1984, Shimony 1986, Hall 2011), but that inference has never been proven.

[4] In Chapter 3 we will show that Bell himself opened the door to such a semantic shift.

Since we have already rejected the general validity of MI above, the immediate consequence is that there would be no fundamental impediment to the existence of *stochastic* and local HVTs, just as to the existence of local and *deterministic* HVTs. The interesting point about stochastic HVTs is that the other two premises OI and PI can be questioned in turn, besides and above MI – *if one associates again the HVs to a background*. Since PI is usually assumed to be inescapable (it would allow superluminal signalling), I focussed in Chapter 2 on OI. There it is argued that it is not an innocent assumption as one so often believes (Ch. 2). These points corroborate the findings we summarized above; they point to other mechanics of correlation that might offer a local solution for Bell’s theorem (Ch. 2). In Chapter 2 I gathered these solutions under the term ‘supercorrelation’ to distinguish them from superdeterminism – the latter being a solution that is still possible but very different, both physically and philosophically.

Let us now conclude on Bell’s theorem, and put our findings in the larger context of the ‘determinism – indeterminism’ debate. The received view in the quantum philosophy and physics communities is that, in view of the BI and its experimental violation, either locality or determinism have to be given up ((i) or (ii) above). Since locality is an axiom of relativity theory, or a direct consequence of it, the orthodox conclusion is that determinism (in the strict sense (2) but even in the sense (4)) has to be given up. “There are no HVs” summarizes this position. According to a widely held belief, Bell’s theorem would allow to definitively confirm the indeterminacy of (quantum) nature, as proclaimed by the Copenhagen interpretation since the 1920s.

If we are right, our findings allow to contest this conclusion. Besides locality and the existence of HVs there is another assumption needed to derive the BI, namely measurement independence. In experiments it may be this condition that is violated through a local mechanism, not the assumption of local determinism (the existence of local HVTs). In other words, we will conclude in Chapters 2 and 3 that local HVTs are in principle possible. We will come back to this conclusion in Chapter 6, the Epilogue of this thesis.

But then the determinism - indeterminism debate is undecided. Actually it suffices to recall that superdeterminism is a solution to Bell's theorem in order to come to this conclusion (Ch. 2 and 6).

At least from a philosophical point of view the situation is now different. It seems we are back at where we started. As will be emphasized in Chapters 2 and 6, indeterminism is not proven, it is at most one of the possible metaphysical interpretations of quantum mechanics - a hypothesis among others. Things are then not decided in the laboratory. The debate belongs again to philosophy, from where it originated.

**4. The thesis in a nutshell: Part II (Chapters 4 and 5)**

If the situation is such, it is legitimate to inquire whether meaningful philosophical arguments exist that favor either determinism or indeterminism. As we saw, 'indeterministic' physical events are probabilistic events. Therefore one possible approach is to seek for indications in probability theory. The interpretation of probability will be the subject of Chapters 4 and 5 of this thesis; the link with the question of determinism will be made in Chapter 6, the Epilogue. As we will argue there, it seems that arguments exist that favor determinism. Of course, it will be impossible to prove such claims within an accepted physics or mathematics theory. But the aim is to show that they are coherent with a model for the interpretation of probability (Ch. 4-5).

Needless to say, philosophical interpretations of the determinism-indeterminism dichotomy that considerably differ from ours are possible. For instance, from a purely pragmatic point of view the straightforward interpretation is to favor indeterminism. For many physical phenomena and in particular quantum phenomena we only have, to date, a probabilistic theory. And indeed, besides positivism and operationalism there are modern philosophical theories that embrace indeterminism (cf. e.g. van Fraassen 1991, Gauthier 1992, 1995). But on the other hand, in view of the results presented in Chapters 2-3, there is no physical impediment anymore to consider any probability, quantum or classical, as emerging from a deterministic background. To start with, probability theory does not prohibit such an assumption: it is fully silent about this point. Indeed, one of its fathers, Laplace, believed that any probability is only a tool we need because of our ignorance of hidden causes. And the examples in classical physics in which probabilistic behavior and probabilities can very well be traced back to deterministic laws, are numerous (Chapter 6 gives examples). From a philosophical point of view one may already ask: if some probabilities result from deterministic processes, why not all of them ? Why two categories (deterministic events and probabilistic events) ? To put these and other questions in what I hope is an illuminating perspective, I will start in Chapter 4 from a detailed interpretation of the notion of probability.

If a personal anecdote is allowed, when one of my professors, Yvon Gauthier, proposed me to investigate the notion of probability my first reaction – stemming from my training as a physicist – was one of surprise. Hadn't I learned during my mathematics classes that basically 'everything' about probability is said by Kolmogorov's simple axioms ? And isn't the calculus of probability applied every day by maybe millions of students and professionals in mathematics, statistics, physics, engineering and a host of other fields ? The only vague memory I had of something slightly disturbing, was that we learned *two* interpretations of probability, the classic interpretation of Laplace for chance games (the eternal urn pulling etc.), and the frequency interpretation basically 'for anything else'. "Why two, not one ?" may have been a fleeting thought that crossed my mind in those days – before hurrying back to my calculator. Having studied since the theories of interpretation, the foundational questions, and especially the paradoxes of probability theory, I believe now it is indeed a wonderful and an extremely subtle topic, at the interface of mathematics, philosophy and physics. It is not without reason that probability theory has been called "the branch of mathematics in which it is easiest to make mistakes" (Tijms 2004). Most textbooks on the calculus or the interpretation present a list of riddles and paradoxes; paradoxes on which sometimes even the fathers of modern probability theory disagreed. (One example is Bertrand's paradox, investigated in Chapter 4.) The problem is clearly not the calculus itself, but the question how to apply the calculus to the real world – i.e. how to interpret probability (another way to put the question is: *what really is a probabilistic system* ?). Recalling footnote 1, probability theory as a theory (T) of real-world events is the conjunction of an axiomatized part (P) and a theory of how to apply the rules, i.e. the interpretation I: T = P ⋃ I. There seems to be a surprisingly wide variety of paradoxes in the sciences and in philosophy that derive from this lack of a precise interpretational part I, i.e. a definition of probability beyond as something that satisfies Kolmogorov's axioms. It has been argued that this extra-mathematical, philosophical part of probability theory deserves a larger part even in the curriculum of scientists (Tijms 2004); I can only agree with it (and I hope the following chapters will provide arguments for this idea).

Needless to say, in the philosophy of science community the subtlety of probability theory is well-known and intensively investigated since a long time (cf. e.g. Fine 1973, von Plato 1994, Gillies 2000). Interestingly, reading Kolmogorov's, Gnedenko's and von Mises' works shows that these fathers of the modern calculus were also fascinated by the interpretation and the philosophy of probability. For instance, they followed the development of quantum mechanics with a keen interest, realizing that probability theory governs a wider and wider set of natural phenomena. It is delightful to plunge into the works of these masters and witness them thinking about their discoveries; I believe they remain highly up-to-date in the debate on the foundations of probability. In the present thesis I will remain close to the usual interpretation of probability in physics, namely the frequency interpretation, which is often attributed to Richard von Mises (1928/1981, 1964). It should however be noted that another model, namely the subjective or Bayesian interpretation, gains importance in the community of quantum physicists and philosophers. Bayesianism comes in many flavors, from very to lightly subjectivist. In Chapter 5, I will moderately criticize the more subjectivist variant as applied to quantum mechanics. But maybe more importantly, I will propose a way to bridge the frequency and the more moderate subjective interpretations, e.g. Jaynes' version (1989, 2003). So in the end these models may not be that different. My goal will be to show that an adequate frequency interpretation is a powerful tool to solve problems.

In some more detail, in Chapter 4 a detailed definition of probability will be proposed. My starting point was to find a definition that encompasses the classic interpretation of Laplace and the frequency interpretation in its bare form; so to make explicit what these definitions seem to implicitly contain. The next step was to investigate as many paradoxes of probability theory I could find and to verify whether the detailed definition could solve these.

At this point it is useful to remember that in physics *any* experimental probability is determined, measured, as a relative frequency. Of course theories may predict probabilities as numbers (such as the square of a modulus of a wave function, a transition amplitude, etc.) that are not obviously ratios or frequencies; but to *verify* these numbers one always determines relative frequencies. If the numbers do not correspond to the measured frequencies, the theory is rejected. Therefore one may look for the meaning of the notion of probability in the way it is verified - the essential idea of the verificationist principle of meaning of the logical empiricists. The definition proposed in Chapter 4 goes somewhat further than von Mises' interpretation, but can be seen as fully deriving from it. It appears to also be close to a definition given by Bas van Fraassen I[5] (1980). Recall that T = P ⋃ I, with T = probability theory as a theory 'about the world', P = the axiomatized part, and I the interpretation. Von Mises developed both a calculus P and an interpretation I. However his 'calculus of collectives' is much more complex than Kolmogorov's set-theoretic approach, and is, as far as I know, almost never used anymore (of course it leads numerically to the same results as Kolmogorov's calculus). So we do not use the calculus of collectives; our calculus (P) is Kolmogorov's. But we do follow von Mises for most of his interpretation (I), which is fully lacking in Kolmogorov's theory. A 'collective' is an (in principle infinite) series of experimental results of a probabilistic experiment, e.g. a series of (real) outcomes of a die throw: {1, 6, 2, 3, 3, 4,….}.

The essential points which our detailed frequency model highlights are, succinctly, the following. First it is argued that probability rather belongs to *composed systems* or *experiments*, not to things and events per se. The same event can have a very different probability in different conditions. It appeared useful to identify and distinguish these conditions, and include them in the definition; this is our main point. These conditions comprise primarily 'initializing' and 'observing' / 'measurement' conditions. Equivalently, one may partition a probabilistic system in trial system (the die), initializing system (the randomizing hand), and observer system (the table + our eye). It is argued that probability belongs to this composed system; changing one subsystem may change the probability. Simple as it may be, such a partitioning seems to allow to solve paradoxes, as argued in Chapters 4-5.

It also appears that this model allows to make a strong link between the interpretation of probability and the interpretation of quantum mechanics. Such a link cannot really be a surprise: quantum systems are probabilistic systems. However, some mysterious elements of the Copenhagen interpretation seem to become quite transparent under this link, as explained especially in Chapter 5. One of the most disturbing, at least remarkable, elements of the

[5] Based on a study of how probability is used in physics, van Fraassen presents in his (1980, pp. 190 – 194) a logical analysis of how to link in a rigorously precise way experiments to probability functions. The author gives as a summary of his elaborate model following definition (p. 194): "The probability of event A equals the relative frequency with which it would occur, were a suitably designed experiment performed often enough under suitable conditions." Our model gives in particular a further analysis of what the 'suitable conditions' are. (Bas van Fraassen told us however he is not in favor anymore of his original interpretation of probability.)

Copenhagen interpretation is the role of the ubiquitous ‘observer’ – in other words the ‘measurement problem’. Why does the ‘observer’ causes the collapse of the wave function, whereas other physical systems such as the environment leave the wave function in a state of superposition ? It will be argued (Ch. 5) that a quantum system is not different here from any classical probabilistic system: the probability of any probabilistic system is determined, not by the mind of an observer, but by the ‘observer system’ and its physical interaction with the trial object. If that is true much of the mystery of the collapse of the wave function seems to disappear. By the same token our model for probability will allow to address several recent claims that were made concerning the interpretation of quantum mechanics (Ch. 5). Next, it will be able to interpret a crucial passage in the notoriously difficult answer by Bohr to EPR. Finally – and this I found particularly surprising – it appears that a detailed interpretation à la von Mises allows to better understand the commutation or complementarity relations of quantum mechanics – considered one of its paradigmatic features. In order to calculate a joint probability between two events A and B von Mises had stressed that the collectives for A and B need to be ‘combinable’, i.e. it must be possible to measure them simultaneously. But that mirrors what Bohr claims about physical quantities (operators) A and B: these can only exist simultaneously if their commutator vanishes, $[A,B] = 0$, i.e. if they can be measured simultaneously. As an example, position x and momentum $p_x$ are complementary since $[x, p_x] = i\hbar \neq 0$. Now the idea that a joint probability distribution for A and B only exists if both quantities can be measured simultaneously is not particular to quantum mechanics: it seems to hold for any probabilistic system, at least if interpreted à la von Mises (Ch. 5). From this point of view, quantum mechanics seems to make precise those prescriptions that are already implicit in probability theory *as a physical theory* (of course it does much more than that; but only things that are allowed by probability theory).

It is hoped that Chapters 4-5 provide a more precise idea of what probabilistic systems are. On the basis of the interpretation of probability that will be exposed in these chapters, it seems that a link can be made with the ‘determinism – indeterminism’ debate that was studied in Chapters 2-3. Since this link forms a convenient way to conclude this thesis, I propose to present these final arguments for the hypothesis of determinism in the last Chapter, the Epilogue.

## References.

# Chapter 2

# Bell’s Theorem: Two Neglected Solutions

**Abstract**. Bell’s theorem admits several interpretations or ‘solutions’, the standard interpretation being ‘indeterminism’, a next one ‘nonlocality’. In this article two further solutions are investigated, termed here ‘superdeterminism’ and ‘supercorrelation’. The former is especially interesting for philosophical reasons, if only because it is always rejected on the basis of extra-physical arguments. The latter, supercorrelation, will be studied here by investigating model systems that can mimic it, namely spin lattices. It is shown that in these systems the Bell inequality can be violated, even if they are local according to usual definitions. Violation of the Bell inequality is retraced to violation of ‘measurement independence’. These results emphasize the importance of studying the premises of the Bell inequality in realistic systems.

## 1. Introduction.

Arguably no physical theorem highlights the peculiarities of quantum mechanics with more clarity than Bell’s theorem [1-3]. Bell succeeded in deriving an experimentally testable criterion that would eliminate at least one of a few utterly fundamental hypotheses of physics. Despite the mathematical simplicity of Bell’s original article, its interpretation – the meaning of the premises and consequences of the theorem, the ‘solutions’ left - has given rise to a vast secondary literature. Bell’s premises and conclusions can be given various formulations, of which it is not immediately obvious that they are equivalent to the original phrasing; several types of ‘Bell theorems’ can be proven within different mathematical assumptions. As a consequence, after more than 40 years of research, there is no real consensus on several interpretational questions.

In the present article we will argue that at least two solutions to Bell’s theorem have been unduly neglected by the physics and quantum philosophy communities. To make a self-contained discussion, we will start (Section 2) by succinctly reviewing the precise premises on which the Bell inequality (BI) is based. In the case of the deterministic variant of Bell’s theorem these premises comprise locality and 'measurement independence' (MI); for the stochastic variant they are MI, 'outcome independence' (OI) and 'parameter independence' (PI) [4-6]. These hypotheses lead to the BI, which is violated in experiments. Therefore rejecting one of these premises corresponds to a possible solution or interpretation of Bell's theorem - if it is physically sound. In Section 3 we will succinctly review well-known positions, which can be termed 'indeterminism' (the orthodox position) and 'nonlocality' (in Bell's strong sense), and give essential arguments in favor and against them. We believe this is not a luxury, since it seems that some confusion exists in the literature: popular slogans such as 'the world is nonlocal', 'local realism is dead', 'the quantum world is indeterministic' are not proven consequences of a physical theory, but metaphysical conjectures among others - or even misnomers. It is therefore useful to clearly distinguish what is proven within a physics theory, and what is metaphysical, i.e. what is not part of physics in the strict sense. Actually, all solutions to Bell's theorem appear to be a conjunction of physical and metaphysical arguments.

The first position that will be investigated here in more detail (Section 4), and that is usually termed total or 'superdeterminism' is, although known, rarely considered a serious option (notable exceptions exist [6-11]). The negative reception of this interpretation is based on arguments of 'free will' or conspiracy, which are however heavily metaphysically tainted. We will argue that rejection of determinism on the basis of these arguments is in a sense surprising, since it corresponds to a worldview that has been convincingly defended by scholars since centuries; and especially since it is arguably the simplest model that agrees with the facts. (The Appendix gives a condensed overview of the history of this position, where a special place is given to Spinoza.) Its main drawback however – for physicists – is that it seems difficult to convert into a fully physical theory.

In Section 5 it will be argued a fourth solution exists, which could be termed 'supercorrelation', and which does not have the latter disadvantage – it is essentially a physical model. In order to investigate its soundness, we will study highly correlated model systems, namely spin lattices. It will be shown that in a Bell-type correlation experiment on such lattices the Bell inequality can be strongly violated. Yet these systems are 'local' according to usual definitions [1, 12]. This violation of the Bell inequality will be retraced to violation of MI. It will be argued that a similar 'supercorrelation' may happen in the real Bell experiment. This will lead us to the conclusion that the premises on which Bell's theorem is based, such as MI, OI and PI, are subtle, and that it is highly desirable to study them in realistic physical systems, not just by abstract reasoning.

Two words of caution are in place. The first is that we will not try to elaborate here a realistic hidden variable theory for quantum mechanics, which seems a daunting task. We are concerned with the much more modest question whether such theories are possible. We are well aware that this implies quite some speculation; but in view of the crucial importance of Bell's theorem for physics (and philosophy) such efforts seem justified, especially if arguments can be backed-up by physical models. Second, there exist many highly valuable contributions to the present field, both experimental and theoretical. It would be far outside the scope of this article to review these works, even a small part of them. We could only refer to the texts that were most relevant for the present findings. Let us however start by paying tribute to Bell himself: his texts [1-3] remain landmarks of clarity, simplicity, and precision.

## 2. Assumptions for deriving the Bell Inequality (BI).

For following discussion it will prove useful to distinguish the deterministic and stochastic variant of Bell's theorem. Within a deterministic hidden variable theory (HVT), the outcomes $\sigma_1$ and $\sigma_2$ (say spin) of a Bell-type correlation experiment are supposed to be deterministic functions of some 'hidden variables' (HVs) $\lambda$, i.e.

$$\sigma_1 = \sigma_1(a,\lambda) \text{ and } \sigma_2 = \sigma_2(b,\lambda), \tag{1}$$

where a and b are the left and right analyzer directions. (In the following $\lambda$ may be a set or represent values of fields; the HVs may be split in $\lambda_1$, $\lambda_2$ etc.: all these cases fall under Bell's analysis.) Recall that (1) assumes '*locality*'[6]: $\sigma_1$ does not depend on b, and $\sigma_2$ not on a. In Bell's original 1964 article [1] it is assumed that the mean product $M(a,b) = < \sigma_1.\sigma_2 >_{a,b}$ can be written as

$$M(a,b) = < \sigma_1.\sigma_2 >_{a,b} = \int \sigma_1(a,\lambda).\sigma_2(b,\lambda).\rho(\lambda).d\lambda. \tag{2}$$

In the most general case however the probability density $\rho(\lambda)$ in (2) should be written as a conditional density $\rho(\lambda|a,b)$ [4-6]. Indeed, it is essential to realize that from (2) the BI can only be derived if one also supposes that

$$\rho(\lambda|a,b) = \rho(\lambda|a',b') \equiv \rho(\lambda) \text{ for all relevant } \lambda, a, b, a', b' \qquad \text{(MI)}, \qquad (3)$$

a condition usually termed 'measurement independence' (MI) [5-6,11]. This hypothesis expresses that λ is stochastically independent of the variable couple (a,b), for all relevant values of a, b and λ. There are of course good reasons to suppose that (3) indeed holds. In an experiment with sufficiently rapidly varying analyzer settings [13], creating a spacelike separation between the left and right measurement events, it would seem that the value of λ determining $\sigma_1$ on the left cannot depend on the simultaneous value of b on the right (similarly for $\sigma_2$ and a) – at least if one assumes Bell's relativistic locality (see former footnote). So this argument says that λ cannot depend on both a and b, i.e. that MI in (3) holds as a consequence of locality.

---

[6] According to Bell's original [1], a HVT is local iff 1) the force fields the theory invokes are well-localized (they drop off after a certain distance, therefore (1) can be assumed even in an experiment with static settings); and 2) it does not invoke action at-a-distance, i.e. it invokes only influences that propagate at a (sub)luminal speed, in particular between the 'left' and 'right' part of the experiment. Notice this corresponds to an extremely mild locality condition: any known physical system satisfies it.

Before critically analyzing MI in Sections 4 and 5, let us already observe that it has never been rigorously proven that Bell's locality necessarily implies MI. One may well wonder whether this view captures all cases, and whether MI can be violated even in local systems. Shimony, Horne and Clauser [14] observed that 'measurement dependence', i.e. stochastic dependence of λ on (a,b), could in principle arise if the (values of the) variables λ, a, and b have *common causes* in their overlapping backward light-cones. More generally, it is maybe conceivable that local correlations exist between λ and (a,b) at the moment of measurement which are a remnant of their common causal past; this is a more general variant of the argument in [14] to be discussed in Section 5. The fact is that the counterargument of Shimony et al. against (3) seems to have had little impact in the literature. It has been discussed in a series of articles [14] by Shimony, Horne, Clauser and Bell, as reviewed in [15]. All come to the conclusion that (3) should be valid on the basis of a 'free will' argument. This position seems largely dominant till date. According to this position, if λ would depend on a and b, then a and b should depend on λ, due to the standard reciprocity of probabilistic dependence. But the values of a and b can be *freely* chosen by one or even two experimenters; how then could they depend on HVs λ that moreover determine the measurement outcomes $\sigma_1$ and $\sigma_2$ ? Ergo, MI must hold. However, we will prove in Section 5.1. that this 'free will' argument does not hold.

In sum, assuming (1), the existence of local deterministic HVs, locality, and (3), MI, one derives the Bell-CHSH [16] inequality:

$$X_{BI} = M(a,b) + M(a',b) + M(a,b') - M(a',b') \leq 2 \qquad (4)$$

by using only algebra. Many have summarized that (4) follows from the assumptions of 'HVs and locality' or from 'local HVs'. But, as we showed above, this phrasing is only valid if locality implies MI.

After Bell's seminal work, Clauser and Horne [12], Bell [2] and others extended the original theorem to *stochastic* HVTs. In such a HVT $\sigma_1$ and $\sigma_2$ are probabilistic variables, for which one assumes (instead of (1)) that

$$P(\sigma_1|a,\lambda),\ P(\sigma_2|b,\lambda) \text{ and } P(\sigma_1,\sigma_2|a,b,\lambda) \text{ exist.} \quad (5)$$

If $\sigma_1$ and $\sigma_2$ are stochastic variables one has now (instead of (2)) that

$$M(a,b) = < \sigma_1.\sigma_2 >_{a,b} = \sum_{\sigma_1=\pm1} \sum_{\sigma_2=\pm1} \sigma_1.\sigma_2.P(\sigma_1,\sigma_2|a,b), \quad (6)$$

where $P(\sigma_1,\sigma_2|a,b)$ is the joint probability that $\sigma_1$ and $\sigma_2$ each have a certain value (+1 or -1) given that the analyzer variables take values a and b (all this in the Bell experiment). Assuming (5), the existence of HVs, and exactly the same condition (3) (MI) as before, it follows from (6) that

$$M(a,b) = < \sigma_1.\sigma_2 >_{a,b} = \sum_{\sigma_1=\pm1} \sum_{\sigma_2=\pm1} \sigma_1.\sigma_2. \int P(\sigma_1,\sigma_2|a,b,\lambda).\rho(\lambda).d\lambda. \quad (7)$$

To derive the Bell inequality (4) one has now to make two supplementary assumptions [4-6], usually termed 'outcome independence' (OI) and 'parameter independence' (PI), which are defined as follows:

$$P(\sigma_1|\sigma_2,a,b,\lambda) = P(\sigma_1|a,b,\lambda) \text{ for all } (\lambda,\sigma_1,\sigma_2) \qquad \text{(OI)}, \quad (8)$$

$$P(\sigma_2|a,b,\lambda) = P(\sigma_2|b,\lambda) \text{ for all } \lambda \text{ and similarly for } \sigma_1 \qquad \text{(PI)}. \quad (9)$$

Using (8-9) one derives from (7) that

$$M(a,b) = \sum_{\sigma_1=\pm1} \sum_{\sigma_2=\pm1} \sigma_1.\sigma_2. \int P(\sigma_1|a,\lambda).P(\sigma_2|b,\lambda).\rho(\lambda).d\lambda, \quad (10)$$

from which the same BI as before (Eq. (4)) follows by using only algebra. Note that the original work by Clauser and Horne [12] assumed the so-called 'factorability' condition

$$P(\sigma_1,\sigma_2|a,b,\lambda) = P(\sigma_1|a,\lambda).P(\sigma_2|b,\lambda) \qquad \text{for all } (\lambda,\sigma_1,\sigma_2), \quad (11)$$

which is however simply the conjunction of (8) and (9). Clauser and Horne justified their assumption (11) by stating that it is 'reasonable locality condition'. Let us note that Einstein locality manifests itself in (11) by the fact that $P(\sigma_1|a,\lambda)$ does not depend on b; similarly for $P(\sigma_2|b,\lambda)$ [12, 2]. Since then the factorability condition (11) seems to have become in the literature the *definition* of locality in stochastic systems. (However, even if (11) or OI and PI may be found 'reasonable' in a Bell experiment with spacelike separation between the left and right measurements, one may doubt their general validity – e.g. for the same reasons for which MI may be questioned. We will come back to this point in Section 5.)

Thus, for stochastic HVTs the Bell-CHSH inequality (4) follows from the assumption of MI, OI and PI. Actually, it appears that all known derivations of generalized Bell inequalities are based on assumptions equivalent to (or stronger than) OI, MI and PI [6]. A more generally known phrasing is that the BI (4) follows from the assumption of 'HVs and locality'. But again, here it must be assumed that locality implies MI, OI and PI; an unproven hypothesis.

**3. The obvious solutions to Bell's theorem: Indeterminism (S1) and Nonlocality (S2).**

In the present Section we will have a brief look at well-known positions which may be adopted with respect to Bell's theorem and the experimental results. We do however not claim to review all admissible solutions. This exercise has been done before (see e.g. [2,3,8,10,17]), but it still seems useful to highlight some pitfalls; simply recognizing that *all* solutions to Bell's theorem have both a physical and a metaphysical component will already prove helpful. Let us first summarize the discussion of Section 2 in its most precise and presumably least controversial manner. In the case of deterministic HVTs, we saw that the BI (4) follows from the following assumptions or conditions (C1-C3):

*The existence of deterministic HVs* (see Eq. (1)) (C1)

*Locality* (see footnote Sec. 2) (C2)

*Measurement independence (MI)* (see (3)). (C3)

In the case of stochastic HVTs, the BI (4) is derived based on following hypotheses:

*The existence of stochastic HVs* (see Eq. (5)) (C4)

*Measurement independence (MI)* (see (3)) (C3)

*Outcome Independence (OI)* (see (8)) (C5)

*Parameter Independence (PI)* (see (9)). (C6)

Since in the Bell experiment the BI (4) is violated (e.g. if the particle pair is in the singlet state), and quantum mechanics vindicated, one simply infers that at least one of the assumptions (C1 – C3) must be false; and that at least one of the conditions (C3 – C6) must be false. Rejection of one particular of these assumptions corresponds to one of the admissible interpretations or solutions of Bell's theorem, if it is physically meaningful. If one can legitimately assume that MI, OI, and PI follow from locality then one can resume Bell's theorem in the following condensed way[7]:

*Local HVTs (deterministic or stochastic) are impossible.* (B1)

This is indeed the phrasing that could resume the work of Bell [1-3] and many others since. Since (B1) is so popular we will first have a closer look at it, but it is important to remember that the conditions (C1-C3) and (C3-C6) are a more precise (and more recent) starting point to analyze Bell's theorem [4-6]; they are at the basis of further solutions. Also recall that (B1) does not only apply to Bell's original spin correlation experiment, but to any entangled state of two quantum systems.

In particular, (B1) says that nature, or in any case a broad class of correlated quantum phenomena, cannot be described by a theory that is *both* local *and* more complete than quantum mechanics. Several articles helpful to understand the full scope of (B1) have derived Bell's theorem without *explicitly* invoking hidden variables [18-20]. Instead of hidden variables they assume that the physical properties $\sigma_1$ and $\sigma_2$ measured in Bell's experiment *have an objective value even before the measurement:* a hypothesis generally termed 'realism'. These works have thus led to following popular variant of (B1):

*Local realistic theories are impossible.* (B2)

Since the assumptions leading to the conclusions (B1) and (B2) lead to exactly the same type of mathematical inequality (4), it is only logical to suppose that also (B1) and (B2) are equivalent. This is indeed the case, at least for deterministic HVTs[8]. But since (B1) also includes stochastic HVTs, it is more general and more precise than (B2); (B1) would thus explain what 'realistic' in (B2) really means. Moreover, it should be noted that the term 'realistic' as used by the community of quantum philosophers and physicists might give rise to confusion. In the original meaning of the word, as used by the broad community of philosophers, 'realism' is the hypothesis that *the physical world (e.g. physical properties and their values) exists independently of the human mind* [21-22]. Bohr was doubtlessly a realist in the latter, original sense [23]; but he was as surely a non-realist in the sense used in (B2). According to Bohr, the measurement *apparatus* determines the values of quantum properties $\sigma_i$, not the human mind [24]. For these reasons we will rely in this Section on Bell's original phrasing (B1), excelling in clarity, rather than on (B2).

[7] (B1) takes the (almost ideal) experimental results into account. Here we do not consider certain so-called loopholes linked to the fact that real Bell experiments would not be 100% faithful tests of the theorem (these loopholes seem to become more and more unlikely).

[8] This may be shown as follows. In a deterministic HVT the physical properties $\sigma_1$ and $\sigma_2$ have a value before each instance of measurement (since $\sigma_i$ is determined by, i.e. a function of, $\lambda$); conversely, in a realistic theory $\sigma_1$ and $\sigma_2$ have a value even before measurement and one can, implicitly or explicitly, index that value by a hidden variable or index.

Some authors have concluded from the observation that quantum mechanics rules out 'any' local HVT (understood: both deterministic and probabilistic), that the conflict arises because of the locality condition alone: it would then be proven by Bell's theorem (and the experiments) that quantum systems are necessarily nonlocal[9]. This interpretation of Bell's theorem is widespread (see e.g. [25] and [8] Chap. 6). However, it is crucial to realize that it is not a proven result: it is one possible interpretation (see S2 below), and certainly not the only admissible position (unless one redefines 'nonlocal' of course). Moreover, it appears that the nonlocality that is invoked by these authors is of a strange kind: it cannot be used for superluminal signal exchange [8]. Alain Aspect and Asher Peres, for instance, term the photon pair of the Bell experiment a 'single non-separable' or a 'single indivisible, nonlocal' object ([25], [8] Chap. 6). But does this really help to understand the correlations in the Bell experiment ? In Bell's work nonlocality straightforwardly refers to superluminal interactions or signals; but not so in the latter interpretation of nonlocality. It seems that Peres attempts, in certain texts (e.g. [8] Chap. 6), to explain quantum nonlocality by the intuitive idea that in 'one indivisible object', if one part 'feels' something, any distant part of the same object immediately also 'feels' something. But this is not a description of what happens in any normal solid object, in which an influence (a force) exerted on one part propagates to any other part at subluminal speed.

Obvious interpretations of Bell's theorem, then, are an immediate consequence of (B1).

*S1. The Standard Solution ('indeterminism')*. The orthodox position, probably adopted by a majority of physicists, is to reject Bell's 'hidden variable' hypothesis (C1 and C4). So according to this position *there are no hidden variables completing quantum mechanics* (for the Bell experiment and other entangled systems), *not even in principle*. As argued above, rejection of deterministic HVs amounts to rejecting a special kind of 'realism', more precisely to rejecting the thesis *that the values of the quantum properties $\sigma_1$ and $\sigma_2$ (of Bell's*

[9] Here 'nonlocal' cannot mean 'entangled' because entanglement is of course not proven by Bell's theorem. In information-theoretic texts 'nonlocal' is often synonymous to 'violating the BI', but in our discussion the term is obviously not used in that way.

*experiment) exist even before their measurement*. This is the orthodox position because it has in essence been anticipated by Bohr decades before Bell's discovery [1] – it is part of the original Copenhagen interpretation ! [23-24, 8] Indeed, in 1935 Bohr had dismissed the EPR paradox by invoking this position. As we read it, Bohr's anti-EPR argument [24] can be summarized as follows: measurement brings observables into being through an inevitable interaction with an observing system; if two observables cannot be measured simultaneously, they do not exist simultaneously. Bohr might have similarly argued that the values of the quantum properties $\sigma_1$ and $\sigma_2$ of Bell's experiment do not exist before measurement, are not determined, i.e. Bell's theorem (B1) or (B2) is not valid.

(It seems that there is one small cloud that stains this perfect picture, namely the fact that this argument à la Bohr does not immediately explain why *stochastic* HVTs would be inconceivable. But advocates of S1 could invoke a remarkable result by Fine [26], proving the equivalence between the existence of stochastic and of deterministic HVTs. And we will see further that Bohr might have resorted to a second argument, linked to the 'contextuality' of the Bell experiment, i.e. the importance of considering the whole experimental set-up including the *two* analyzer settings a,b: see e.g. the first footnote in Section 5.1.)

The main arguments in favor of the Standard Position are, it seems to us, the following. First, it saves locality in Bell's sense. And of course, it is part of the Copenhagen interpretation which has proven itself countless times since its conception. Here one may however observe that the Copenhagen interpretation contains theses that have mainly a physical content (such as Born's rule) which indeed are admirably confirmed by experiment, but also metaphysical theses – and rejecting HVs obviously belongs to these extra-physical hypotheses. Rejecting any yet to discover HVs reminds of a slogan as "we talk about what we can measure or calculate; the rest does not exist" – a slogan which summarizes an axiom of the positivist philosophy, by which Bohr may well have been influenced [27, 2]. In sum, in a sense position S1 espouses well the daily practice of mainstream physics; *in a sense* it seems to make extra-physical commitments that are minimal. However, it is not part of physics in the strict sense. Proponents of other positions may very well continue to inquire about the perfect correlations of the EPR/Bell experiment. Probability theory does not prohibit that probabilities (which these correlations are) are considered as resulting from underlying causal mechanisms. The examples of physical systems in which probabilistic behaviour can be retraced to deterministic laws, are countless. And in a different sense, S1 and the Copenhagen interpretation become quite spectacular, and metaphysically heavily loaded. Indeed, S1 can be restated as follows: when, at the moment of measurement of a quantum property σ like spin (in the Bell experiment) one obtains a certain value (+1 or -1), this value is the result of *absolute hazard*, in the sense that it will never be possible, not even in principle, to explain this result. No theory can ever be constructed going beyond the statistical predictions of quantum mechanics (only P(σ) exists). In other words: we are in the presence of events that have no cause; the microscopic world is full of such 'indeterministic' events; and quantum mechanics is the 'final' theory for such events (in the sense stated). (Note that what S1 claims for the properties measured in a Bell experiment, the Copenhagen interpretation claims for *all* quantum properties.)

*S2. The Non-Standard Solution ('nonlocality' in Bell's strong sense)*. The non-standard approach is to conclude from (B1) that locality (in Bell's sense) is violated in nature, i.e. that superluminal influences (forces) exist. Needless to say, this is a speculative, unorthodox solution since it violates relativity theory. However, it is possible in principle ([2], [8] p. 171). A well-known example of a non-local HVT is Bohm's theory [28]. Note that it seems that of all the solutions reviewed here S2 and S4 below are the only ones that could in principle be proven. Also, the superluminal force field needed for S2 could be ultraweak and dynamically enhanced by nonlinear dynamics, as shown in [29]. An ultraweak force field may have, till date, escaped from detection.

It will be no surprise that for many people both S1 and S2 remain unsatisfactory, for the reasons stated. In the next Sections we will investigate two further positions, which aim at avoiding the 'unpleasant' features of S1 and S2.

## 4. First Neglected Solution: Superdeterminism (S3).

That 'total determinism' or 'superdeterminism' (S3) offers a solution to the Bell impasse has been observed by a few physicists, soon after [1], starting by Bell himself [2, 14]. Besides in interesting analyses by authors as Brans [7], Peres [8], 'tHooft [9], Khrennikov [10] and recently Hall [6,11], S3 has until now been considered a completely implausible solution. However, the debate may be unduly unbalanced, as we will argue now.

Whereas S1 and S2 accept Bell's original no-go theorem (B1), total determinism (S3) questions its derivation; or rather, starts from the more precise analysis (C1-C3), and recognizes that MI (C3) is an essential assumption of any derivation of the BI. In short, total determinism assumes 1) that *any* event, property or variable is determined in the sense (1), *including the choices 'a' and 'b' of the analyzer settings* in a Bell experiment; and 2) that there is, when going back in time, a contraction of the 'causal tree' (the collection of all events) to a small space-time region – which is nothing more than the hypothesis of the Big-Bang. If this view is correct, it immediately follows that $\sigma_1$ and $\sigma_2$ must be determined by some cause $\lambda$, but also that a, b and $\lambda$ should have *common causes* (since the 'world cone' converges to say a point). Now, as we saw in Section 2, in that case $\lambda$ may stochastically depend on (a,b) and MI in (3) does not necessarily hold, even in a fully local world. This point has first been made by Shimony et al. [14], it seems. If MI (C3) does not hold, Bell's no-go theorem cannot be derived, i.e. there may be local HVTs reproducing the quantum statistics.

Thus, solution S3 considers the world as 'superdeterministic', in that even our choices of parameter settings are determined and *in principle* linked to virtually all other physical properties through a retracting world cone. Because this position seems in contradiction with a classic conception of 'free will' [15], Bell termed S3 a 'mind-boggling' or 'conspiratorial' option [2]; David Mermin speaks of 'the most paranoid of conspiracy theories' [19]. These negative verdicts seem to have had a lasting influence on the community of quantum physicists and philosophers. But since these arguments are not strictly part of physics, but of philosophy, it seems inappropriate to disconnect the discussion from the most relevant philosophical theories on the matter (in the remainder of this Section and in the Appendix we try to provide at least a rudimentary introduction). In particular, it deserves to be emphasized that above verdicts, presented as patently commonsensical, neglect a worldview that is cogent, well-documented and widespread[10] outside the quantum community. Determinism is at the heart of philosophical debate since millennia. One of the philosophers who in our personal opinion defended determinism best is Spinoza, who put it at the basis of his system [30-31]. According to Spinoza (and a very considerable part of all philosophers having studied the question) determinism is not antagonistic to free will, just to a simple conception of it.

Let us go once more over the arguments. On the one hand it is just normal that Bell [1] started from (implicitly) assuming measurement independence, for at least two reasons. First, it seems that the practice of physics is only possible because separated subsystems can be described by physical parameters (such as $\lambda$, a, b) that belong *only* to particular subsystems, and not to all systems; at any rate, 'systemic' descriptions are customary in physics, if not the only possible ones. More importantly, in the Bell experiment the analyzer positions a and b can be chosen and set by an experimenter. How then could they be determined by other variables - variables that moreover would also fix properties of the particle pairs ? That seems too much of a violation of free will - at least, such is the dominant position in the Bell literature. However, from another point of view, assuming MI is not innocent. Indeed, recall that one of the initial motivations of Bell, just as of Einstein, was to investigate whether quantum mechanics could be made deterministic, as classical theories. Now start, just as Bell, from the hypothesis that certain quantum properties $\sigma$, intervening in the Bell experiment, are deterministic (determined, caused, by yet hidden variables). *If one wants to reason within the simplest, most economic model*, then one should also assume that *all other* physical properties are determined (a worldview with both deterministic and indeterministic quantities needs two categories). Still within this most simple worldview or ontology, human beings are also determined physical beings: their actions can be described – in principle, not in practice – by deterministic properties. In short: all events (or systems), whether of animate or inanimate origin, are caused by previous events; which are caused in turn by still earlier events, etc.. Now, if one takes the Big-Bang theory into account, it would appear that the idea of a universal causal 'branching' or 'network' between events, originating at the Big-Bang and connecting (almost) 'anything to anything', is a quite natural conclusion. In sum, measurement independence appears to be in contradiction with a simple ontology, the mentioned total or superdeterminism (actually, determinism would suffice as a term). As is well-known, free will and probabilities are explained in this model as 'emerging' due to our unavoidable lack of knowledge of all causal factors. On this view free will is a perception, imparted by our obvious *feeling* to be free – a feeling that surely is immensely functional but still might be an illusion.

[10] Total determinism seems to be a popular philosophy. For what it is worth, here is the result of a little survey we did, one among about 20 physicists, experts of the foundations of quantum mechanics, one among about 20 philosophy students. In both cases, about 40% of participants said to be in favour of determinism, 60% in favour of indeterminism. (In a third group (30 p.), after a defense of determinism, the ratio was rather inversed.) These surveys were casual and have of course no pretension to definitiveness.

So, what seems at first glance, from our daily point of view of free agents, a paranoid conspiracy theory, becomes from another point of view a quite reasonable hypothesis. As recalled in the Appendix, this point of view has been advocated since centuries, and probably millennia, by countless philosophers and scientists. In this context, it seems that terming S3 'conspiratorial' is rash. Authors who call S3 'conspiratorial' seem to do so because they look at this position from following angle: an operator (or an automat) doing the Bell experiment would be determined to set, in function of each particle pair, the polarizers in just these positions so as to make the result coincide with the quantum prediction. But this seems an anthropocentric point of view. Determinism basically only says that every event happens according to deterministic laws – full stop. Then, if we do an experiment and obtain a result obeying quantum laws these events all *must* be determined *and* linked (see above); in other words quantum mechanics is an effective theory (as many others) having at least in principle a subjacent explanation. This seems a direct consequence of logic and the above simple hypotheses, not of conspiracy[11]. As so often, conspiracy is in the eye of the beholder. All depends from which assumptions one starts.

In conclusion, a third position (S3) w.r.t. Bell's theorem is conceivable, namely, in short, the assumption of total determinism, implying a (local) violation of measurement independence (C3). This position saves locality, does not violate any known physical law, and leaves open the possibility to complete quantum mechanics in the EPR/Bell sense, as in Eq. (1) or (5). It points however to a *very different kind of non-locality*, namely a universal connectedness of virtually all systems (including human beings), due to a receding world cone. Of course, if S3 would be true, the truly mind-boggling thing would be that quantum mechanics and Bell's theorem allow us to discover this ancient and universal link between virtually all objects; and to corroborate a millennia old philosophy. We believe that the main argument in favor of superdeterminism is that it corresponds to the simplest worldview, based on the fewest concepts (and should one not adopt the simplest theory agreeing with the facts ?). Its main drawback is that it may be difficult to directly transpose it in a physical theory. (At least this is Bell's position, see his last article in [14]. The reason invoked is that theories that describe both our choices *and* Bell experiments by explicitly exhibiting common parameters are doubtlessly impossible. Other people are not impressed [9].)

We believe however that a solution exists (S4 below) that is essentially physical, i.e. that may be more easily backed-up by a new physical theory.

---

[11] A well-known theory on conspiracy theories [32] proposes following ground for people believing in conspiracy. In short: extraordinary effects call for extraordinary causes. In the face of events or 'coincidences' that are *perceived* as formidable, people would have a tendency to look for formidable explanations: a conspiracy by higher powers (or simply the powerful). Now, exactly this theory [32] might apply to people calling S3 a 'conspiracy theory' (!): they perceive S3 as too formidable to be true, and believe that only higher powers – a conspiracy – can explain what S3 proposes. In this context, see also Spinoza [30], who denounced fallacious reasoning of a quite similar type. He analyzed in particular the case of his contemporary fellows, who, in view of the perfection of the world and the quasi-infinite potential of harmonious interaction it offers, concluded that it surely must have been made for them by a higher power. But according to Spinoza's determinism, *and modern biology*, nature and mankind evolved in a lawful way so as to *necessarily* be in some kind of harmony – no divine plan is needed. In sum: anthropocentric reasoning is widespread and often wrong.

## 5. Second Neglected Solution: Supercorrelation (S4).

### 5.1. Measurement dependence through past interaction (S4a).

For convenience we will collect *two* potential solutions under the term 'supercorrelation': it will be seen they both invoke stronger (or other) correlations between the system variables $\sigma_1$, $\sigma_2$, $\lambda$, a, b than the other positions do. More precisely, we propose here solutions that refute MI in (C3) *but not through superdeterminism*; or OI in (C5).

Needless to say, abstract reflection on HVTs is perilous – we know by definition almost nothing about them, and it is difficult not be guided by some preconceptions. Many people will have very different ideas about what they may look like. In the present Section I am guided by some early attempts to construct realistic HVTs for certain aspects of quantum mechanics, in which the HVs are (values of) *fields*. (As far as I know these are at present the most promising candidates.) For instance, in [33-35] the essential HV is a stochastic zero-point field that imparts Brownian motion to quantum particles, from which on average the standard quantum statistics would emerge. Ref. [34] mentions as one of its sources of inspiration recent and spectacular experiments by Couder et al. [36], in which quantum behavior (e.g. double slit interference) is reproduced by macroscopic particles, namely oil droplets. The latter are excited by an external field (the vibration of an oil bed) imparting Brownian motion to the droplets. There seems to be a common denominator in these theories [33-35] and experiments [36], a kind of 'contextuality', namely the fact that the precise shape of the (zero-point) field ($\lambda$ for us, see further) depends on the 'context', i.e. the boundary conditions of the *whole* experimental set-up including the parameters of all, even remote, detectors. For instance, the experiments [36] impressively show that the wave field guiding a particular oil droplet through slit 1 is determined by the geometry of *both* slits; a feature also present in the HVT for the quantum version of double slit interference of Ref. [34]. If a similar contextual $\lambda$-field would exist in the Bell experiment, then MI in (3) may *not* be satisfied in general – the field $\lambda$ may well depend on (a,b) by a similar mechanism as above – at least if the detector settings are static. Note that this violation of MI (at the moment of measurement) would come about not through common causes between $\lambda$ and (a,b), i.e. through superdeterminism, but through an earlier interaction between the field $\lambda$ and the analyzers (a and b). Let us emphasize that the experiments of Couder et al. show that one would be ill-advised to consider such a contextuality as far-fetched: it has now been proven to exist even in macroscopic systems, *and* to lead to quantum-like behavior, including tunneling and quantization of angular momentum [36].

What if in a Bell experiment a spacelike separation between the left and right measurement events is imposed, as is the case in the most sophisticated experiments [13, 25] ? Then MI seems harder to overcome, but not as untouchable as often believed[12]. First of all, note that these experiments [13, 25] do not really use changing analyzer directions (e.g. rotating analyzers), as Bell repeatedly advocated, starting in his first paper [1]. They achieve spacelike separation between measurement events by using random switches and sufficiently retarding the choice between two *static* analyzer directions (a and a' on the left, b and b' on the right). But then it seems conceivable, within the kind of contextual theories just mentioned, that a hidden field accompanying or describing the particles exhibits four modes each depending on a left and right analyzer direction, so labeled by (a,b), (a',b), (a, b') and (a', b'); and that when the random switches of the experiment chose e.g. (a,b) they select the λ-mode depending on (a,b) (similarly for the 3 other modes) – a resonance phenomenon. In other words, it seems that in this experiment the λ-field may depend on (a,b), i.e. that MI in (3) may be violated, notwithstanding the spacelike separation. As we will see further this argument applies a fortiori to OI. On the above view, it is essential that experiments be done with one polarizer on both sides, each changing its polarization direction rapidly enough (ideally by rotation). This may however be difficult to realize.

Before looking at a model system, I believe it is important to emphasize that 'measurement dependence (through past interactions)' (or 'contextuality') is fully compatible with a precise application of probability theory. As is well known, the interpretation of probability is a surprisingly subtle topic debated among all fathers of probability theory, such as Laplace, Kolmogorov, von Mises etc., and countless scholars since [10]. In recent studies, following von Mises, it has been stressed that 1) probabilities belong to experiments and not to objects or events per se, and 2) that any probability depends *at least in principle* on the 'context' including *all* detector settings of the probabilistic experiment [10, 37]. (This is a position that Bohr held regarding quantum systems, but that arguably holds also for classical probabilistic systems [37]). According to this analysis, then, ρ(λ) in (2) and (7) should in principle be considered as $\rho(\lambda|a,b)$. This seems the essential idea of several papers criticizing Bell's theorem (see e.g. [10, 38-39] and references therein)[13]. In the present paper we focus on physical arguments on how such a measurement dependence can come about.

[12] Note added in the reviewed version of the thesis (Feb. 2014): this case will be treated in detail in Chapter 3. This paragraph looks now dubious to me.

Indeed, it seems essential to demonstrate the above ideas in existing physical systems. Theoretical work on the justification or rejection of MI, OI and PI has until now been restricted to mathematical considerations and information-theoretic toy models (for a recent review, see [6]). As far as we know no realistic physical systems have been investigated. The initial question of this Section was: could there be a local HVT for the Bell experiment that violates MI but that *does not invoke common causal factors for λ, a, b*, so that doesn't rely on superdeterminism ? Although we cannot provide a full-blown (and not ad-hoc) HVT, we can exhibit a physical system, well-known from classical statistical mechanics, that violates the BI, that is local, and in which MI is violated without superdeterminism, i.e. even if we assume free will in the usual sense. That local (and classical) HV systems can strongly violate the BI, is a surprise in itself.

We are looking for model systems that are strongly correlated, since for those it is a priori clear that MI, OI and PI (Eq. (3), (8-9)) may not necessarily hold. We want to perform a Bell-type correlation experiment on such a system and verify whether the BI holds. One of the most studied models in statistical mechanics is the Ising model (originally proposed to investigate ferromagnetic phase transitions in low-dimensional electron systems [40-41]). Its Hamiltonian is:

$$H(\theta) = -\sum_{i,j} J_{ij}.\sigma_i.\sigma_j - \sum_i h_i.\sigma_i. \qquad (12)$$

Here the $J_{ij}$ represent the interaction between spin i ($\sigma_i = \pm 1$) and spin j and are responsible for 'cooperative' behaviour and long-range correlations (positive $J_{ij}$ induce lowering of the energy if the spins are aligned). $J_{ij}$ ranges over first neighbours of N particles (electrons, ions, …) sitting on a D-dimensional lattice. The $h_i$ are local magnetic fields and θ is the N-spin configuration $\{\sigma_1, \sigma_2, \ldots, \sigma_N\}$, occurring according to the usual Boltzmann distribution. Note that this is really a classical system (in quantum versions of (12) the $\sigma_i$ are Pauli matrices, see (17)) but that in some anisotropic magnetic materials, when only one spin direction (z) matters, (12) coincides with the quantum description. Also note that the 'interaction' between the spins is of course not a direct spin-spin interaction; the first term in (12) arises through a Coulomb potential combined with the Pauli exclusion principle[14]. Finally, the $\sigma_i$ do not even need to be spins (they can represent atomic occupation in a crystal or a lattice gas, deviation from equilibrium position in a network of springs, etc.): the Ising Hamiltonian is ubiquitous in physics.

[13] It seems noteworthy that this is an argument that even Bohr might have liked: he often stressed that a quantum system or phenomenon includes the whole measurement set-up, due to the complementary nature of such arrangements.

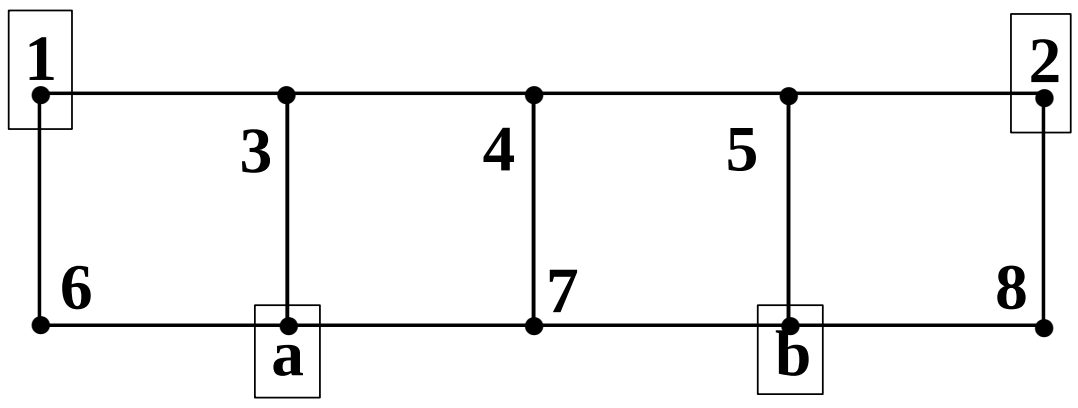

Fig. 1. 10 spins on a square lattice.
Each node contains a spin $\sigma_i$ (i = a,b,1,…,8), which can be up or down.

Let us consider a square lattice (Fig. 1) of N = 10 spins (of electrons, ions,…) interacting with first neighbours only. At given temperature $1/\beta$ a configuration $\theta = \{\sigma_a, \sigma_b, \sigma_1,\dots, \sigma_8\}$ occurs with the Boltzmann probability (or configuration probability) $P(\theta)$ given by:

$$P(\theta) = e^{-\beta H(\theta)} / Z, \text{ with } \beta = 1/kT \text{ and } Z = \Sigma_\theta\, e^{-\beta H(\theta)}, \text{ the partition function.} \qquad (13)$$

It is then well-known and straightforward to calculate that this system shows *full pairwise correlation* in the sense that

$$P(\sigma_i=\varepsilon, \sigma_j=\delta) \neq P(\sigma_i=\varepsilon).P(\sigma_j=\delta) \quad \text{for all } i,j \leq N = 10 \text{ and } \varepsilon, \delta = \pm 1. \qquad (14)$$

As an example, $P(\sigma_i=+1, \sigma_j=-1)$ is calculated as $\Sigma_\theta P(\theta)$ where the sum runs over the $2^8$ 10-spin configurations $\theta$ in which $\sigma_i=+1$ and $\sigma_j=-1$. Since each term in the sum involves the energy $H(\theta)$ given by (12) the calculation is easily done by numerical simulation. Similarly, in the following any probability $P(\eta)$ with $\eta$ an m-spin configuration ($m \leq 10$) is calculated by:

$$P(\eta) = \sum_{\theta(\eta)}^{2^{10-m}} P(\theta), \qquad (15)$$

where the sum runs over the $2^{10-m}$ 10-spin configurations $\theta(\eta)$ that 'contain' $\eta$.

We are interested in producing an analog of the Bell experiment in which quantum probabilities as $P(\sigma_1,\sigma_2|a,b)$ are completed or explained by additional variables $\lambda$, by

[14] The quantum treatment shows that the $J_{ij}$ correspond to the exchange integrals $\int \psi^*_{ab}.V.\psi_{ba}.d^3\mathbf{x}_1.d^3\mathbf{x}_2$, with V the Coulomb potential and $\psi_{ab}(\mathbf{x}_1,\mathbf{x}_2) = \psi_a(\mathbf{x}_1).\psi_b(\mathbf{x}_2)$ (where $\psi_{a(b)}(\mathbf{x}_{1(2)})$ are the single electron eigenfunctions located at $\mathbf{x}_1$ and $\mathbf{x}_2$ respectively) ([41] Chap. 7).

summing over probabilities as in (5) – remember that is what we ask of a stochastic HVT[15]. Consider then an ensemble of 10-spin lattices as in Fig. 1, all at the same temperature, and measure for each lattice the value (±1) of $\sigma_1$, $\sigma_a$, $\sigma_2$, $\sigma_b$ (here $\sigma_a$ and $\sigma_b$ mimic the analyzer variables a and b of the Bell experiment; $\sigma_3, \sigma_4, \ldots, \sigma_8$ can be considered hidden variables)[16]. Measurement on a large ensemble allows to determine the joint probabilities $P(\sigma_1=\varepsilon, \sigma_2=\delta \mid \sigma_a,\sigma_b)$ ($\varepsilon,\delta = \pm1$); these probabilities can be determined for any of the 4 possible couples (a,b) ≡ $(\sigma_a,\sigma_b) = (\pm1,\pm1)$ by postselecting 4 subsensembles from the total run – exactly as in the real Bell experiments. With these probabilities the average product M(a,b) = $< \sigma_1.\sigma_2>_{a,b}$ for the 4 couples (a,b) can then be determined by using Eq. (6). And finally, putting a≡b≡+1 and a'≡b'≡ –1, the quantity $X_{BI}$ in (4) can be experimentally determined (and calculated) for each parameter set ($J_{ij}$, $h_i$) (we always take $\beta = 1$, a common value [40]).

Note that this system is local in the usual sense [1, 12]. It is local in Bell's sense [1] because there is no interaction (and a fortiori no superluminal interaction) between the 'left' and 'right' sides of the lattice, i.e. between $(\sigma_1,\sigma_a)$ and $(\sigma_2,\sigma_b)$. Indeed, the interactions $J_{ij}$ range only over first neighbors ($J_{ij}$ is taken = 0 otherwise). Importantly, it is also local in the sense that the Clauser-Horne factorability condition (11) is satisfied, with $\lambda = (\sigma_3,\sigma_4,\ldots,\sigma_8)$ or any subset thereof, as is easily calculated by using (15). Recall that (11) serves as the usual definition of locality in stochastic systems. Further, it is easy to calculate that in a lattice in which there *is* an interaction $J_{ij} \neq 0$ between some 'left' or 'right' spins (e.g. between $\sigma_1$ and $\sigma_b$), also the locality condition (11) fails to hold [42].

It is then surprising, to some point, that in this Bell-type experiment on a local (and classical) system the BI (4) can be strongly violated. Violation occurs for wide ranges of system parameters $h_i$, $J_{ij}$, as also for a variety of other 1D and 2D geometries we investigated. By numerical simulation one finds e.g. for the values $h_i$=1 (all i), and $J_{ij}$=1.4 (first neighbours), that $X_{BI} = 2.24 > 2$ (the value 2.24 is at least a local maximum). If one allows the magnetic fields $h_i$ to vary over the sites (keeping left-right symmetry), the BI can be violated to a much higher degree. For instance, for $h_1=h_2=h_6=h_8= 1.9$, $h_3=h_4=h_5=h_a=h_b= 0.4$, $J_{ij} = 2.0$ (these are realistic values [40]), $X_{BI} = 2.883$, which can be compared to $2\sqrt{2} \approx 2.83$, the value for the singlet state.

[15] Also, it plays no role whether these λ are classical or quantum-like, a case we will investigate elsewhere [42], as explicitly mentioned by Bell (see [2]).
[16] If one wants to push the analogy with the Bell experiment further, suppose that Alice measures $\sigma_1$ and $\sigma_a$, and Bob $\sigma_2$ and $\sigma_b$.

If the BI is violated in a probabilistic HV system, at least one of the conditions MI, OI, and PI must be violated. Since the factorability (11) holds, OI and PI hold (as can also be calculated independently); therefore the only remaining 'resource' for violation of the BI is measurement dependence. By applying again (15), one indeed immediately finds that

$$P(\sigma_\lambda|\sigma_a,\sigma_b) \neq P(\sigma_\lambda|\sigma_{a'},\sigma_{b'}) \quad \text{for any } \sigma_\lambda \text{ and any } (\sigma_a,\sigma_b) \neq (\sigma_{a'},\sigma_{b'}) \tag{16}$$

where $\sigma_\lambda \equiv (\sigma_3,\sigma_4,\ldots\sigma_8)$ or any subset thereof ($\sigma_i = \pm1$). So MI in Eq. (3) does not hold, as is not really a surprise in this highly correlated system satisfying Eq. (14).

Thus this system indeed offers an example of measurement dependence *without superdeterminism* (without common causes between a, b, λ); so, if one prefers, compatible with a usual conception of free will. If this is not clear already, it can be proven as follows. In the above thought experiment Alice and Bob extract the 4 correlation functions M(a,b) needed to determine $X_{BI}$ from one total run. But if one explicitly assumes that they can *control* the value of $\sigma_a$ and $\sigma_b$, so set these spins to +1 or -1 according to their free choice, and that they perform 4 *consecutive* experiments to determine the M(a,b), then it is clear that these 4 correlation functions are identical to the ones of the first experiment (both cases can of course be explicitly calculated). Therefore also in this second experiment the BI will be violated. In sum, this proves that correlation of λ with (a,b) through local interaction (see (16)) can arise also without assuming superdeterminism. This is important, since as we stated in Section 2, according to the accepted view [14-15, 6] violation of MI is dismissed by observing that it would be a superdeterministic fact, violating free will. This result shows once more that the conditions of probabilistic independence (MI, OI, PI) that are the premises of the BI, should be considered with extreme caution. Of course our model system seems to best apply to a Bell experiment with static analyzers; one may still wonder whether in an experiment with varying analyzer settings MI could be violated. But as we argued in the beginning of this Section, we conjecture that the existing experiments [13, 25] do not exclude that MI is violated[17]. For instance, a resonance phenomenon may create measurement dependence through the interaction of 'something' (λ, say a field) with both analyzers.

In sum, according to this solution (S4a), local correlations at the moment of measurement are a remnant of past correlations persisting in time. Note that correlations that persist in time are ubiquitous in nature. For briefness one could term this position '(local) measurement dependence through past interactions', a variant of what we termed supercorrelation.

[17] Note added in the reviewed version of the thesis (Feb. 2014): this case will be treated in detail in Chapter 3.

### 5.2. Outcome dependence (S4b).

Let us continue to focus on the stochastic variant of Bell's theorem. Then, as always starting from the premises (C3-C6), the next potential solution would be based on refuting outcome independence (C5); an option that would be interesting if it is compatible with Einstein locality. Note that also parameter independence (C6) could be questioned, but it is generally believed that rejecting PI amounts to a strong nonlocal effect ('non-local signaling') [4-6]. Let us now argue that OI could be violated in local physical systems. First, locality is usually supposed to imply the factorability condition (11) and OI; but recall that it has never been proven that Bell's and Einstein's locality [1] necessarily imply OI. Are there systems that are local in Bell's sense [1] (footnote Sec. 2) but violate OI ? Closer inspection of the definitions of MI, OI and PI in (3), (8) and (9) shows that there is a relevant difference between MI and PI on the one hand, and OI on the other. As we saw in detail above, there is some justification in accepting MI and PI in experiments with spacelike separation (but see our counterarguments above). Both MI and PI involve conditional probabilities in which one of the conditioning parameters (a or b) is supposed to be irrelevant because it is chosen in a space-like separated part of the experiment. However, OI invokes a very different kind of conditional independence: here it is $\sigma_1$ that is supposed to be irrelevant for $\sigma_2$ or v.v. (given $\lambda$,a,b). But then the argument justifying MI and PI is not helpful for OI: correlation between $\sigma_1$ and $\sigma_2$ may very well be independent of the spacelike separation between the choices of a and b. Next, in the case of the singlet state of the Bell experiment, correlation between $\sigma_1$ and $\sigma_2$ (given $\lambda$,a,b) is to be expected *a fortiori*, since $\sigma_1$ and $\sigma_2$ are governed by a conservation law. Such correlation may well exist and persist over time; the time of choice of a,b seems to have nothing to do with it. In sum, it is somewhat mysterious why OI is so generally believed to necessarily hold in local systems. Maybe because one believes that such correlations would get wiped out. But we don't know much about sub-quantum reality, and countless correlations in nature do persist almost indefinitely.

Needless to say, there is still a long way to go from this remark to the construction of a full-blown local HVT that reproduces the quantum correlations (and e.g. violates OI).

However it seems again possible to corroborate above arguments by a Bell-type thought experiment on a known physical system. Indeed, consider again a spin lattice as in Fig. 1 (containing N≥ 10 lattice points), but this time described by the 'quantum Ising' Hamiltonian [43]:

$$\mathbf{H} = -\,J \textstyle\sum_{ij} \hat{\sigma}_{z,i}\hat{\sigma}_{z,j} - h.J \sum_{i}\hat{\sigma}_{x,i}. \qquad (17)$$

This Hamiltonian describes other magnetic materials than those described by (12) [43]. Here the $\hat{\sigma}_z$ and $\hat{\sigma}_x$ are Pauli spin matrices; the first sum runs over all nearest neighbour pairs; and h is a dimensionless coupling constant. As we will show elsewhere [42], performing on this system the Bell-type correlation experiment described in Section 5.1, one finds that the BI can be violated under very broad conditions. (The HV are here the eigenvalues of $\hat{\sigma}_z$ on the sites other than 1,2,a,b [42].) One also finds that *none* of the conditions OI, MI and PI holds here. Yet this system is local in Bell's original sense [1], as any known real system is (the interactions J are truncated after first neighbours and there is obviously no superluminal interaction between parts). Note that the system does not even have a symmetry between $\sigma_1$ and $\sigma_2$; in which case violation of OI is expected a fortiori.

Based on above arguments, we conjecture that another solution to Bell's theorem exists, namely 'local outcome dependence' (S4b), which we classified as a form of supercorrelation. Realistic local HVTs should exhibit a dynamical mechanism, e.g. mediated by a field (possibly along the lines of [33-35]), that creates such a supercorrelation and reproduces the quantum correlations.

### 5.3. Supercorrelation as an intermediate position between Bohr's and Einstein's.

Let us end this Section with a brief remark, linking in a way the four positions S1-S4. From large parts of what we said above, it is clear that both superdeterminism and supercorrelation point to a causal 'connectedness' between particles (objects) and their space-time environment. Superdeterminism includes 'everything' in the environment, supercorrelation only typical experimental arrangements. That is why the latter seems the more physical solution. As we already noted in Sections 3 and 5.1, this reminds us of the well-known 'contextuality' or 'holism' of the Copenhagen interpretation and of Bohm's theory; hence the link between the four positions. Interestingly, this connection between S1-S4 seems corroborated by what Einstein [44] maintained about the EPR paradox, long after its publication. Here is what the great man said (quoted by Bell in [2]):

"If one asks what, irrespective of quantum mechanics, is characteristic of the world of ideas of physics, one is first of all struck by the following: the concepts of physics relate to a real outside world. […] It is further characteristic of these physical objects that they are thought of as arranged in a space-time continuum. An essential aspect of this arrangement of things in physics is that they lay claim, at a certain time, to an existence independent of one another, provided these objects are situated in different parts of space. The following idea characterizes the relative independence of objects far apart in space (A and B): external influence on A has no direct influence on B. […]"

"There seems to me no doubt that those physicists who regard the descriptive methods of quantum mechanics as definitive in principle would react to this line of thought in the following way: they would drop the requirement […] for the independent existence of the physical reality present in different parts of space; they would be justified in pointing out that the quantum theory nowhere makes explicit use of this requirement. I admit this, but would point out: when I consider the physical phenomena known to me, and especially those which are being so successfully encompassed by quantum mechanics, I still cannot find any fact anywhere which would make it appear likely that [that] requirement will have to be abandoned. I am therefore inclined to believe that the description of quantum mechanics […] has to be regarded as an incomplete and indirect description of reality, to be replaced at some later date by a more complete and direct one."

In a sense, then, supercorrelation is a position intermediate between Einstein's and Bohr's (or at least the position that Einstein attributes to faithful fans of the Copenhagen interpretation): it assumes that there indeed may be an interdependence between physical realities in far apart places; at the same time it aims at explaining this interdependence in a 'realist's' way à la Einstein.

In sum, the four interpretations S1-S4 have a striking element in common (they all deal with the connectedness of things); but they favor very different physical theories and very different philosophies to explain this connectedness.

**6. Conclusion**.

We reviewed here some of the admissible interpretations of Bell's theorem, paying attention to some lesser-known aspects. The most precise starting point to look for solutions to Bell's theorem are the premises (C1-C3) and (C3-C6) which are the minimal assumptions to derive the BI, in respectively a deterministic and stochastic setting [4-6]. Besides the orthodox interpretation (indeterminism) and a well-known non-standard solution (nonlocality, i.e. the existence of nonlocal influences in Bell's strong sense), we investigated two rather neglected solutions, termed here 'superdeterminism' and 'supercorrelation'. Superdeterminism rejects MI 'through common causes between λ and (a, b)'; supercorrelation rejects OI or MI 'through past interaction'. *All* these solutions have physical and metaphysical components; they *all* have something mind-boggling about them. Superdeterminism is often considered implausible by the community of quantum physicists and philosophers [2, 14-15, 19], but strictly on the basis of extra-physical arguments linked to 'free will' or conspiracy. These criticisms can be questioned simply because they are metaphysical. It was argued that superdeterminism is a solution that does not violate any known physical law, that is based on the simplest ontology, and that has therefore a strong philosophical appeal. It was also shown that supercorrelation has the additional advantage that it is more easily backed-up by physical arguments.

Indeed, we emphasized the importance of investigating the conditions MI, OI and PI in realistic physical systems, since it appears extremely difficult to assess their validity by logic or mathematics alone. It was shown that in a correlation experiment on certain spin lattices the Bell inequality can be strongly violated; yet these systems are local according to usual definitions [1,12]. This surprise was explained by the fact that these systems violate MI. They do so even if one assumes free will; one almost always supposes that MI must hold because of free will [14-15]. Consequently we argued that violation of MI through local interaction cannot be excluded on the basis of existing Bell experiments.

In conclusion, the Ising lattices, the early HVTs for double-slit interference we mentioned [33-35], and the remarkable experiments by Couder et al. [36] all have something in common, namely a strong correlation between all system variables. Since in such strongly correlated systems the BI may be violated, we hope that other such systems will be investigated.

The ultimate goal of this program would be to devise a realistic HVT for the singlet state, and beyond. Until the day such a genuine HVT explaining quantum mechanics will be confirmed by experiments, the different interpretations of Bell's theorem are likely to remain with us. In the meanwhile it seems best to take indeterminism (S1) not for more than what it is, namely one of several possible hypotheses – not an axiom of a physics theory. Other interpretations just pursue a usual scientific program, namely to explain striking probabilistic behavior, here the EPR-Bell correlations.

Acknowledgements. I would like to thank, for many stimulating discussions, Gilles Brassard, Mario Bunge, Emmanuel M. Dissakè, Henry E. Fischer, Yvon Gauthier, Gerhard Grössing, Andrei Khrennikov, Marian Kupczynski, Jean-Pierre Marquis and Eduardo Nahmad.

## Appendix. Determinism in the history of philosophy; Spinoza's system.

In particular concerning the issue of determinism discussed in Section 4, there is a strong interrelation between physics and philosophy. Rejecting (total) determinism (S3) amounts, in a sense, to a dramatic discontinuity in the history of western thought (which is of course not a proof of determinism). The following is a highly condensed and selective introduction to the history of this position, an introduction unfortunately but inevitably biased by personal preference.

Actually, consulting general encyclopedia of philosophy would suffice to see that determinism and free will belong to the most hotly debated topics of philosophy. Virtually all well-known philosophers – and countless scholars from other fields - have written about the topic [21, 45]. The debate is millennia old: among the first known western philosophers who defended determinism were Leucippus and his pupil Democritus (5$^{th}$ cent. BC), who stated that *everything happens out of necessity, not chance*. Democritus, often called the 'Father of Science', is most famous for having elaborated a detailed and incredibly modern-looking atomic theory. It is, in this context, a fascinating question to inquire on what basis Democritus could conjecture so precociously the existence of ultimate and indivisible constituents of matter, governed by laws. According to Sextus Empiricus he did so on the basis of empirical observations, such as the fact that certain substances dissolve in water in constant ratios, that certain physical and biological components degrade but also regenerate, etc.; but also of theoretical principles - namely the principle of determinism (everything has a cause) and the related idea of 'nihil ex nihilo' (nothing comes from nothing). The latter idea has been retraced to Parmenides (6$^{th}$ cent. BC), but might be much older, since it was generally accepted in Greek antiquity. It is not exaggerated to say, we believe, that these ideas are among the very few founding postulates of science and philosophy.

Since the ancient Greeks, some of the philosophers who defended determinism (understood: total determinism) were Avicenna, Spinoza, Leibniz, Locke, Kant, Schopenhauer, Laplace, Russell, Einstein, S. Hawking - to name a few. Needless to say, the views of these philosophers on determinism may differ in certain respects; but the key ingredient clearly remains. As already mentioned, the majority of philosophers who believed in determinism did not believe in 'free will' in the usual sense (but not all of them, see [45]). Aristotle was one of the early advocates of indeterminism (the action of irreducible chance or randomness). Leibniz' name is forever linked to his celebrated 'principle of sufficient reason', stipulating that everything must have a reason. Kant famously elected the thesis that all events have a cause one of his 'synthetic a priori principles'.

In the above list I believe Baruch Spinoza (1632 – 1677) deserves a special mention. I find Spinoza's defense of determinism particularly attractive and powerful, since he puts determinism at the very basis of a systematic theory of the world and of human action [30-31]. (Needless to say, strong philosophical theses can in general not be proven, but they can acquire cogency if they are part of full-blown theories that explain things. The more the theory explains, the more convincing the founding premises are.) In Spinoza's principal work, the Ethics, constructed as a deductive system based on axioms, determinism is omnipresent from the start [30-31]. Moreover, it is simple and radical. One typical example is Spinoza's Proposition 29 of the Ethics, Part 1: "Nothing is fortuitous in Nature; everything is determined by the necessity of Nature to exist and produce effects in a given manner." In other places Spinoza illustrates this thesis by stating that every individual action of any human being is as determined, as necessary to happen, as it is necessary that the sum of the angles of a triangle is 180°. In sum, within Spinoza's philosophy human free will is an illusion, or rather, should be redefined (which however does not bring Spinoza to fatalism, but to a wonderful pro-active ethical theory).

Even this very selective review will illustrate that 'measurement independence', a necessary assumption of all Bell theorems, cannot be considered trivial, as is so often done. In many well-known, popular, and solid ontologies, such as Spinoza's, it does not hold. As already said, in one sense the indeterminism of S1 represents a formidable discontinuity in the history of science. S1 implies that a property σ acquires during measurement a certain value (+1 or -1) based on no 'reason' (cause) whatsoever; and that no theory ever will be able to provide such a reason (i.e. new physical parameters of which σ will appear to be a function, or that determine the probability P(σ)). But the history of scientific discovery is the history of finding explanations of phenomena that are only random at first sight. S1 says: we can stop our search for explanations here, forever. And yet, S1 may be the right interpretation. The moral is inevitable: a good dose of agnosticism seems in place.

As a last remark, notice that for a true determinist also supercorrelation (S4) can be considered compatible with the principle of determinism, the hypothesis that everything has a cause. Indeed, suppose that a theory would exist agreeing with supercorrelation. If that theory would predict some probabilities then these may of course be supposed to result from hidden causes not part of the theory. Probability theory does not prohibit such an assumption. Indeed, one of its fathers, Laplace, believed that any probability is only a tool we need because of our ignorance of hidden causes. And the examples in physics in which probabilistic behavior can very well be retraced to deterministic laws, are countless.

Thus full determinism (S3) remains possible *at least as a philosophical theory*. S4 is (more easily) subject to scrutiny as a part of a physical theory. And indeed, some will find that S3 or S4 explain the 'connectedness' of things in a less mysterious way than S1, the Copenhagen interpretation, which essentially just accepts it.

From the point of view of philosophy, one further remarkable point is that S3 offers a solid scientific basis for theories such as Spinoza's; a possible link with oriental philosophies will not have escaped from the attention of experts. We refer to Spinoza's work [30] to remind the reader that there a subtle view on free will and human interaction is exposed. On a very personal note, it may therefore be utterly relevant for the formidable problems western society faces: the latter is based on a belief in the virtually unrestricted freedom of the individual. Which may be too simple a picture.

# Chapter 3

# On the possibility that background-based theories complete quantum mechanics

*Abstract*. We investigate the possibility that sub-quantum theories based on a background field complete the quantum predictions of the Bell experiment. It appears that such 'background-based' theories violate one of the premises of Bell's theorem, namely 'measurement independence'; Bell's no-go result therefore does not apply to them. Spin-lattices are one model that can serve as an example of such background-based theories, as already shown in former work by numerical simulation. Here we first present analytical results on 1-D and 2-D spin-lattices. Then we show that measurement independence can be violated in a much broader context than spin-lattices, also in the most advanced, dynamic Bell experiments. Finally we argue that it might be possible to test such background-based models by a straightforward extension of existing experiments.

## 1. Introduction.

The question of completing quantum mechanics by introducing additional variables explaining the quantum probabilities, is posed in a particularly clear manner by Bell's theorem (Bell [1964, 1981]). Bell's Inequality (BI) is derived from a small number of assumptions or conditions, which are assumed to describe any reasonable local hidden-variable theory (HVT) completing quantum mechanics. A detailed analysis of the stochastic version of Bell's theorem was provided by Jarrett ([1984]). He derived a set of conditions which entail the BI, and which we will term here 'outcome independence' (OI), 'parameter independence' (PI) and 'measurement independence' (MI) following a terminology of Shimony ([1986]). Let us define them as follows, staying close to Bell's original notation:

$$P(\sigma_1|\sigma_2,a,b,\lambda) = P(\sigma_1|a,b,\lambda) \text{ for all } (\lambda,\sigma_1,\sigma_2) \qquad \text{(OI)}, \qquad (1a)$$

$$P(\sigma_1|a,b,\lambda) = P(\sigma_1|a,\lambda) \text{ for all } \lambda \text{ and similarly for } \sigma_2 \qquad \text{(PI)}, \qquad (1b)$$

$$\rho(\lambda|a,b) = \rho(\lambda|a',b') \equiv \rho(\lambda) \text{ for all } (\lambda,a,b,a',b') \qquad \text{(MI)}. \qquad (1c)$$

Here $\rho$ is the probability distribution of the hidden-variable set $\lambda$; a and a' are values of the left analyzer angle ($\alpha$), and b and b' of the right analyzer angle ($\beta$); and $\sigma_1$ and $\sigma_2$ are the left and right measurement results, say spin values. Most authors consider OI, PI, MI to be a minimal set of premises from which the BI follows (see e.g. Hall [2011]). OI and PI are generally believed to follow from locality; in conjunction they form the factorizability condition for local HVTs first proposed by Clauser and Horne ([1974]). A detailed analysis of both these conditions can be found in (Butterfield [1992], Dickson [2007]).

The third condition, measurement independence, seems to have attracted less scrutiny. Only few researchers have questioned whether MI really must hold in all local systems, both in the philosophy and physics communities; but a few authors have (cf. e.g. Khrennikov [2008], Weinstein [2009], Nieuwenhuizen [2009], Hall [2010, 2011]). The status of inviolability of MI seems essentially based on an argument of 'free will' or 'no conspiracy'. First note that a mathematically more precise notation for an expression as $\rho(\lambda|a,b)$ is:

$$\rho(\lambda|a,b) \equiv \rho_\Lambda(\lambda|\alpha=a,\beta=b), \qquad (2)$$

with $\rho_\Lambda$ the conditional probability density of the variable set $\Lambda$[18]. The set of variables $\Lambda$ (taking values $\lambda$) may contain discrete or continuous variables; we might split them up in 'left', 'right' or common variables; all such cases are subsumed in (1c). Now MI is a condition of stochastic independence. It is deemed 'obvious' because violating it would mean that the HVs $\lambda$ depend on (a,b), which means by standard rules of probability calculus that the analyzer angles (a,b) depend on the HVs $\lambda$. But (a,b) can be freely or randomly chosen in experiments – so how could these angles depend on the $\lambda$ (variables which moreover determine the probabilities for the left and right outcomes) ? Ergo, MI must hold, unless one accepts a conspiratorial world. The latter conclusion was reached quite early, in the beginning of the debate on Bell's theorem (cf. the review by Shimony et al. [1993], in particular the debate between Bell and Shimony on the matter, leading to the now standard view).

But what if the $\lambda$ would be stochastic variables describing a background medium in which the particles move, and which interacts with the particles and the analyzers ? If we split the $\lambda$ in 'left' and 'right' variables ($\lambda_1$ and $\lambda_2$; in general these are sets), where $\lambda_1$ describes the medium in the close environment of the left analyzer, and similarly for $\lambda_2$ on the right, then one suspects that $\lambda_1$ may very well be correlated with a and $\lambda_2$ with b. Such correlation could arise, it seems, simply due to the physical interaction of the analyzers with their local environment, here the surrounding '$\lambda$-medium'. In other words,

$$\rho(\lambda_1,\lambda_2|a,b) \neq \rho(\lambda_1,\lambda_2|a',b') \text{ in general, i.e. for some values of } (\lambda_1,\lambda_2,a,b,a',b'). \qquad (3)$$

[18] Moreover we will designate the variables by $\lambda$ instead of $\Lambda$, as is also common practice.

A similar suspicion has already been expressed, namely by Butterfield in ([1992], p. 58). Butterfield briefly considers the case that the $\lambda$ are not associated with the state of the particles at emission (as usual) but with the state at impingement on the analyzers, just before measurement. The author states: "The only doubt we might have about this second way of thinking of $\lambda$ is that [the probability measure] pr's independence of a and b is more questionable. One cannot expect the pair's state, and so its measure pr, to be unaffected by the settings once the particles interact with their apparatus. And when exactly does this interaction begin? When is 'just before the measurement'?" (Butterfield [1992], p. 58; the measure pr corresponds to $\rho$ above). Here we will take this remark seriously and go one step further: we will associate the $\lambda$ not with the particles but with a medium in the close neighbourhood of the analyzers. This applies better to a case study involving spin-lattices we recently presented elsewhere (Vervoort [2013]), as recalled below. Of course, associating the $\lambda$ with a background medium rather than with the particles may not have been Bell's initial intention[19]. But that is not really important for us: we are interested in the possibility of local HV models completing the Bell probabilities – i.e. models that allow to calculate the experimental probabilities $P(\sigma_1,\sigma_2|a,b)$, $P(\sigma_1|a)$ etc. by summing over probabilities as $P(\sigma_1|a, \lambda)$ etc.

Eq. (3) amounts to violation of MI, in any model that contains the variables $\lambda_1$ and $\lambda_2$. Since MI is a premise of the BI, Bell's no-go result would trivially not apply to any HV model satisfying Eq. (3). There is therefore an obvious interest in investigating whether a general physical framework or system exists that violates MI as envisaged in (3), in particular one that does so *by relying only on local interactions* – bona fide physical forces that are Lorentz-invariant and spatially localized. Clearly, violation of MI is not in itself sufficient for violation of the BI and reproduction of the quantum probabilities of the Bell experiment. But it is important to note that Hall has recently shown that violation of MI (or 'measurement dependence', MD) is the strongest 'resource' for reproducing the quantum probabilities – stronger than violation of OI or PI (Hall [2010, 2011]). And mathematical and information-theoretic models have been constructed that reproduce the quantum probabilities based on violation of MI only, first by Brans ([1988]), then by others (for recent reviews, see Hall [2011], Di Lorenzo [2012]).

[19] At least not in his first paper ([1964]), where he explicitly associates the $\lambda$ with the particle pair. However, in the considerably more recent paper (Bell [1981]), where he derives the BI in a stochastic setting, Bell explicitly states that he aims now at a much more general framework (cf. p. 52). It seems quite clear that now the $\lambda$ do not need to describe the particles (only). Bell says e.g. that "$\lambda$ denotes any number of other variables that might be relevant" (p. 55), and on p. 56: "It is notable that in this argument nothing is said about the locality, or even localizability, of the variable $\lambda$" (Bell [1981]).

Thus in the following we evidently not reject Bell's theorem as a mathematical result, which obviously follows from its premises. But we will argue that at least one of these premises, namely MI, is not a reasonable assumption for a wide class of local HV models *if the HVs describe a background medium*.

In a former article (Vervoort [2013]) we showed that in spin-lattices, described by the usual spin-1/2 Ising Hamiltonian, MI can be violated[20]. We used numerical simulation to obtain these results, as is the usual method for 2-D lattices (see e.g. Yeomans [1992]). In a Bell-type thought experiment on spin-lattices two spins ($\sigma_1$ and $\sigma_2$) take the role of the spins of the Bell-particles in the original Bell experiment, and two spins ($\sigma_a$ and $\sigma_b$) take the role of the analyzer angles (a and b). Now, the intermediate spins can be considered as forming a stochastic medium that interacts with the Bell-particles and the analyzers. Violation of MI in spin-lattices is therefore an instance of Eq. (3); in other words spin-lattices seem a simple physical realization of the 'background-based' HV models envisaged in Eq. (3). We further showed in Vervoort [2013] that also the BI may be violated in spin-lattices, for certain lattice configurations and parameter values. Yet spin-lattices are local in the sense that they are based on a Hamiltonian that explicitly only includes local, near-neighbour interaction. It seems therefore not surprising that we found that the lattices we investigated satisfied the Clauser-Horne factorizability condition (the conjunction of OI and PI), the usual locality criterion for stochastic systems. Thus violation of MI appears to be a sufficient resource for violation of the BI in these systems.

The present paper has three main purposes. First we will present analytical results on 1-D and 2-D spin-lattices. This appears not a luxury since reactions of readers have convinced us that the above results remained puzzling to many, and that clarification and indeed proofs are necessary. Analytical results will deliver some precise new insights (and confirm the results of (Vervoort [2013])). Next we will extend some of the claims of the latter article to a framework that is much broader than spin-lattices, and that applies besides to static experiments also to dynamic ones. Finally we will propose an experiment to test the models considered here. In some detail, the article is organized as follows. In Section 2 we will calculate MI, OI and PI analytically for 1-D and 2-D spin-lattices. Next, we will prove that violation of MI in spin-lattices is fully compatible with free will, and should not be considered a form of 'superdeterminism': this also appears a matter of calculation. However the most important result concerns the extension of results of our [2013] to the case of *dynamic* experiments (Section 3). In the most relevant Bell experiments the analyzer directions are quickly and randomly switched, with the aim of 'imposing' locality, more precisely OI and PI. Now spin-lattices can only serve as a model for, or an illustration of, background-based theories that apply to *static* experiments, as shown in Section 3. If background-based theories want to be credible candidates for completing quantum mechanics, then they should also apply to dynamic experiments and be capable, as a minimal criterion, to violate MI in these experiments. In Section 3 we will argue they can; we will claim that violation of MI appears to be possible basically for *any* background-based theories *even* in the most advanced experiments. To this end, we will have to analyze the sophisticated experiment performed by Scheidl et al. [2010]. To the best of our knowledge the latter experiment is the only one that aims at imposing *simultaneously* OI, PI and MI. In Section 4 we will suggest that it might be possible to test this class of background-based theories by a straightforward extension of existing experiments. Finally, we will make a link with existing models that invoke, it seems, a background medium (cf. e.g. De la Peña and Cetto [1996], Bacciagaluppi [1999, 2005], Khrennikov [2011], Grössing [2012] and references therein).

[20] In this reference we presented an overview of the possible solutions to Bell's theorem, and argued that the stochastic variant of the theorem might admit solutions based on correlations that are stronger than allowed by OI, PI and MI (a generic solution we termed 'supercorrelation'). Here we will propose a physical mechanism by which such strong correlations can occur, namely the interaction of the Bell particles and analyzers with a background.

It may be instructive to say a word about the context of the present investigation. We actually started our research by investigating whether the Clauser-Horne factorizability condition always holds in realistic physical systems (that can serve as HV models for a Bell experiment). Now, it is well-known that there is a close link between the Clauser-Horne condition and Reichenbach's common cause principle (see e.g. Hofer-Szabó, Rédei and Szabó [1999, 2002]). And some authors have expressed the suspicion that Reichenbach's common cause principle may be violated in *strongly correlated* systems, such as flocks of birds (Arntzenius [2010]). A natural question is then: what about strongly correlated systems from statistical mechanics ? Maybe the most studied system of statistical mechanics is the Ising spin-lattice (cf. any book on statistical mechanics or phase transitions, e.g. Feynman [1988], Yeomans [1992]). The Ising-lattice is an archetypical example of a highly correlated system in which long-range collective phenomena, such as phase transitions, occur. Such systems seem therefore, a priori, interesting testing-grounds to study the common cause principle and the conditions OI, PI (the Clauser-Horne condition) on which Bell's theorem is based. To our surprise, it appeared that not the Clauser-Horne condition is violated in these spin-lattices, but MI.

## 2. Spin-lattices

### 2.1. Spin-lattices and the Bell inequality

Let us first consider a thought experiment on a 2-D spin-lattice; the lattice is a variant of one investigated numerically in (Vervoort [2013]). Suppose Alice and Bob perform a Bell-type experiment on an ensemble of spin-lattices as schematized in Fig. 1.

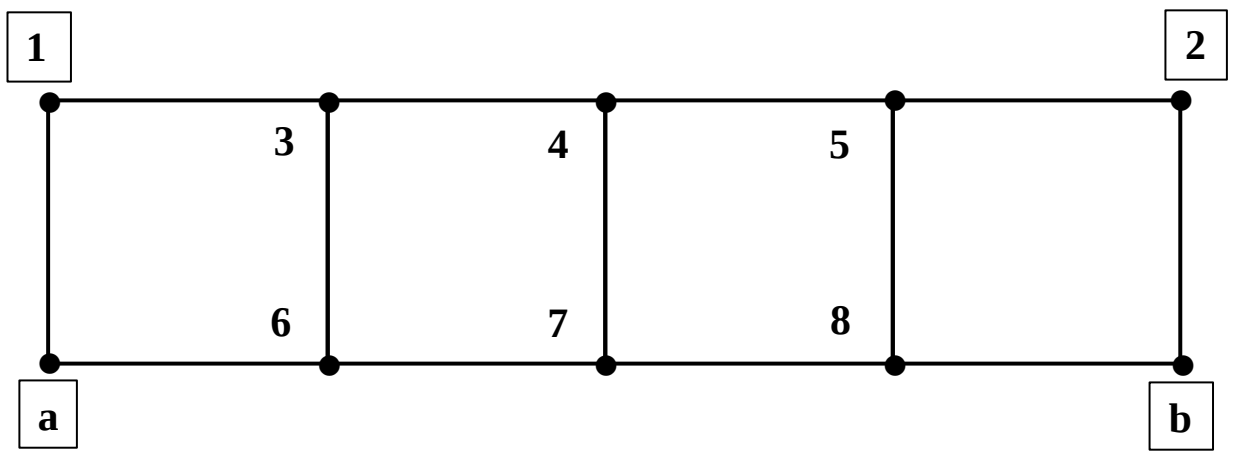

FIGURE 1. 10 spins on a lattice

10 ions or electrons sit on a lattice, each having a spin $\sigma_i$ (=±1 for i =a, b, 1,…, 8). Such two-dimensional lattices are an approximate description of a variety of systems that exist in nature; the archetypical case occurs in magnetic layers (cf. Yeomans [1992]). Suppose the system Hamiltonian is the usual Ising Hamiltionian:

$$H(\theta) = -\sum_{i,j} J_{ij}.\sigma_i.\sigma_j - \sum_i h_i.\sigma_i. \qquad (4)$$

Here θ is a 10-spin configuration ($\sigma_a,\sigma_b,\sigma_1,\ldots \sigma_8$), the $h_i$ are local magnetic fields, and the $J_{ij}$ are the interaction constants between spin-i and spin-j, as usual assumed to be zero between non-nearest neighbours. In other words, *the Ising model explicitly relies only on local* *interaction between nearest neighbours*[21]. Notice that if $J_{ij}$ is positive, the physically relevant case, then the total energy H in (4) decreases if spin-i and spin-j are aligned, i.e. have the same value. Spins seem to 'feel' whether their neighbors are aligned or not; this leads to collective modes and long-range correlations. It is interesting to note that a Hamiltonian of the form (4) also describes a variety of purely classical phenomena: the $\sigma_i$ can represent atomic occupation in a 'lattice gas' or a crystal, deviation from equilibrium position in a network of springs, etc. (Yeomans [1992]). The probability of a given spin configuration (at fixed temperature 1/β) is the usual Boltzmann probability:

$$P(\theta) = e^{-\beta H(\theta)} / Z, \text{ with } Z = \Sigma_\theta\, e^{-\beta H(\theta)}, \text{ the partition function.} \qquad (5)$$

Assume, then, that Alice and Bob share an ensemble of such lattices, each in thermal equilibrium at the fixed temperature 1/β. Suppose Alice has the means to measure the spin (±1) on nodes 1 and a, and Bob on 2 and b, for each element of the ensemble. They can then empirically determine joint probabilities as $P(\sigma_1,\sigma_2|\sigma_a,\sigma_b) \equiv P(\sigma_1=\varepsilon_1, \sigma_2=\varepsilon_2\ |\sigma_a=\varepsilon_a, \sigma_b=\varepsilon_b)$ (all $\varepsilon_i = \pm1$) simply by sitting together and counting relative frequencies over the ensemble. These 16 probabilities are the only ones needed to verify the BI, with $(\sigma_a,\sigma_b)$ taking the role of (a,b) and $\lambda \equiv \sigma_\lambda \equiv (\sigma_3,\sigma_4,\ldots\sigma_8)$ (or any subset of this set), as recalled below (cf. Eq. (6-7)). For the following it is essential to realize that Alice and Bob can do two equivalent experiments (Ex1 and Ex2) to determine the needed probabilities, exactly as in real Bell experiments. Either (Ex1) they 'postselect' 4 sub-ensembles out of one long run, each sub-ensemble corresponding to one of the 4 possible couples of $(\sigma_a, \sigma_b)$-values. Then they determine $P(\sigma_1,\sigma_2|\sigma_a,\sigma_b)$ within each sub-ensemble by counting relative frequencies. If these exceptional experimenters would have sufficiently sophisticated technological means to *control* $\sigma_a$ and $\sigma_b$, i.e. set $\sigma_a$ and $\sigma_b$ to either +1 or -1 at their free choice, they can do 4 consecutive experiments, each corresponding to a fixed value of $\sigma_a$ and $\sigma_b$ (Ex2). The dynamics of the system (Eq. (4) and (5)) remains unchanged in this second experiment. Therefore all relevant probabilities are identical in both experiments. In view of the importance of this point for the following, we prove it explicitly in Appendix 1. In sum, the value of the BI is the same whether the system evolves fully 'on its own', or whether Alice and Bob freely set two of the spins ($\sigma_a$ and $\sigma_b$).

Indeed, all probabilities just mentioned can be calculated. The Bell Inequality reads:

---

[21] Note that the interaction $J_{ij}$ is of course not a direct spin-spin interaction, which does not exist. The interaction is mediated through a force, in the case of magnetic Ising lattices the Coulomb potential. This is well explained in (Feynman [1988]).

$$X_{BI} = M(a,b) + M(a',b) + M(a,b') – M(a',b') \leq 2. \quad (6)$$

For our experiment:

$$M(a,b) = <\sigma_1.\sigma_2>_{a,b} = \sum_{\sigma_1=\pm1}\sum_{\sigma_2=\pm1} \sigma_1.\sigma_2.P(\sigma_1,\sigma_2|\sigma_a,\sigma_b)$$

$$= P(+,+|\sigma_a,\sigma_b) + P(–,–|\sigma_a,\sigma_b) – P(+,–|\sigma_a,\sigma_b) – P(–,+|\sigma_a,\sigma_b). \quad (7)$$

Thus we can calculate $X_{BI}$ in (6) if we define a≡b≡+1 and a'≡b'≡ –1. In (7) we have:

$$P(\sigma_1,\sigma_2|\sigma_a,\sigma_b) = \frac{P(\sigma_1,\sigma_2,\sigma_a,\sigma_b)}{P(\sigma_a,\sigma_b)} \equiv \frac{P(\eta_1)}{P(\eta_2)}, \quad (8)$$

where $\eta_1$ is a 4-spin configuration and $\eta_2$ a 2-spin configuration. Any probability P(η) with η an m-spin configuration (m≤10) is given by:

$$P(\eta) = \sum_{\theta(\eta)}^{2^{10-m}} P(\theta), \quad (9)$$

where the sum runs over the $2^{10-m}$ 10-spin configurations θ(η) that contain η. P(θ) is the Boltzmann factor in (5), involving the Hamiltonian (4). Thus we find:

$$P(\sigma_1,\sigma_2 \,|\, \sigma_a,\sigma_b) = \frac{\sum_{\theta(\eta_1)}^{2^6} e^{-\beta H(\theta)}}{\sum_{\theta(\eta_2)}^{2^8} e^{-\beta H(\theta)}} . \quad (10)$$

Probabilities (10) can easily be computed by numerical simulation since they just involve sums over Boltzmann terms. They are also analytically tractable if one assumes that all $J_{ij}$ are equal (all $J_{ij}$ = J) and all $h_i$ = 0, as we will now show. Below the sums $\sum_{i,j}$ run over the 13 first-neighbour pairs (i,j) = (1,a), (1,3), (a,6),…, (2,b) as one reads on Fig. 1. In the sum $\sum_{\sigma_3,\sigma_4..\sigma_8}$ all variable spins ($\sigma_3,\sigma_4,\ldots,\sigma_8$) run over the values +1 and -1. The numerator of Eq. (8) or (10) becomes now:

$$Z.P(\eta_1) = \sum_{\theta(\eta_1)}^{2^6} e^{-\beta H(\theta)} = \sum_{\sigma_3\sigma_4..\sigma_8} e^{\beta J\sum_{i,j}\sigma_i\sigma_j} = \sum_{\sigma_3\sigma_4..\sigma_8}\prod_{i,j} e^{\beta J\sigma_i\sigma_j}$$

$$= \sum_{\sigma_3\sigma_4..\sigma_8}\prod_{i,j}[\cosh(\beta J\sigma_i\sigma_j)+\sinh(\beta J\sigma_i\sigma_j)]$$

$$= \sum_{\sigma_3\sigma_4..\sigma_8}\prod_{i,j}[\cosh(\beta J)+\sigma_i\sigma_j\sinh(\beta J)] = \sum_{\sigma_3\sigma_4..\sigma_8}(\cosh(\beta J))^{13}\prod_{i,j}[1+\sigma_i\sigma_j\tanh(\beta J)]$$

$$= \sum_{\sigma_3\sigma_4..\sigma_8}\alpha\prod_{i,j}[1+K.\sigma_i\sigma_j]$$

$$= \alpha(1+K.\sigma_1\sigma_a)(1+K.\sigma_2\sigma_b)\sum_{\sigma_3\sigma_4..\sigma_8}(1+K.\sigma_1\sigma_3)(1+K.\sigma_a\sigma_6)...(1+K.\sigma_5\sigma_2)(1+K.\sigma_8\sigma_b)$$

$$= \alpha(1+K.\sigma_1\sigma_a)(1+K.\sigma_2\sigma_b) \times$$

$$\sum_{\sigma_3\sigma_4..\sigma_8}[1+K(\sigma_1\sigma_3+\sigma_a\sigma_6+...)+K^2(\sigma_1\sigma_3\sigma_a\sigma_6+\sigma_1\sigma_3^2\sigma_6+...)+...+K^{11}\sigma_1\sigma_a\sigma_3^3\sigma_4^3\sigma_5^3\sigma_6^3\sigma_7^3\sigma_8^3\sigma_2\sigma_b]\cdot \quad (11)$$

Here α and K are defined as follows: $\alpha = (\cosh(\beta J))^{13}$ and $K = \tanh(\beta J)$. Grouping the terms in powers of K, one sees that the only non-vanishing terms are those in which all $\sigma_i$ appearing as indices ($\sigma_3, \sigma_4, \ldots, \sigma_8$) are squared. The lowest-order terms in which this happens are $K^3\sigma_1\sigma_3^2\sigma_6^2\sigma_a$ and $K^3\sigma_2\sigma_5^2\sigma_8^2\sigma_b$. These terms correspond to a path linking the nodes 1-3-6-a and 2-5-8b respectively (cf. Fig. 1); the power of K corresponds to the number of steps or segments in the path. To identify all non-vanishing terms, we thus have to count 1) all direct[22] paths linking nodes 1 and 2, 1 and a (and 2 and b), 1 and b (and 2 and a) and a and b; 2) all closed loops (such as 3-4-7-6-3); and all products of such paths that have no segments in common (such as 1-3-6-a and 4-5-8-7-4). This is a straightforward though somewhat tedious procedure leading to:

$$Z.P(\eta_1) = \alpha(1+K.\sigma_1\sigma_a)(1+K.\sigma_2\sigma_b).2^6 \times$$

$$\times \{1+(K^3+K^5+2K^7)(\sigma_1\sigma_a+\sigma_2\sigma_b)+(K^4+3K^6)(\sigma_1\sigma_2+\sigma_a\sigma_b)+$$
$$+(K^6+3K^8)\sigma_1\sigma_2\sigma_a\sigma_b+(3K^5+K^7)(\sigma_1\sigma_b+\sigma_2\sigma_a)+2K^4+K^6\}. \quad (12)$$

Following the same procedure we find for the denominator:

$$Z.P(\eta_2) = \sum_{\sigma_1\sigma_2..\sigma_8} e^{\beta J\sum_{i,j}\sigma_i\sigma_j} = \alpha\sum_{\sigma_1\sigma_2..\sigma_8}\prod_{i,j}[1+K.\sigma_i\sigma_j]$$

$$= \alpha.2^8.[1+\sigma_a\sigma_b(K^4+10K^6+5K^8)+4K^4+3K^6+5K^8+3K^{10}]. \quad (13)$$

Thus we obtain the desired expression for $P(\sigma_1,\sigma_2 | \sigma_a,\sigma_b)$ via Eq. (8), and $X_{BI}$ via (6-7). For instance,

$$P(+,+|+,+) = \frac{(1+K)^2[2K^3+4K^4+8K^5+8K^6+6K^7+3K^8]}{2^2[1+5K^4+13K^6+10K^8+3K^{10}]}, \quad (14)$$

implying that for a lattice with homogeneous interactions $J_{ij} = J = 1$ and $\beta = 1$, $P(+,+|+,+) = 0.95$ (J=1=β are common values used in simulations, cf. Yeomans [1992]). $X_{BI}$ is a combination of such terms, cf. Eq. (6-7); it is a complex expression that can numerically be determined for given parameters (β,J). For instance, for J=1=β, $X_{BI} = -0.667$, a value that satisfies the BI. This complex expression can also be easily simplified in the weak-interaction limit K << 1. In that case one finds using Eq. (12-13) and (6-7): $X_{BI} \approx -2\ K^2$, satisfying the BI.

[22] A direct path is one not intersecting itself.

It is safe to verify formulas as (12-14) by a computer program that computes probabilities directly as sums of Boltzmann terms as in Eq. (10); such a program needs few lines. Moreover, such numerical calculations reveal (cf. our [2013]) that if we introduce varying interaction constants $J_{ij}$ and local magnetic fields $h_i \neq 0$, we can strongly violate the BI, for a wide range of parameter values for β, $h_i$, $J_{ij}$ (we can maintain left-right symmetry in the lattice, such that e.g. $J_{a6} = J_{b8}$). For instance for β = 1, $h_i \in \{-1, 1, 3\}$, $J_{ij} \in \{1, 2, 3, 4\}$ we find that $X_{BI} = 2.87$ at its local maximum[23]. This value may be compared to $2\sqrt{2} \approx 2.83$, the value for the singlet state in the real Bell experiment.

## 2.2. Spin-lattices and Measurement Independence

If the BI is violated, at least one of the conditions MI, OI, PI does not hold. It appears that the resource for violation of the BI in our experiment on the lattice of Fig. 1 is violation of MI, as we will now prove. To verify MI in Eq. (1a) analytically, we can use the same method of calculation as used above, if we again assume all $J_{ij} = J$ and all $h_i = 0$. One finds, for $\lambda \equiv \sigma_\lambda \equiv (\sigma_3, \sigma_4, \ldots \sigma_8)$:

$$P(\lambda \mid a,b) \equiv P(\sigma_3,\sigma_4,...,\sigma_8 \mid \sigma_a,\sigma_b) = \frac{\sum_{\sigma_1\sigma_2} e^{\beta J \sum_{i,j}\sigma_i\sigma_j}}{\sum_{\sigma_1\sigma_2..\sigma_8} e^{\beta J \sum_{i,j}\sigma_i\sigma_j}} = \frac{\sum_{\sigma_1\sigma_2}\prod_{i,j}[1+K.\sigma_i\sigma_j]}{\sum_{\sigma_1\sigma_2..\sigma_8}\prod_{i,j}[1+K.\sigma_i\sigma_j]}$$

$$= \frac{\prod_{i,j\neq 1,2}[1+K.\sigma_i\sigma_j]\sum_{\sigma_1\sigma_2}(1+K\sigma_1\sigma_a)(1+K\sigma_1\sigma_3)(1+K\sigma_5\sigma_2)(1+K\sigma_2\sigma_b)}{\sum_{\sigma_1\sigma_2..\sigma_8}\prod_{i,j}[1+K.\sigma_i\sigma_j]}$$

$$= \frac{2^2.[1+K^2(\sigma_a\sigma_3+\sigma_b\sigma_5)+K^4\sigma_a\sigma_b\sigma_3\sigma_5]\prod_{i,j\neq 1,2}[1+K.\sigma_i\sigma_j]}{\sum_{\sigma_1\sigma_2..\sigma_8}\prod_{i,j}[1+K.\sigma_i\sigma_j]}$$

---

[23] In detail, one obtains $X_{BI} = 2.87$ for following numerical values: $h_1 = 3$, $h_3 = h_4 = 1$, $h_6 = h_a = -1$ (and identical values at symmetric nodes on the right); $J_{1a} = J_{13} = 2$, $J_{36} = J_{34} = 1$, $J_{47} = J_{67} = 4$, $J_{6a} = 3$ (and identical values for symmetric interactions).

$$= \frac{2^2.(1+K^2\sigma_a\sigma_3)(1+K^2\sigma_5\sigma_b)\prod_{i,j\neq 1,2}[1+K.\sigma_i\sigma_j]}{\sum_{\sigma_1\sigma_2..\sigma_8}\prod_{i,j}[1+K.\sigma_i\sigma_j]}. \quad (15)$$

The term in $K^2$ in the numerator (second last line) corresponds to the two paths in which $\sigma_1$ or $\sigma_2$ are squared (namely a-1-3 and b-2-5); the term in $K^4$ to the product of these two paths; there are no other non-zero terms in the sum over $\sigma_1$ and $\sigma_2$. Using (13) for the denominator, we obtain:

$$P(\sigma_3,\sigma_4,...,\sigma_8 \,|\, \sigma_a,\sigma_b) = \frac{(1+K^2\sigma_a\sigma_3)(1+K^2\sigma_b\sigma_5)(1+K\sigma_a)(1+K\sigma_b)\prod_{i,j\neq 1,2,a,b}[1+K.\sigma_i\sigma_j]}{2^6[1+\sigma_a\sigma_b(K^4+10K^6+5K^8)+4K^4+3K^6+5K^8+3K^{10}]}. \quad (16)$$

This clearly implies that MI is violated, except for the trivial case K=0 (i.e. J=0). For instance:

$$P(+++...+ \,|\, \sigma_a,\sigma_b) = \frac{(1+K^2\sigma_a)(1+K^2\sigma_b)(1+K\sigma_a)(1+K\sigma_b)(1+K)^7}{2^6[1+\sigma_a\sigma_b(K^4+10K^6+5K^8)+4K^4+3K^6+5K^8+3K^{10}]}. \quad (17)$$

Thus one immediately sees that P(+++…+ | +,+) ≠ P(+++…+ | –,–), or numerically for β=J=1 (i.e. K = 0.762): 0.973 ≠ 0.0012. Again, numerical calculations can also be done by a computer algorithm that evaluates the sums over Boltzmann terms (first line in (15)).

In conclusion, in the lattice of Fig. 1 MI is always violated, for all non-trivial parameter values of β and J. Extensive numerical simulations have shown that this conclusion remains valid when one introduces different interactions $J_{ij}$ and local fields $h_i$ over the nodes; and that it also holds for other 1-D and 2-D structures. We prove the latter claim for an arbitrary N-spin chain in Appendix 2. The latter calculation shows that MI is only asymptotically satisfied for N = ∞.

To quantify to which degree a HV model for given {h, J} violates MI we use a measure introduced by Hall in [2010, 2011]. We term this parameter for self-explaining reasons 'measurement dependence' (MD):

$$MD = \sup_{(a,a',b,b')} \int d\lambda . \,|\, \rho(\lambda \,|\, a,b) - \rho(\lambda \,|\, a',b') \,|\, . \quad (18a)$$

Here $\sup_{(X)}(Y)$ indicates the maximum value of Y when varying the parameters X over all their values. Thus one sees that MD = 0 is equivalent to MI. We analogously define 'Outcome Dependence' (OD) and 'Parameter Dependence' (PD) (cf. Hall in [2011]):

$$OD = \sup_{(a,b,\lambda)} \sum_{\sigma_1,\sigma_2} |\, P(\sigma_1,\sigma_2 \,|\, a,b,\lambda) - P(\sigma_1 \,|\, a,b,\lambda).P(\sigma_2 \,|\, a,b,\lambda) \,| \quad (18b)$$

$$\text{PD} = \sup_{(a,a',b,\sigma_2,\lambda)} |P(\sigma_2 | a,b,\lambda) - P(\sigma_2 | a',b,\lambda)| \quad (18c)$$

As an example, for the parameter set given in the former footnote, MD = 1.99 (its maximum possible value is 2), indicating that, with these parameters, we are likely close to the absolute maximum for $X_{BI}$ in the model (cf. Hall [2011]).

Recall that while MI is always violated (MD ≠ 0), the BI only is for certain ranges of parameter values. Numerical simulations show that in 2-D structures the tendency is the same as in 1-D structures (cf. Appendix 2): for fixed parameters {h, J}, MD decreases with increasing size of the lattice; in parallel $X_{BI}$ decreases. This is not really a surprise (cf. Hall [2011], Vervoort [2013]): the correlation in the system decreases – which can also be simulated by decreasing the interactions strengths $J_{ij}$.

Note that violation of the BI is not the point of most importance for us, but violation of MI. (As we will see in the next Section, the Ising Hamiltonian is quite generally applicable also to dynamical systems; but the Boltzmann probability only holds for static systems. So we obviously do not claim that a 10-spin lattice is a realistic description of a sub-quantum reality for real Bell experiments – even if it violates the BI.) The essential point is that spin-lattices are an example of physical systems in which MI is violated *without superdeterminism*, i.e. without common causes between the λ and (a,b), so compatible with free will. Recall that in the case of real Bell experiments, it is generally inferred that violation of MI implies that the freely chosen parameters (a,b) would be 'causally determined' by some HV λ - by a superdeterministic or conspiratorial process. It is usually believed this contradicts any reasonable conception of 'free will', and that one therefore should assume that MI necessarily holds in a Bell experiment. However, for the above thought experiment this inference does clearly not apply. Indeed, we have shown above (cf. text above Eq. (6)) that all relevant probabilities of the experiment are identical when the system evolves 'on its own' and when free experimenters intervene on a and b; the proof is given in Appendix 1. So MI and the BI can be violated in spin-lattices in an experiment in which Alice and Bob are manifestly free[24]. The conclusion is that violation of MI in this system does not arise through 'superdeterminism' or 'conspiracy' or 'absence of free will', but through an ordinary

[24] This implies that we also have in these models that P(a,b|λ) ≠ P(a,b|λ') in general, again even if Alice & Bob may choose to set (a,b) in whatever sequence, with whatever frequency, they fancy. The point is that one should not understand this as a manifestation of a causal determination of the freely chosen (a,b) by λ. An infinity of such systems exist. Think e.g. of P(x|T) with x = half-life of a nucleus, T = experimental temperature (suppose that a few discrete values of x and T are sampled). If one performs 1000 experiments measuring x at different T's, P(T|x) ≠ P(T) in general even if one chooses T freely.

Coulomb interaction between neighboring nodes[25], much as envisaged in Eq. (3). We will show in Section 3 that essentially the same mechanism can arise in a dynamical system in which the two Bell particles move and the analyzers are switched.

### 2.3. Spin-lattices and locality

Importantly, in our former article [2013] we had found by numerical simulation that OI and PI are always exactly satisfied, for any parameter set, implying that the system is local in the sense of Clauser-Horne (cf. e.g. Shimony [1986]). Indeed, the conditions OI and PI in (1a-b) are equivalent to the Clauser-Horne factorizability condition:

$$P(\sigma_1,\sigma_2 \,|\, \lambda,a,b) = P(\sigma_1 \,|\, \lambda,a).P(\sigma_2 \,|\, \lambda,b) \quad \text{for all } (\lambda,\sigma_1,\sigma_2). \tag{20}$$

Let us here verify this condition for the case $J_{ij} = J$ and $h_i = 0$, as applied to Fig. 1. One finds, using a result obtained in (15) in the denominator:

$$P(\sigma_1,\sigma_2 \,|\, \lambda,a,b) \equiv P(\sigma_1,\sigma_2 \,|\, \sigma_\lambda,\sigma_a,\sigma_b) = \frac{e^{\beta J \sum_{i,j} \sigma_i\sigma_j}}{\sum_{\sigma_1\sigma_2} e^{\beta J \sum_{i,j} \sigma_i\sigma_j}} = \frac{\prod_{i,j}[1+K.\sigma_i\sigma_j]}{\sum_{\sigma_1\sigma_2}\prod_{i,j}[1+K.\sigma_i\sigma_j]}$$

$$= \frac{\prod_{i,j}[1+K.\sigma_i\sigma_j]}{2^2.[1+K^2(\sigma_a\sigma_3+\sigma_b\sigma_5)+K^4\sigma_a\sigma_b\sigma_3\sigma_5].\prod_{i,j\neq 1,2}[1+K.\sigma_i\sigma_j]}$$

$$= \frac{(1+K\sigma_1\sigma_a)(1+K\sigma_1\sigma_3)(1+K\sigma_5\sigma_2)(1+K\sigma_2\sigma_b)}{2^2.[1+K^2(\sigma_a\sigma_3+\sigma_b\sigma_5)+K^4\sigma_a\sigma_b\sigma_3\sigma_5]}$$

$$= \frac{(1+K\sigma_1\sigma_a)(1+K\sigma_1\sigma_3)(1+K\sigma_5\sigma_2)(1+K\sigma_2\sigma_b)}{2^2.(1+K^2\sigma_a\sigma_3)(1+K^2\sigma_5\sigma_b)}. \tag{21}$$

On the other hand:

$$P(\sigma_1 \,|\, \sigma_\lambda,\sigma_a) = \frac{\sum_{\sigma_2\sigma_b} e^{\beta J \sum_{i,j} \sigma_i\sigma_j}}{\sum_{\sigma_1\sigma_2\sigma_b} e^{\beta J \sum_{i,j} \sigma_i\sigma_j}} = \frac{\sum_{\sigma_2\sigma_b}\prod_{i,j}[1+K.\sigma_i\sigma_j]}{\sum_{\sigma_1\sigma_2\sigma_b}\prod_{i,j}[1+K.\sigma_i\sigma_j]}$$

[25] If one contests this conclusion then one should accept that superdeterminism is a phenomenon that can arise in experiments on ordinary spin-lattices.

$$= \frac{\prod_{i,j\neq 2,b}[1+K.\sigma_i\sigma_j]\sum_{\sigma_2\sigma_b}(1+K\sigma_5\sigma_2)(1+K\sigma_8\sigma_b)(1+K\sigma_2\sigma_b)}{\prod_{i,j\neq 1,2,b}[1+K.\sigma_i\sigma_j]\sum_{\sigma_1\sigma_2\sigma_b}(1+K\sigma_1\sigma_a)(1+K\sigma_1\sigma_3)(1+K\sigma_5\sigma_2)(1+K\sigma_8\sigma_b)(1+K\sigma_2\sigma_b)}$$

$$= \frac{(1+K\sigma_1\sigma_a)(1+K\sigma_1\sigma_3)(1+K^3\sigma_5\sigma_8)}{2.(1+K^2\sigma_a\sigma_3+K^3\sigma_5\sigma_8+K^5\sigma_a\sigma_3\sigma_5\sigma_8)} = \frac{(1+K\sigma_1\sigma_a)(1+K\sigma_1\sigma_3)(1+K^3\sigma_5\sigma_8)}{2.(1+K^2\sigma_a\sigma_3)(1+K^3\sigma_5\sigma_8)}$$

$$= \frac{(1+K\sigma_1\sigma_a)(1+K\sigma_1\sigma_3)}{2.(1+K^2\sigma_a\sigma_3)}. \tag{22}$$

By symmetry we then immediately also have:

$$P(\sigma_2 \,|\, \sigma_\lambda,\sigma_b) = \frac{(1+K\sigma_2\sigma_b)(1+K\sigma_2\sigma_5)}{2.(1+K^2\sigma_b\sigma_5)}. \tag{23}$$

Thus indeed Eq. (20) and OI and PI are satisfied. The fact that the lattice satisfies Clauser-Horne factorizability could be expected, or is at least in full agreement with the fact that we forced the interactions to be local: we took the interaction constants $J_{ij}$ = J = 0 beyond nearest-neighbors. Interestingly, if one artificially introduces a delocalized 'left-right' interaction, e.g. $J_{12}$ or $J_{1b} \neq 0$, calculation shows that the Clauser-Horne factorizability is *not* satisfied. This is for instance the case in the system of Fig. 2.

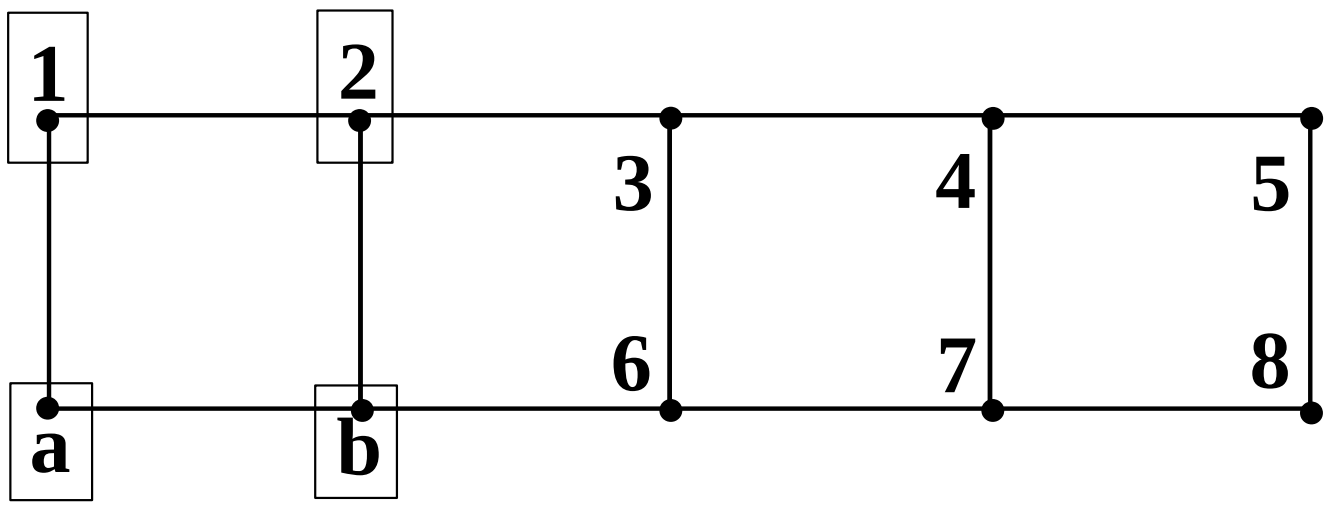

**FIGURE 2**. A lattice with delocalized left-right interactions.

Let us take here interactions $J_{ij}$ that are $\neq 0$ for first *and* second neighbours. If for instance all $h_i$=1, $J_{ij}$=1 (first neighbours), $J_{ij}$=0.5 (second neighbours), then one finds $X_{BI}$ = 2.32, MD = 0.03, PD = 0.78, and OD = 0.15 (cf. definitions (18)). In other words, none of the conditions OI, PI, MI holds.

We reached identical conclusions for a variety of 2-D lattices that are small enough to be numerically tractable. To further confirm our conclusions, we analytically investigated an arbitrary 1-D lattice in Appendix 2. This exercise is also useful to treat the case N ⟶ ∞.

## 3. Generalization. Violation of MI in existing, dynamic Bell experiments.

Based on the above result on Ising lattices, an obvious question arises. Besides the particular case of spin-lattices, does a general theoretical framework or a general physical system exist for which measurement independence is violated – while yet being local ? If such a model or system exists, it trivially escapes from Bell's no-go theorem; the Bell inequality presupposes MI. We already argued above on intuitive grounds that such a general class of MI-violating theories might indeed exist, namely those models that involve a background medium or field that interacts with the Bell particles and analyzers. We found first support for the idea in Ising-like systems: these can be seen as an example of a stochastic medium (described by the set $\sigma_\lambda$) interacting with $\sigma_1$, $\sigma_2$, and a and b ($\sigma_a$ and $\sigma_b$).

In this Section we will show that the intuitive reasoning leading to Eq. (3) is still valid for the most advanced dynamic Bell experiment (cf. Scheidl. et al. [2010]), which tries to impose not only MI but also OI and PI by switching analyzer settings. It is essential to explicitly show that background-based theories can survive dynamic Bell tests, since in the case of spin-lattices we only considered static Bell experiments. To this end, we will construct an elementary model for the experiment of Scheidl et al. [2010] that violates MI in a local and non-superdeterministic way.

The goal of the experiment of Scheidl et al. [2010] was to simultaneously close, for the first time, various loopholes, essentially the locality and 'freedom-of-choice' loopholes. As the authors clearly explain: "The locality loophole arises when Alice's measurement result can in principle be causally influenced by a physical (subluminal or luminal) signal from Bob's measurement event or Bob's choice event, and vice versa. The best available way to close this loophole is to space-like separate every measurement event on one side from both the measurement [outcome independence] and setting choice [setting independence] on the other side" (Scheidl et al. [2010]). In other words, in this manner OI and PI would be imposed in the experiment. This loophole was already addressed in older experiments, such as the celebrated experiments of Aspect et al. [1982] and Weihs et al. [1998], but not simultaneously with the freedom-of-choice loophole. "Experimentally, the freedom-of-choice loophole can only be closed if Alice's and Bob's setting values are chosen by random number generators and also if the transmission of any physical signal between their choice events and the particle pair emission event is excluded, i.e., these events must be space-like separated […]" (Scheidl et al. [2010]). The authors assume that in this way MI necessarily holds in the experiment. In the present context it is however crucial to note that their reasoning seems to presuppose – and this is of course common practice – that the HVs *describe the particles at the source*. Summarizing, the essential experimental conditions of the experiment of Scheidl et al. [2010] were the following:

E1) The measurement events were SLS;

E2) The left (right) measurement event was SLS from the choice of b (a);

E3) The setting choice events were SLS;

E4) The emission event and the setting choice events were SLS.

Spacelike separation between relevant events was obtained by fast and random switching of the settings, and/or by the fact that in some reference frame the events happened simultaneously (SLS is invariant under Lorentz transformation). In particular, condition E4), aiming at imposing MI, was realized as follows: the bit corresponding to a setting choice, determined by a random number generator on average at the time of emission of a pair (hence SLS), was fed into a delay line and subsequently activated the analyzer (an electro-optic modulator) at the arrival time of the particles (details in Scheidl et al. [2010]). The switching frequency was 1 MHz, implying that the analyzers kept their settings on average during 1μs. Based on the results of Scheidl et al. [2010], doubtlessly an experimental *tour-de-force*, it is now generally believed that the locality loophole is closed, thanks to E1)-E2), and also the vital and newly addressed freedom-of-choice loophole, due to E4).

Let us consider a slightly simplified experiment mimicking the experiment of Scheidl et al. [2010] to a sufficient degree (we do not need to be concerned with the fact that the original experiment used an asymmetric configuration and therefore an asymmetric delay line: Alice was closer to the source than Bob, cf. Scheidl et al. [2010]).

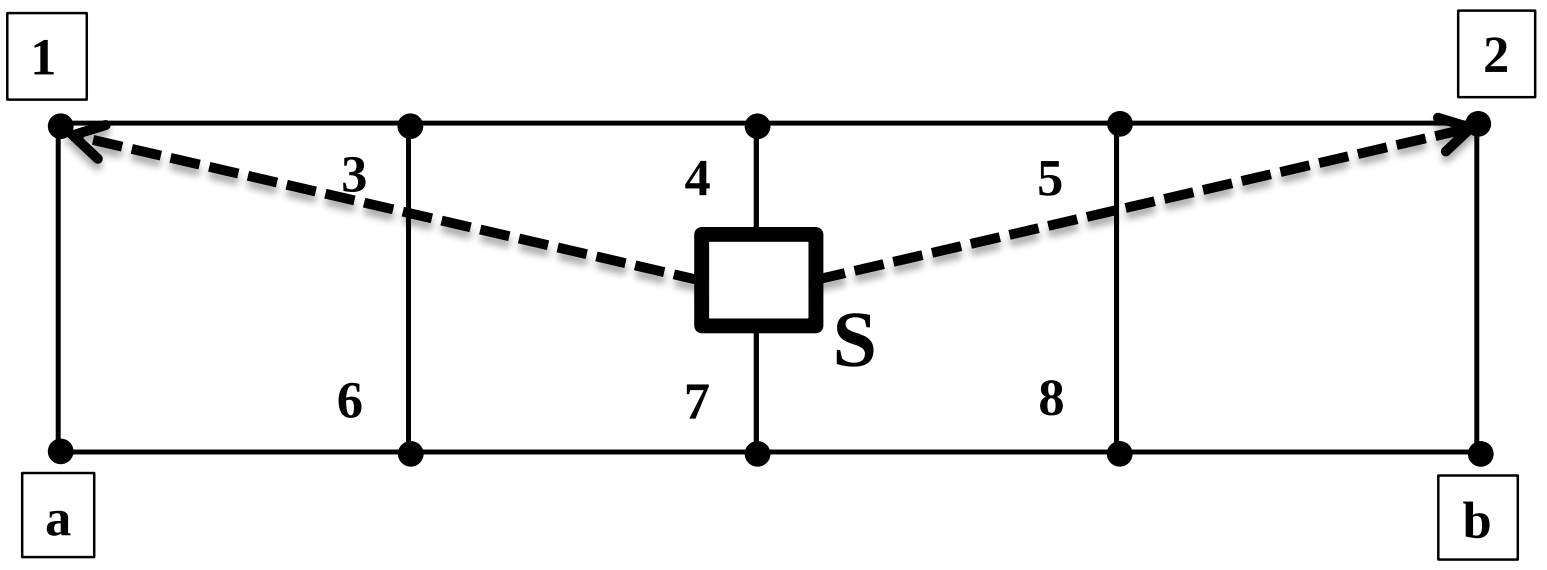

FIGURE 3. An experiment mimicking the experiment of Scheidl et al. [2010].
Particles 1 and 2 leave a source S and are measured at $t_0$ when arriving at points 1 and 2.

In our simplified experiment, the two particles leave a source S at a speed close enough to the speed of light, following the dashed trajectories in Fig. 3. Alice and Bob use random number generators to choose, at a frequency of 1 MHz, between two values for the analyzer angles a and b (located at nodes a and b). They measure the spins $\sigma_1$ and $\sigma_2$ at time $t_0$, the moment the particles arrive at nodes 1 and 2, which are close to nodes a resp. b, since particles and analyzers interact (we might have drawn nodes a and b closer to 1 and 2 respectively, but that doesn't play any role for the following). In order to impose the relevant SLS in the S-frame, it is sufficient to require that Alice's and Bob's number generators determine a and b *at the moment the pair leaves S (t = 0)*, and that the electronic signals corresponding to these choices are delayed and applied to the analyzers during a 1μs interval around $t_0$, exactly as in Scheidl et al. [2010]. Then we have precisely the conditions of SLS as used in the experiment of Scheidl et al. [2010], including those needed for OI and PI, as one easily verifies.

Suppose now that following hidden reality underlies the experiment. Our model is based on three simple hypotheses H1-H3. (Some may find that spelling out these hypotheses in detail is almost redundant; nevertheless it will reveal helpful, e.g. when we will look for a way to test the model.)

H1) In essence, our HV model will *only* assume that particles and analyzers interact with a 'background field' or 'background medium' (particles) characterized by HVs $\lambda$. For definiteness, suppose the particles or the field are lead-out on a lattice as in Fig. 3. This will allow to draw a parallel with the preceding Section (but we definitely do not need the lattice for our conclusion). Thus suppose that at nodes 3, 4,…8 of the lattice sit particles that are characterized by $\lambda_3$, $\lambda_4$,… $\lambda_8$; the $\lambda_i$ are stochastic parameters. Also assume that all $\lambda_i$ can only take two values (±1) – call the $\lambda_i$ 'generalized spins'. Note that stochastic properties taking two discrete values are not rare in classical physics: it occurs e.g. in lattice gases, where a site can be occupied or not, or in hydrodynamic systems, where vortices can turn in two directions only (vorticity = ±1), etc..

H2) The particles 1 and 2 are characterized by their spins $\sigma_1$ and $\sigma_2$ *and* HVs $\lambda_1$ and $\lambda_2$; the analyzers are characterized by $\lambda_a$ and $\lambda_b$, besides by a and b. Suppose that $\lambda_a$ and $\lambda_b$ are determined by a and b by a simple functional relationship: $\lambda_{a(b)}$ = a(b). Similarly, $\sigma_1$ and $\sigma_2$ are determined by $\lambda_1$ and $\lambda_2$ by $\sigma_{1(2)} = \lambda_{1(2)}$. Thus $P(\sigma_1,\sigma_2|a,b) = P(\lambda_1,\lambda_2|\lambda_a,\lambda_b)$ for all parameter values. So when Alice or her random number generator sets her analyzer to angle a, $\lambda_a$ assumes a value equal to a. If we also assume the convention that a and b only take the values ±1, which is enough for verifying the BI, then all the $\lambda_i$ only take values ±1.

H3) All $\lambda_i$ (i = a,b,1,…,8) interact, via some Hamiltonian, with close (say 1st) neighbours only: the interaction is local.

As an example, one could suppose that at the time of measurement ($t_0$) the interaction between the $\lambda_i$ is described by a generalized Ising Hamiltonian (n = 10):

$$H(\lambda_1,\lambda_2,\ldots\lambda_n) = c_0 + \sum_{i=1}^{n} c_{1i}\,\lambda_i + \sum_{i,j=1}^{n} c_{2ij}\,\lambda_i.\lambda_j + \sum_{i,j,k=1}^{n} c_{3ijk}\,\lambda_i.\lambda_j.\lambda_k + \ldots \qquad (25)$$

Expression (25) is just a Taylor expansion of a general Hamiltonian H($\lambda_1$, $\lambda_2$,… $\lambda_n$) depending on n 'spin' DOF. The Ising Hamiltonian corresponds to the 2 lowest-order 'spin'-dependent terms in (25), with $\lambda_i = \sigma_i$, $c_{1i} = -h_i$, $c_{2ij} = -J_{ij}$. Stochastic lattice gases are also described by these two terms[26], so (25) is appropriate also for a dynamical system. In both the latter examples one only considers close neighbour interaction, as we do (H3). Now, if one could also assume that, at the moment of measurement $t_0$, the configuration probability of a 10-'spin' configuration $\theta = (\lambda_1,\lambda_2,\ldots\lambda_n)$ is given by a Boltzmann probability as in (5), then the results of Section 2 would have proven that in our HV system MI and the BI can be violated (all equations are identical, if one neglects the $c_{3ijk}$ in (25) as usual). However in the present experiment Alice and Bob switch $\lambda_a$ and $\lambda_b$ at a high speed and therefore the Boltzmann probability is not applicable, since the latter expression presupposes thermal equilibrium. *The point is we do not need the Boltzmann probability neither the form of the Hamiltonian for our argument.* It seems the only observation we need to make is the following: independently of the Hamiltonian, at the time of measurement ($t_0$) $\lambda_a$ and $\lambda_b$ *can have made causal contact with nearest neighbours* ($\lambda_6$ and $\lambda_8$ respectively, cf. Fig. 3). Indeed, in the experiment of Scheidl et al. [2010] the settings remained constant for a duration of the order of 1μs, which corresponds to a causal range R = 300 m. Therefore, if the 1st neighbours of nodes a and b, i.e. nodes 6 and 8, are positioned closer than R (suppose this is the case), then $\lambda_a(\lambda_b)$ and $\lambda_6(\lambda_8)$ are time-like separated. If the HV $\lambda_6(\lambda_8)$ can have interacted with $\lambda_a(\lambda_b)$, then in general we will have that:

$$P(\lambda_6, \lambda_8|\lambda_a,\lambda_b) \neq P(\lambda_6, \lambda_8|\lambda_{a'},\lambda_{b'}),$$

[26] Here $\lambda_i$ is the occupation number of site i ($\lambda_i$ = 0 or 1), $c_{1i}$ the local chemical potential and $c_{2ij}$ an interaction potential (typically a Lennard-Jones potential).

$$\Rightarrow P(\lambda_3, \lambda_4,\ldots, \lambda_8|\lambda_a,\lambda_b) \neq P(\lambda_3, \lambda_4,\ldots, \lambda_8|\lambda_{a'},\lambda_{b'}),$$

$$\Rightarrow P(\lambda_3, \lambda_4,\ldots, \lambda_8|a,b) \neq P(\lambda_3, \lambda_4,\ldots, \lambda_8|a',b'), \qquad (26)$$

because of assumption H2). (Note this is formally exactly what we calculated for spin-lattices, cf. text above and under Eq. (17).) But Eq. (26) means that the simple HV model based on H1-H3 can violate MI, also in a dynamic experiment as in Scheidl et al. [2010]. Note that violation comes again about through physical interaction between neighbour nodes. Of course, nothing in our reasoning hinges on the fact that the HVs $\lambda_i$ are spread-out on a lattice; the essential assumption simply is that the analyzers must have time, during a switching period, to interact with nearby particles (a nearby background medium). In a general field vocabulary our point is thus particularly straightforward: the analyzers may interact with 'something' (say a λ-field) which interacts in turn with the particles. The first interaction may create a correlation between the analyzer settings and the field-values in their vicinity, from which our conclusion follows.

One way to resume is the following: *an experiment as that of Scheidl et al. [2010] can decouple the left and right wings, but not the analyzers from their nearby environment*. Therefore it appears that for this experiment a wide class of local HVTs is conceivable in principle, namely models involving a background medium or field that locally interacts with particles and analyzers (the essence of H1-H3 above), and which as a consequence do not satisfy MI[27].

Needless to say, there is still a long way to go from this conclusion to the construction of a realistic HVT, one that reproduces all quantum probabilities. But recall that Hall has recently shown that MD (violation of MI) is the strongest resource for reproducing the quantum probabilities, as compared to OD or PD (cf. Hall [2011]). And mathematical models have been proposed that reproduce the quantum probabilities through MD only (for recent reviews, see Hall [2010, 2011], Di Lorenzo [2012]). Note that the latter models are mathematical or information-theoretic models. Here we propose a *physical* interpretation for violation of MI, moreover one that is compatible with free will.

## 4. Interpretation, and closing the loophole.

[27] The full utility of our detailed analysis based on H1-H3 should become clear in a moment, when we will try to close the loophole we exploit.

The origin of the discrepancy between our conclusion and the one of Scheidl et al. [2010], Aspect et al. [1982], Weihs et al. [1998] etc. is not necessarily a surprise, and may simply be related to what one accepts as 'HVs'. One standardly interprets HVs as describing intrinsic properties of the particles. The particles 'carry' the HVs with them from source to analyzer. This is obviously a legitimate and certainly the first idea that comes to mind when one wants to explain statistic outcomes of spin measurements by HVs. It corresponds to Bell's initial intuition (in Bell [1964]) and it seems to have remained a basic assumption in the whole Bell literature. (But things are very subtle: we believe that Bell himself has opened the debate towards a much more general interpretation of the HVs in his article [1981], where he treats the stochastic variant of his theorem – cf. footnote 2.) As announced from the beginning, we have assumed here a different starting point; in the above model (H1-H3) the HVs describe a stochastic property of a background medium.

As a consequence our model exploits the freedom-of-choice loophole without superdeterminism (in our model the angles a and b are not determined by the $\lambda$; a and b can be freely and/or randomly set *and* the setting event is SLS from the emission event). As we saw above, one almost always assumes (cf. standard discussions as in Bell [1981], Shimony [1993], Dickson [2007], Hall [2011], Di Lorenzo [2012]) that if SLS is imposed as in the experiment of Scheidl et al. [2010], superdeterminism is the only possible explanation of measurement dependence, i.e. the assumption that "the settings are not chosen independently from the properties of the particle pairs" (Scheidl et al. [2010]). But we have argued here this is not the case if the HVs describe a background medium, starting from an investigation in spin-lattices.

Our interpretation in terms of a stochastic background is not only motivated by the above results, but also by recently developed 'sub-quantum' theories that aim at explaining certain aspects of quantum mechanics. Indeed, a series of models have been published that are all based on a stochastic 'zero-point field' or 'vacuum field' (see e.g. De la Peña and Cetto [1996], Bacciagaluppi [1999, 2005], Khrennikov [2011], Grössing [2012] and refs. therein). In these models quantum particles as electrons etc. interact with a background field, and adopt as a consequence a Brownian motion from which the quantum statistics is assumed to arise. Recently it was claimed that double-slit interference and entanglement can be explained in this manner (Grössing [2012]). In the latter reference the link was drawn between these zero-point field theories and recent and spectacular experiments performed on classical systems that mimic quantum properties (Couder et al. [2005, 2006]). In these experiments oil droplets show quantum-like behavior (including double-slit interference, tunneling and quantization of angular momentum), *also* by interaction with a background field, namely an external vibration. The wave field generated by the external vibration (which would be the λ-field for us) creates a correlation between the outcomes ($\sigma_1$ and $\sigma_2$ above) and the 'context' of the experiment (i.e. the geometry of the oil bed and slits, playing the role of the analyzers above). In the same way in our model the λ-medium couples with the analyzers and the particles, leading to a strong correlation between the variables of the experiment (on left and right sides).

These different results involving different (classical) descriptions thus all seem to put a stochastic background field on the stage. From our analysis follows that such descriptions for the emergence of quantum behavior from a classical stratum, might indeed be 'allowed' – if they would violate MI and therefore bypass Bell's no-go result. Of course, further research should determine the exact link between what we termed 'background-based theories' and these sub-quantum theories.

Importantly, it might be possible to experimentally close the loophole we exploit, or rather to test the hypothesis of a MI-violating background. (We are well aware that the following proposal is speculative. But shouldn't such proposals be part of philosophy of physics ? – especially if they are so tempting as in this case.) In principle the means is to increase the switching frequency of the analyzer settings. Let us briefly develop this point, because if correct, we believe it applies to the whole class of background-based models. If the (left) polarization direction in a Bell experiment is switched rapidly enough in the time interval (0, $t_0$) between angles a and a', one expects that from a critical frequency range on a λ-field or medium *will at most experience a smeared-out or averaged influence from these two analyser directions (a and a')*. In other words ρ(λ) in (1c) may depend on (a,a',b,b') but not just on (a,b). *But that is equivalent to MI*: the same distribution for λ applies to the 4 subensembles for (a,b); the BI can again be derived. One can look at things in a slightly different manner: if the switching frequency increases above a limit the causal range R of the setting (a) will become too small to still influence the λ that determine $\sigma_1$; then again ρ(λ) cannot depend on (a,b)[28] (cf. the detailed reasoning in Section 3). If this argument is correct, an experiment with high enough switching frequency would allow to discriminate between quantum mechanics and background-based theories; it seems the same reasoning holds for all such background-based theories. Of course, it seems difficult or impossible to predict at which frequencies this decoupling occurs. But a first estimation is maybe possible: if one assumes that the coupling breaks down when the causal range R becomes of the order of the typical length of the polarizers (or of the minimum polarizer length that still can be used to measure polarization), say 10 cm, one finds a decoupling frequency of the order of a few GHz. Even if such an experiment represents a technological challenge, it may be close to being feasible[29].

[28] In spin-lattices this can be simulated by letting the interaction constants between spin a and spins λ (and between b and λ) tend to zero. Indeed, such a decoupling leads asymptotically to MI, as can be calculated.

**5. Conclusion**.

We explored the in-principle possibility that HV models involving a background medium or field can complete quantum mechanics. We argued that if in a Bell experiment this background locally interacts with the Bell particles and analyzers, MI (measurement independence) can be violated – in a model that includes variables describing the medium in the neighbourhood of the analyzers. Since MI is a premise of the BI, such a model would trivially escape from Bell's no-go theorem, even if it involves only local, bona fide physical interactions. It seems there is therefore no in-principle prohibition against background-based theories that locally complete quantum mechanics.

We explored here in detail 1-D and 2-D spin-lattices, introduced in a former article (cf. Vervoort [2013]). These can serve as a simple example of an MI-violating system based on a stochastic background. In particular we analytically proved that MI can be violated in these systems without relying on superdeterminism (in a static Bell experiment). We also showed that these systems are local in the Clauser-Horne sense – in agreement with the fact they are based on first-neighbour interaction. Next we argued in some detail that even for the most advanced dynamic Bell experiments background-based theories can violate MI in a local, non-superdeterministic way. Again, in hindsight this may not really be a surprise, given the semantic shift we operate – namely to associate the HVs with a background, rather than with the Bell-pair. Finally we conjectured that such background-based theories might be tested by upgraded versions of existing experiments.

[29] As communicated by Gregor Weihs.

Of course, there is no a priori guarantee that it is possible *in practice* to construct background-based theories reproducing the whole of quantum mechanics. But on the other hand such theories seem to be an appealing concept for understanding the ever mysterious 'quantum contextuality' or 'quantum holism'. And maybe some prototypes of these theories already exist (cf. De la Peña and Cetto [1996], Bacciagaluppi [1999, 2005], Khrennikov [2011], Grössing [2012] and references therein). The philosophical, conceptual implications of the admissibility of such background-based theories seem capital.

Acknowledgments. I would like to thank, for stimulating discussions, G. Bacciagaluppi, G. Brassard, Y. Gauthier, Y. Gingras, L. Hardy, G. Grössing, G. Hofer-Szabó, A. Khrennikov, J.-P. Marquis. I am grateful to Johannes Kofler for having drawn my attention to the experiment discussed here in detail (Scheidl et al. [2010]), and to Lucien Hardy for having pointed out an error in a preceding version. Needless to say, remaining errors are entirely the author's responsibility.

## Appendix 1. 'Free will' and violation of MI in spin-lattices.

Here we calculate the 16 probabilities $P(\sigma_1,\sigma_2|\sigma_a,\sigma_b) \equiv P(\sigma_1=\varepsilon_1,\ \sigma_2=\varepsilon_2\ |\sigma_a=\varepsilon_a,\ \sigma_b=\varepsilon_b)$ (all $\varepsilon_i = \pm1$) needed to verify the BI for the experiment of Section 2. More precisely, we compare two experiments that Alice and Bob can do to determine these probabilities (Ex1 and Ex2 of Section 2).

Ex1) In this experiment the ensemble evolves fully 'on its own' (there is no intervention of Alice and Bob on any of the spins). Alice just measures $\sigma_1$ and $\sigma_a$, Bob $\sigma_2$ and $\sigma_b$ for the whole ensemble of lattices. From this set of results they 'postselect' 4 sub-sets, each sub-set corresponding to one for the 4 possible couples of ($\sigma_a$, $\sigma_b$)-values. They then compute the probabilities $P(\sigma_1,\sigma_2|\sigma_a,\sigma_b)$ by counting relative frequencies within the sub-sets. They will find following values:

$$P(\sigma_1,\sigma_2|\sigma_a,\sigma_b) = \frac{P(\sigma_1,\sigma_2,\sigma_a,\sigma_b)}{P(\sigma_a,\sigma_b)} = \frac{P(\eta_1)}{P(\eta_2)}$$

$$= \frac{\sum_{\theta(\eta_1)}^{2^6} e^{-\beta H(\theta)}}{\sum_{\theta(\eta_2)}^{2^8} e^{-\beta H(\theta)}}, \tag{A1}$$

where we use the results (8-10) derived in the main text.

Ex2) A second way to determine the $P(\sigma_1,\sigma_2|\sigma_a,\sigma_b)$ is available to Bob and Alice if they can intervene on $\sigma_a$ and $\sigma_b$. If they have sufficient technological means to control $\sigma_a$ and $\sigma_b$ they can do 4 consecutive experiments each corresponding to a given value of $\sigma_a$ and $\sigma_b$. In that case they will find:

$$P*(\sigma_1,\sigma_2|\sigma_a,\sigma_b) = \frac{P*(\sigma_1,\sigma_2,\sigma_a,\sigma_b)}{P*(\sigma_a,\sigma_b)} = \frac{P*(\eta_1)}{1}, \tag{A2}$$

where the asterisk reminds us that the probability is determined in an experiment in which $\sigma_a$ and $\sigma_b$ have a given value. With Eq. (5) of the main text we have:

$$P*(\eta_1) = \sum_{\theta(\eta_1)}^{2^6} P*(\theta) = \sum_{\theta(\eta_1)}^{2^6} \frac{e^{-\beta H(\theta)}}{Z^*}. \tag{A3}$$

Here the partition function $Z^*$ is the sum over all Boltzmann terms given that $\sigma_a$ and $\sigma_b$ are fixed, i.e.:

$$Z^* = \sum_{\theta(\eta_2)}^{2^8} e^{-\beta H(\theta)}. \tag{A4}$$

Thus comparing (A2-A3) and (A1) proves that the probabilities $P(\sigma_1,\sigma_2|\sigma_a,\sigma_b)$ determined in Ex1 and Ex2 are identical. As shown in the main text, this was to be expected. Also, this nicely reflects what happens in real Bell experiments. Finally, it is clear that any other relevant probability, such as $P(\lambda|\sigma_a,\sigma_b) \equiv P(\sigma_\lambda|\sigma_a,\sigma_b)$ is also identical in both experiments: the partition function Z* in Ex2 always corresponds to $P(\sigma_a,\sigma_b)$, the denominator in Ex1. Ergo, the BI and MI can be violated in an experiment compatible with free will (Ex2).

**Appendix 2. 1-D spin-lattices: test of MI, OI and PI.**

Here we investigate a 1-D spin-lattice with N+2 spins, as in Fig. 4. We assume all $J_{ij}$ = J and all $h_i$ = 0, which makes analytical calculations tractable.

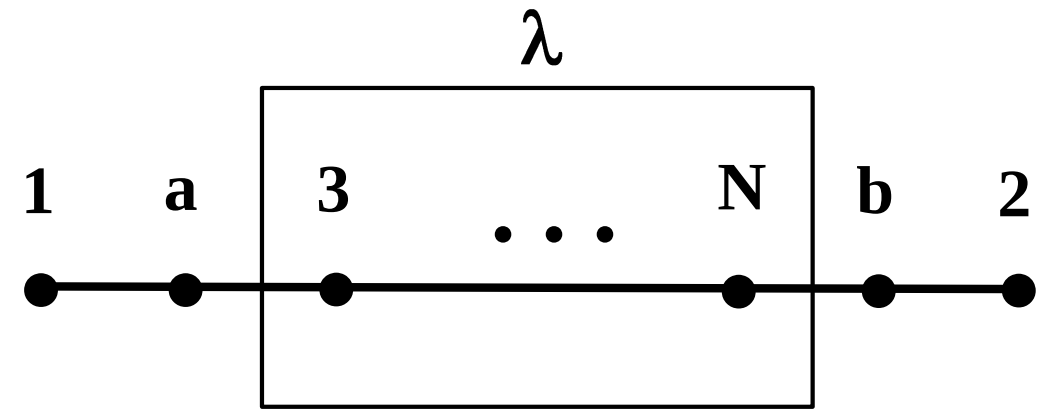

FIGURE 4. 1-D chain of N+2 spins.

To verify whether MI holds, we calculate P(λ | a,b) for $\lambda \equiv \sigma_\lambda \equiv (\sigma_3,\sigma_4,\ldots\sigma_N)$. First:

$$P(\lambda|a,b) \equiv P(\sigma_\lambda|\sigma_a,\sigma_b) = \frac{P(\lambda,a,b)}{P(a,b)} = \frac{P_1}{P_2}. \qquad \text{(A5)}$$

Introducing analogous definitions of α and K as in the 2-D case (cf. main text), we now have:

$$P_1 = \frac{1}{Z}\sum_{\sigma_1\sigma_2} e^{\beta J\sum_i \sigma_i\sigma_{i+1}} = \frac{1}{Z}\sum_{\sigma_1\sigma_2}(\cosh(\beta J))^{N+1}\prod_i[1+\sigma_i\sigma_{i+1}\tanh(\beta J)]$$

$$= \frac{1}{Z}\sum_{\sigma_1\sigma_2}\alpha\prod_i[1+K.\sigma_i\sigma_{i+1}]$$

$$= \frac{1}{Z}\alpha(1+K\sigma_a\sigma_3)(1+K\sigma_3\sigma_4)...(1+K\sigma_{N-1}\sigma_N)(1+K\sigma_N\sigma_b)\sum_{\sigma_1\sigma_2}(1+K\sigma_1\sigma_a)(1+K\sigma_2\sigma_b)$$

$$= \frac{1}{Z}\alpha(1+K\sigma_a\sigma_3)(1+K\sigma_3\sigma_4)...(1+K\sigma_{N-1}\sigma_N)(1+K\sigma_N\sigma_b).2^2 \qquad \text{(A6)}$$

The last step follows from the fact that in the sum $\sum_{\sigma_1\sigma_2}$ both $\sigma_1$ and $\sigma_2$ run over +1 and -1. For $P_2$ we find likewise:

$$\frac{Z}{\alpha}P_2 = \sum_{\sigma_1\sigma_2\sigma_3..\sigma_N}\prod_i[1+K.\sigma_i\sigma_{i+1}]$$

$$= \sum_{\sigma_1\sigma_2\sigma_3..\sigma_N}\left\{1+K(\sigma_1\sigma_a+\sigma_a\sigma_3+...)+K^2(\sigma_1\sigma_a^2\sigma_3+\sigma_1\sigma_a\sigma_3\sigma_4+...)+...+K^{N-1}(\sigma_a\sigma_3^2\sigma_4^2...\sigma_{N-1}^2\sigma_N^2\sigma_b+...)+...\right\}$$

(A7)

Grouping the terms in powers of K, one sees that in only one term all $\sigma_i$ appearing as indices are squared, namely in $K^{N-1}.\sigma_a\sigma_3^2\sigma_4^2...\sigma_{N-1}^2\sigma_N^2\sigma_b$. All other terms vanish, so that we obtain:

$$P_2 = \frac{1}{Z}\alpha\sum_{\sigma_1\sigma_2\sigma_3..\sigma_N}(1+K^{N-1}\sigma_a\sigma_b) = \frac{1}{Z}\alpha.2^N(1+K^{N-1}\sigma_a\sigma_b). \qquad \text{(A8)}$$

Finally

$$P(\lambda | a,b) = P(\sigma_3,...\sigma_N | \sigma_a,\sigma_b) = \frac{(1+K\sigma_a\sigma_3)(1+K\sigma_3\sigma_4)...(1+K\sigma_{N-1}\sigma_N)(1+K\sigma_N\sigma_b)}{2^{N-2}(1+K^{N-1}\sigma_a\sigma_b)}. \qquad \text{(A9)}$$

Thus in general $P(\sigma_\lambda | \sigma_a,\sigma_b) \neq P(\sigma_\lambda | \sigma_{a'},\sigma_{b'})$. In other words, according to Eq. (1c) MI is violated (MD ≠ 0), except for the trivial case K=0, i.e. J=0, corresponding to a non-interacting lattice. For instance:

$$P(+,+,...,+ | +,+) = \frac{(1+K)^{N-3}(1+K)^2}{2^{N-2}(1+K^{N-1})} \neq \frac{(1+K)^{N-3}(1-K)^2}{2^{N-2}(1+K^{N-1})} = P(+,+,...,+ | -,-). \qquad \text{(A10)}$$

Notice that MD $\longrightarrow 0$ for N $\longrightarrow \infty$. Formulas as (A9-A10) can be checked by a short computer program. Such numerical calculations also show that nothing substantially alters by introducing different interactions $J_{ij}$ between sites and local fields $h_i \neq 0$.

For verifying the Clauser-Horne factorizability condition (OI and PI), we need to calculate:

$$P(\sigma_1,\sigma_2 \,|\, \sigma_\lambda,\sigma_a,\sigma_b) = \frac{e^{\beta J \sum_i \sigma_i \sigma_{i+1}}}{\sum_{\sigma_1\sigma_2} e^{\beta J \sum_i \sigma_i \sigma_{i+1}}} = \frac{\prod_i (1+K.\sigma_i\sigma_{i+1})}{\sum_{\sigma_1\sigma_2}\prod_i [1+K.\sigma_i\sigma_{i+1}]} =$$

$$= \frac{\prod_i (1+K.\sigma_i\sigma_{i+1})}{(1+K.\sigma_a\sigma_3)...(1+K.\sigma_N\sigma_b)\sum_{\sigma_1\sigma_2}(1+K.\sigma_1\sigma_a)(1+K.\sigma_b\sigma_2)} = \frac{(1+K.\sigma_1\sigma_a)(1+K.\sigma_2\sigma_b)}{4}. \quad \text{(A11)}$$

On the other hand,

$$P(\sigma_1 \,|\, \sigma_\lambda, \sigma_a) = \frac{\sum_{\sigma_2\sigma_b} e^{\beta J \sum_i \sigma_i \sigma_{i+1}}}{\sum_{\sigma_1\sigma_2\sigma_b} e^{\beta J \sum_i \sigma_i \sigma_{i+1}}} = \frac{(1+K.\sigma_1\sigma_a)\sum_{\sigma_2\sigma_b}(1+K.\sigma_N\sigma_b)(1+K.\sigma_b\sigma_2)}{\sum_{\sigma_1\sigma_2\sigma_b}(1+K.\sigma_1\sigma_a)(1+K.\sigma_N\sigma_b)(1+K.\sigma_b\sigma_2)}$$

$$= \frac{(1+K.\sigma_1\sigma_a)}{2}, \quad \text{(A12)}$$

and similarly

$P(\sigma_2 \,|\, \sigma_\lambda, \sigma_b) = \dfrac{(1+K.\sigma_2\sigma_b)}{2}$, so that we do satisfy OI and PI, i.e. locality à la Clauser-Horne for an arbitrary N-chain.

# Chapter 4

# A Detailed Interpretation of Probability as Frequency, and its Physical Relevance

**Abstract**. In the following we will elaborate the frequency interpretation of probability of Richard von Mises, and insist on the physical relevance of a precise conceptual understanding of probability. Following von Mises, we will argue that probability can only be defined for events that can be repeated in similar conditions, and that exhibit frequency stabilization. The central idea of the present article is that the mentioned 'conditions' should be well-defined and 'partitioned'. More precisely, we will divide probabilistic systems into object, initializing, and probing subsystem, and show that such partitioning allows to solve a wide variety of paradoxes. Moreover, we will argue that a key idea of the Copenhagen interpretation of quantum mechanics (the determinant role of the observing system) is a direct consequence of a detailed definition of objective probability.

## 1. Introduction.

The first and simplest axiomatic system for probability, published in 1933 by Andrei Kolmogorov (1933/1956), is generally believed to cover all probabilistic systems of the natural, applied and social sciences. At the same time Kolmogorov's theory – the calculus of probabilistic or random systems - does not define what a 'probabilistic / random system' is, and does not provide any interpretation of the notion of probability besides through the mathematical axioms it fulfills (Bunge (2006), p. 100). For an intuitive understanding beyond mathematics of what probability 'is', one thus needs to resort to the ideas of other fathers of probability theory, Laplace, Fermat, Venn, von Mises, and others. Clearly, in developing his theory Kolomogorov was much inspired by these interpretations, so much so that one can read (von Plato (1994), p. 2): "Kolmogorov himself followed a rather vague intuition of frequentist probability, an attitude most likely to be found shared today by probability theorists."

It is likely that also in the broader physics community the most popular interpretation of probability is the frequency model – especially the limiting frequency version due to physicist and mathematician Richard von Mises (1928, 1964). However, in fields as philosophy and the foundations of quantum mechanics other interpretations, in particular the subjective interpretation, are increasingly popular (more on the subjective interpretation in quantum mechanics in, e.g., Caves et al. (2002), Bub (2007); for general references on the different interpretations of probability see e.g. Fine (1973), von Plato (1994), Gillies (2000), Khrennikov (2008)). This recent tendency in itself seems to justify a return to the sources. Indeed, one aim of the present article is to show that von Mises' work is a powerful starting point in the debate on foundations. Here we will therefore stick to the relative frequency interpretation and not analyze other models for the interpretation of probability[30].

At any rate, anyone who tried to construct a precise definition of probability soon realizes that the topic is surprisingly subtle. One challenge is to devise a definition that applies to both chance games *and* natural probabilistic phenomena; in standard mathematics textbooks one gives a separate definition for both cases. At the same time such a definition is highly desirable, for at least two reasons. First, it suffices to have a look at the texts of von Mises (or almost any reference text on the calculus) to realize that without a precise idea of what probability *is*, beyond the formal aspect, any somewhat subtle problem of calculus is likely to be treated in a wrong manner. Von Mises provides many tens and probably hundreds of such flawed calculations by even experts. No surprise, it has been said that "In no other branch of mathematics is it so easy to make mistakes as in probability theory" (Tijms 2004, p. 4). Next, for foundational reasons it is even more important to have a clear idea of the implicit notions, as we will show by exposing a surprisingly close link between the interpretation of probability and of quantum mechanics. The interpretation of probability would thus appear to be a paradigmatic topic at the interface of mathematics, physics, and philosophy.

Along these lines, let us note that besides von Mises another pioneer of probability theory, Kolmogorov, explicitly called for a conceptual (philosophical) analysis of probability (see Kolmogorov 1933/1956 p. 9). This need was further recognized by Kolmogorov's pupil Gnedenko ((1967) p. 23)[31]; it was subsequently addressed by several philosophers and mathematicians (see reviews in Fine (1973), von Plato (1994), Gillies (2000)). The debate on the foundations of probability theory continues. At least one very recent work concludes, as we will do here, that von Mises' theory is highly relevant for interpretational questions (Khrennikov 2008).

[30] We believe that, for our investigation of the physical relevance of the interpretation of probability, the most relevant features of probability are to be found in the real world around us, more so than in our heads.

[31] We found this reference work on probability calculus most helpful in our study. Boris Gnedenko did not only substantially contribute to probability calculus (especially in elaborations of the Central Limit Theorem and statistics); his work manifestly reflects his interest in the foundational issues.

In some detail, the frequency interpretation of probability we propose here differs from von Mises' theory (1928, 1964) in two respects. First, our model is simpler. First note that a real-world (or physical) theory for probability is the union of a calculus (C) and an interpretation (I) stipulating how to apply the calculus to real world events: Th = C ⋃ I (Bunge (2006)). Von Mises' calculus is based on the concept of 'collective' (an in principle infinite series of experimental results) – a calculus that strikes however by its complexity, and that probably no-one would consider using nowadays on a regular basis (see e.g. von Mises' treatment of de Méré's problem, von Mises (1928) p. 58ff). We will not make use of the concept of collective; in particular we believe that it is not necessary to resort to the notion of 'invariance under place selection' to characterize randomness (see next Sections and Appendix 1, but also Gillies (2000) p. 112). Our attitude is pragmatic: few people challenge the idea that the mathematics of probability theory is fully contained in Kolmogorov's calculus – so our calculus (C) is Kolmogorov's. What *is* however controversial is the following question, intimately linked to the interpretation of probability: *what is the exact subject matter of probability theory – to what exactly to apply it?* We will push the answer further than von Mises' classic answer: according to him probability theory treats "mass phenomena and repetitive events" (von Mises (1928 / 1981) p. v), characterized by a set of attributes that form the attribute space. (Anticipating, we will arrive at the conclusion that probability theory applies to a special type of random events, which we will term 'p-random'.) As said, the above question is intimately related to the detailed interpretation of probability. The definition we will propose in Section 3 captures, we believe, the essence of von Mises' interpretational model (I), but in a substantially simpler form.

Besides a simplification, our model is intended as a clarification of the frequency interpretation: we believe it helps to avoid the paradoxes to which probability theory seems so sensitive. The main claim of the present article is that one substantially gains in *partitioning* probabilistic systems into subsystems, namely test object, initiating, and probing subsystem (or 'environment' if one prefers). We will argue that this partitioning allows to solve a wide variety of classic problems (such as Bertrand's paradox), and bring in focus a surprising link between classical and quantum systems.

As said, recently another author, A. Khrennikov, has come to the conclusion that von Mises' work is more than worth a reappraisal (see his (2008)). The author offers in his textbook a detailed and convincing analysis based on the calculus of collectives. A. Khrennikov concludes in particular that well-known criticisms claiming that von Mises' theory (C) lacks mathematical rigor are not really cogent, as we argue for completeness in Appendix 1. Here we will not focus on the formalism but on a conceptual study (I); so we at least are immune against the criticisms relative to C.

## 2. A simple random system: a die. Introductory ideas: frequency stabilization and partitioning.

In order to get a feel for the problems to be solved, let us have a look at a typical probabilistic system, namely a die. When or why is a die 'probabilistic', or 'random' (or rather the throwing events, or outcomes) ? Simply stating in non-anthropocentric terms what a die throw is, already brings to the fore a few important notions. In physical terms a 'die throwing event' or 'die throw' consists of the 3-dimensional movement of a (regular) die, that is 1) characterized by the time evolution of its center of mass and its three Euler angles, 2) *caused* or *initiated* by an 'initiating system' (e.g. a randomizing and throwing hand, or automat), and 3) *probed* by an 'observing / probing system' (e.g. a table) that allows to observe (in general 'measure') an outcome or result R (one up, two up, etc.). It is easy to realize that if we want to use the die as it should, i.e. *if we want to be able to observe or measure a probability* for the different results of the throws, we have to repeat the throws not only by using the same die (or similar regular dies), but also by using the same 'boundary conditions'. Note that we are obviously *not* talking here about the detailed initial conditions of each individual throw, but about the conditions of our random experiment as they can be communicated to someone else ("throw roughly in this manner, probe on a table", etc.). These boundary conditions, then, are related to our throwing, the table, and the environment in general. *Irregular conditions in any one of these three elements may alter the probability distribution*. We can for instance not substantially alter our hand movement, e.g. by putting the die systematically ace up on the table. Nor can we put glue on the table, position it close to our hand, and gently overturn the die on the table while always starting ace up – one of the outcomes could occur much more often than in a ratio of 1/6$^{th}$. Nor can we do some experiments in honey instead of air: again one can imagine situations in which the probabilities of the outcomes are altered[32]. As will be seen further, it will make sense to isolate the mentioned elements, and to consider our random event as involving a *random system* containing three subsystems, namely 1) the (random) *test object* itself (the die), 2) the *initiating system* (the throwing hand), and 3) the *probing* or *observing system* (the table, and let's include the human eye). We will call such a composed random system a 'p-system' or probabilistic system. Just as one can associate subsystems to the random event, one can associate (composed) *conditions* to it, so conditions under which the random experiment occurs, in particular *initiating* and *probing conditions*.

Let us immediately note that instead of 'initiating system' and 'probing system', it can be more appropriate to speak of 'environment', namely in the case of spontaneous or 'natural' probabilistic events (versus 'artificial' ones, as outcomes of chance games, which are created by human intervention). Indeed, many random events occur spontaneously, without any known cause. A spontaneously disintegrating nucleus has probabilistic properties (e.g. its half-life, i.e. the time after which it disintegrates with a probability = 1/2) that are, on usual interpretations, not 'initiated', as in our die throwing, by a physical system or a person. Neither is it 'probed', except when subject to lab experiments. But the nucleus disintegrates spontaneously, according to a given well-defined probabilistic pattern, *only in a well-defined environment*. Changing the environment, e.g. by irradiating the nuclei or by strongly heating them, may very well change the probability distribution in question. In other words, if we want to determine (measure) the half-life of the nucleus, i.e. the probability of disintegration, we have to put it in well-defined ('initializing') conditions of temperature, pressure, etc. and measure its properties in well-defined ('probing') conditions, that scrupulously imitate its natural environment. So also here the initial and final conditions, or environment, re-appear.

By the above partitioning in subsystems we have just rendered the concept of 'conditions' explicit – a concept that is the starting point of, e.g., the reference work of Gnedenko on probability calculus (1967). Gnedenko states (p. 21):

> "On the basis of observation and experiment science arrives at the formulation of the natural laws that govern the phenomena it studies. The simplest and most widely used scheme of such laws is the following: *Whenever a certain set of conditions C is realized, the event A occurs*."

[32] Here is another example from darts. If we want to determine the probability that a given expert (or robot to make it simple) hits a given section of the dartboard, we let him throw say a hundred times, and count relative frequencies. Now, it is obvious that the conditions of the experiment should broadly remain constant: the throwing should not be irregular, the darts should remain the same, and also the dartboard should remain without trickery. Else it may be we cannot even *define* a probability.

And a little further (p. 21):

"An event that may or may not occur when the set of conditions C is realized, is called *random*."

For the well-known physical system of a thrown die, that is now well-defined, it seems not difficult to deduce the conditions for 'randomness' (we will see further that the condition proposed above by Gnedenko is not sufficient). Indeed, in order that the die throws are 'probabilistic' or 'random' or 'stochastic', it is a necessary condition that the *relative frequencies of the results* $R_j$ (j = 1,…, 6), namely the ratios { (the number of events that have result $R_j$) / n }, where n is the number of trials, *converge towards a constant number* when n grows (1/6 for a regular die). If such '*frequency stabilization*' would not occur, one cannot speak of probability[33]. Indeed, if for instance the die sublimates during the experiment in an asymmetric manner (thus gradually losing weight on one side), or if we accompany the die's movement in a structured manner that does not lead to frequency stabilization, we simply cannot attribute a probability to the outcomes. In sum, if there is no frequency stabilization, the die throwing is not probabilistic; on the other hand, if frequency stabilization occurs, the probability of the result $R_j$ is given by $P(R_j)$ = (# events that have result $R_j$) / n for large n. This is, or this is in full agreement with, the frequency interpretation of von Mises and others (von Plato (1994), Gillies (2000), Khrennikov (2008)).

Note that, as formulated above, random systems include deterministic systems. Indeed, in the latter case frequency stabilization will also occur. If we move the die by placing it systematically – deterministically – ace up on the table, the six $P(R_j)$ will trivially converge to a value (namely to 1, 0, 0, 0, 0, 0). We could exclude deterministic cases from our definition of random events / systems by adding that, besides exhibiting frequency stabilization, the events / systems should be unpredictable. Thus 'genuinely' random events show frequency stabilization *and* are unpredictable (or one could replace the latter by another predicate that qualifies the 'unstructured', disordered nature of randomness).

[33] The term 'frequency stabilization' appears in von Mises' work; in our opinion it best captures the property in question.

Before generalizing these ideas, a remark is in place. Von Mises uses in his theory the concept of limiting relative frequency (for the limit $n \rightarrow \infty$)[34]. This limit for n tending to infinity is considered problematic by some authors. Von Mises has however dealt in detail with such objections (we analyze some objections to von Mises' theory in Appendix 1). Among other points he recalls that probability theory is not only a mathematical but also a physical theory (it applies to the physical world); and the concept of infinity is ubiquitous in theoretical physics, even if it is obviously an idealization – a useful one when the formulas correctly predict reality to the desired precision. At any rate, we will in the following *not* define frequency stabilization in terms of convergence in the limit $n \rightarrow \infty$, but 'for n large enough for the desired precision'. We believe however that the limit $n \rightarrow \infty$ does not pose a real problem, but reflects to the contrary a particularly interesting fact of nature. Indeed, at the very basis of the frequency interpretation lies the scientific hypothesis that if certain well-defined experiments (trials) *would* be repeated an infinity of times, certain ratio's (involving outcomes of these trials) *would* converge to a limit. Experiment has shown over and over again that long series (obviously not infinite ones) do show such frequency stabilization. Let us emphasize that these massive empirical findings corroborate the initial hypothesis as strongly as any scientific hypothesis can be confirmed (scientific hypotheses are never confirmed with absolute certainty, neither with absolute precision).

## 3. Generalization, definitions.

Before proposing our definitions, which will essentially be straightforward generalizations of what we learned in the preceding Section, a few remarks are in place.

While studying in some detail our die, we found that the essential feature qualifying the throws as random, or probabilistic, is frequency stabilization. It seems just a natural interpretation of the frequency model to put this characteristic at the centre of the theory. That is indeed what von Mises did (1928, p. 12)[35]. But now an obvious, but neglected, question pops up: are all random systems really random ? At this point of analysis the answer seems not difficult: if frequency stabilization is indeed the essential feature of random systems and events, then it is clear that *not* all so-called random systems are random in the sense of probabilistic or stochastic. *Frequency stabilization is by no means guaranteed for an arbitrary disordered event, system or parameter*. The duration of my Sunday walks looks random, but does this parameter exhibit frequency stabilization ? No, because 1) there are *systematic* variations during certain periods, e.g. during the 'lazy months', 2) with age the walks will systematically tend to become shorter, and 3) at some point they will unfortunately disappear altogether. Or consider a mountain torrent, and define a parameter characterizing its flow, such as the (average) height of its waves or its rate of flow through a given surface. Such a torrent may show such erratic behavior over time, e.g. due to periods of rain or drought, that it seems impossible to define a parameter of which the frequency would stabilize. Atmospheric temperatures in a certain city look random, they may even look statistically random, but they probably don't show frequency stabilization over long periods because of climate changes[36]. What is even less likely to exhibit frequency-stabilized features, is 1) any 'random' system that is subject to human actions, e.g. the chaos on my table, or the random-looking features of a city-plan, etc., and 2) any 'composite' random-looking system, like a car and the house in front of it, or a tree, a stone underneath, and a cloud above. In other words, frequency stabilization happens often[37], but certainly not always. Not all randomness is the randomness that is the object of probability calculus.

[34] Probability is defined as $\lim_{n \rightarrow \infty} n(R_j)/n$, where $n(R_j)$ is the number of events or trials that have result $R_j$ in the total series of n trials, as in above example.

[35] Von Mises writes (1928, p. 12): "It is essential for the theory of probability that experience has shown that in the game of dice, as in all the other mass phenomena which we have mentioned, the relative frequencies of certain attributes become more and more stable as the number of observations is increased."

The above observation justifies the introduction of the concept of 'p-randomness', to be contrasted with 'randomness' and 'non-p-randomness': p-randomness, or probabilistic or structured randomness, is randomness (chance) that is characterized by frequency stabilization (at infinity, or if one prefers 'for large n'). Before one applies probability calculus to an event, one conjectures, or has evidence, that it is p-random - a feature that can only be confirmed by experiment. Note that a technically more familiar way to characterize a parameter or variable R as 'p-random', is to say that it has a 'probability density function', a parameter often symbolized by 'ρ' in physics ($\rho = \rho(R)$).

As stated, the only way to prove the probabilistic nature of systems or events is to perform experiments, and to show that frequency stabilization occurs. Now, we have illustrated in the preceding Section that it seems essential to clarify the *conditions* under which such experiments need to be done. The experiments involve repeated tests, in which the object under study (e.g. the die) should be subjected to repeatable, 'similar' or 'identical' conditions before being probed. It is however a well-known and unsolved problem of the foundations of probability to identify what 'similar' / 'identical' exactly means. How similar is similar (we will propose our answer in C4 below) ? We all know by experience that we do not need a lot of precautions when shaking a die so as to launch it in a 'regular' manner – leading to the probability distribution of a true die. In other words, it seems that in general we have a good idea what these 'similar conditions' are; but that is not always the case, as will be illustrated in the next Section. At this point the idea of partitioning we introduced in the preceding Section proves useful. Indeed, similar tests on similar systems means, in general: similar initiating system (or environment), similar test object, and similar probing system. These *three* subsystems should 'act', in the repeated test sequence, in *similar (enough) ways* in order that a probability can be defined. Alterations or divergences in any of these three subsystems can lead to different probabilities (probability distributions) – the latter may not even be defined, i. e. existing. Remember the sublimating die, or the inadequate randomization etc., which can lead to an undefined (i.e. changing, unstable) probability distribution of the die throws. As an example of the importance of the probing subsystem, recall the example of someone throwing a die on a table while always starting 'ace up', and from a fixed, shallow height. On a normal table the usual flat distribution will obtain; but if the tabletop is covered by a layer of adhesive material, a different distribution may be measured, e. g. showing often 'six up'. Thus, *probability depends - in principle - on the probing subsystem*, just as on the initiating system and the environment. "The probability of throwing a 6 with this die" is in itself meaningless; in principle one has to specify the experimental conditions.

[36] This may however be seen as a limiting case, in which frequency stabilization occurs 'for large n', even if not 'at infinity'. It seems that statistics is still applicable here, which shows the following point: the assumption of statistical stabilization is in general a matter of which precision one accepts.

[37] A particularly beautiful mathematical argument for the ubiquity of probabilistic randomness is provided by the uncrowned queen of probability calculus, namely the Central Limit Theorem.

We now propose a definition that can be derived from von Mises' theory, but that avoids the concept of collective. The simple move we make is the following. 1) Von Mises starts by defining a collective as an (in principle infinite) series of *random* results, and defines probability with respect to a given collective. But we have just argued that probability cannot be attributed to any random series, but only to p-random series. We therefore first define the concept of frequency stabilization or p-randomness: it is logically prior to that of probability. 2) We *partition* the probabilistic system, for pragmatic reasons, namely to avoid misuse of the concept of probability. The task is then to construct a definition that makes the notion of ‘experimental conditions’ explicit, *and* that applies to both chance games and natural probabilistic phenomena.

DEF1. A system or event possesses the property of frequency stabilization IFF

(i) it is possible to repeat n identical (or similar) experiments (with n a number sufficiently large for the desired precision) on the identical object of the system by applying n times identical initial and final actions or conditions on the object (realized by the initiating and probing subsystems, or more generally, by the environment); and

(ii) in this experimental series the relative frequencies of the results $R_j$ (j = 1,…, J) of the corresponding events (detected by the probing subsystem) converge towards a constant number when n grows; more precisely, the ratios { (# events in the series that have result $R_j$) / n } converge towards a constant number when n grows, for all j.

DEF2: Only for an event (or system, or outcome, or variable) that shows frequency stabilization according to DEF1, generated by an experiment in well-defined conditions, the probability of the result or outcome $R_j$ (j = 1,…, J) of the event is defined, and given by (for all j):

$P(R_j)$ = (# events that have result $R_j$) / n, for n large enough for the desired precision.

Following von Mises we conjecture that the above definitions allow to characterize all probabilistic phenomena; i.e. that they cover all events or systems described by probability calculus. To some believers of the frequency interpretation this claim may seem acceptable or even obvious. Others may find that a heavy burden of providing arguments awaits us. Clearly, we cannot present a full review of all probabilistic systems and show in detail how the definitions apply in all these cases. But the cited works of von Mises offer a wealth of such examples. Moreover, under C1 – C6 we provide those arguments and completing observations that seem the most important to us in rendering above conjecture defendable.

C1. It is well-known that the frequencies or ratios defining probability in a model like the above fulfill the axioms of Kolmogorov’s probability calculus.

C2. If one reduces the definitions to their essence, they state that probability *only* exists for systems or events that can be subjected to massively repeated physical tests, occurring under well-defined (and composed) conditions; and that in that case probability is given by a simple ratio or ‘frequency’. The numerical precision of that ratio grows when the number of experimental repetitions grows.

The intimate link between probability and physical experiments, leading to the idea that probability is a property not of objects but of repeatable events, or even better of *repeatable experiments,* is built-in, in a quite trivial manner, in von Mises' concept of collective - a series of experimental results. In the case of physics, this link has been emphasized and analyzed by philosophers as Popper (1957, p. 67)[38] and van Fraassen (1980, Ch. 6) (see also e.g. von Plato (1994)). Careful consideration of how probabilities are calculated outside physics, have convinced us that DEF1 and DEF2 are also applicable to the probabilistic systems of other fields. This seems difficult to prove but we give several examples in the remainder of this chapter.

Note that the definitions can be used both for artificial systems (chance games, typically) and natural systems (behaving in a p-random manner without human intervention); in other words our phrasing allows to see how the same concept applies to such widely different events as a die thrown by a human, a molecule in a gas, and a quantum measurement. Indeed, in the case of chance games the initializing and probing subsystems are easily identified as truly separate systems acted upon, or made, by humans (see C3). In the case of natural probabilistic phenomena (quantum phenomena, diffusion phenomena, growth phenomena, etc.) the initializing and probing systems coincide with – are – the environment.

Note that DEF1 of p-randomness is stated in terms of 'possible' repetitions, or of 'possible' trials, not necessarily actual ones. The modal aspect of probability has been emphasized by van Fraassen (1980). Natural, spontaneous random events are normally not initiated nor measured by humans in repeated experiments. Still, according to the model we advocate, they only 'have' a probability if they occur, or can occur, on a massive scale as could be imitated in the lab, or studied under repeatable conditions. Also for such systems probability only exists *if massively repeated tests are possible, and if in these tests frequency stabilization occurs* (or if massively repeated tests would in principle be possible, and etc.). Note that the probability of such natural events can *only* be revealed by 'artificial' experiments imitating the environment (or by calculations using experimental data, and representing such experiments). But this seems to be nothing more than the scientific paradigm.

C3. Most p-random phenomena or events are characterized by outcomes that are (described by) *continuous* variables, rather than discrete ones. In other words, R will range over an interval of the real numbers, rather than taking discrete values $R_j$, j=1,…, J as in die throws, coin tosses and other chance games. One could observe that these objects used in games are constructed on purpose by humans in order to ‘produce’ a limited number of discrete outcomes – in contrast to natural systems. The latter, like diffusion phenomena, gas kinetics, biological growth phenomena, etc. have outcomes that are continuous, or quasi-continuous. (A notorious exception are certain quantum systems, notably when they are carefully ‘prepared’.) In the case of a continuous p-random variable x (we used R before), one defines in probability theory the probability density function $\rho(x)$, the meaning of which is however defined via the concept of probability. Indeed, in the continuous case $P(R_j) = \rho(R_j).dR$, defining $\rho$ via P as characterized in DEF2. One can thus, for the present purpose, safely treat probability density functions and probability on the same foot. A model that interprets probability for discrete variables, also does the job for the ubiquitous continuous variables. (Also note in this context that one can formally describe discrete cases by a density function $\rho(R) = \Sigma_j P_j.\delta(R-R_j)$, where $\delta(-)$ is the Dirac delta-function: the discrete case is a special case of the more general continuous case.)

[38] Popper (1957, p. 67) famously interprets probability as the *propensity* that a certain experimental set-up has to generate certain frequencies.

C4. DEF1 of p-randomness relies on the notion of ‘identical’ or ‘similar’ conditions and objects. It may thus, as already stated, look suspicious in the eyes of a philosopher. But we will now argue that the definition is sound in the sense that it allows to define p-randomness and probability in a fully objective manner. First, notice that it seems impossible to define ‘similar’ or ‘identical’ in a more explicit manner. One could try in following (partly circular) way: “to the best of common knowledge, or of expert knowledge, one should repeat the ‘same’ event in a manner that allows to speak about the ‘same’ event, and that generates frequency stabilization.” But does this help ? We believe however there is no real problem. Indeed, as defined, *frequency stabilization can be tested for by independent observers; and these observers can come to the same conclusions*. The conditions for doing the ‘frequency stabilization test’ can be communicated for any particular system: “do such-and-such (‘identical’) initial and final actions on such-and-such (‘identical’) systems - and the probabilities $P_j$ will emerge. I found frequency stabilization and the probabilities $P_j$, so if you do the experiment in the ‘right, identical’ conditions, you should find them to.” It would seem that the problematic term ‘identical / similar’ of probability theory can thus be defined in an operationally fully consistent manner.

In sum, the above shows that one could, or should, speak of 'p-identical', or rather 'ρ-identical', events / systems, where ρ is the probability density of the outcomes of the events under study: to test whether a system is probabilistic, and to identify the probability of the corresponding events, one needs to perform a series of experiments on ρ-identical systems (including object, environment, initializing / probing subsystem) – systems that lead to the probability distribution ρ(R) of the results R of the events 'defined' or 'generated' by the system. Upon this view, then, *'identical' is 'identical' insofar a probability distribution ρ emerges*. It will also be clear that in an analogous way one could meaningfully define a ρ-system = ρ-{object, environment, initializing /probing subsystem}. We therefore believe that in the context of probability theory, one should, in principle, speak of ρ-systems, or ρ-identical systems[39].

C5. Does the frequentist model cover the classical interpretation of probability, traditionally used to tackle chance games, urn pulling and the like ? Von Mises (1928/1981, p. 66 ff.) and many modern texts on probability calculus come to this conclusion. We will therefore only succinctly present what we believe to be the essential arguments for an affirmative answer to above question. Notice, first, that our definitions can at least in principle be applied to such chance games. The initializing subsystem is most of the time a 'randomizing hand' (tossing a coin or a die, pulling a card or a colored ball, etc.); the probing subsystem is often simply a table (plus a human eye).

According to the famous expression of Pierre-Simon Laplace, for calculating a probability of a certain outcome, one should consider "events of the same kind" one is "equally undecided about"[40]; within this set, the probability of an outcome is the ratio of "favorable cases to all possible cases". In all reference books on probability calculus, the latter definition is the basis of probability calculations in chance games and urn pulling. Notice now that dice, card decks, urns containing balls, roulette wheels, etc. are constructed so that they can be used to produce *equiprobable* (and mutually exclusive and discrete) *basic*

[39] At this point one might be tempted to include a speculative note. In effect, there seems to exist such a strong link between probability (ρ) and the concept of (identical) 'system' – or 'thing' –, that one may indeed wonder whether we conceive of something being a (simple) 'object' just because it shows frequency stabilization. We saw that composed objects, like a house and the tree in front, or a car and a cloud above, are very unlikely to exhibit frequency stabilization – and indeed we do not conceive of them as simple objects, but as composed ones. (Remember in this context that deterministic systems are just a type of probabilistic systems; obviously they are the first candidates for being associated with 'things'.)

[40] The events thus fulfill the 'principle of indifference' introduced by Keynes.

*outcomes*, i.e. *having all $P_j$ equal, and given by 1/J* (J = the number of basic[41] outcomes). Equiprobability is at any rate the assumption one starts from for making mathematical predictions, and for playing and betting; and indeed Laplace's "events of the same kind one is equally undecided about" would now be termed equiprobable events. Along the lines exposed above, a chance game can thus be seen to correspond to an (artificially constructed) ρ-system with $\rho(R) = \Sigma_j (1/J)\, \delta(R-R_j)$.

It is at this point simple to show that in the special case of equiprobable and mutually exclusive events, the frequentist and classical interpretation lead to the same numerical values of probability. Indeed, within the frequentist model, $P(R_j) = \frac{n.\frac{1}{J}}{n} = 1/J$ (the numerator n / J = the number of $R_j$-events among n (>>J) exclusive and equiprobable events each having a probability 1/J). Thus the result, 1/J, is equal to the prediction given by Laplace's formula (1 favorable case over J possible cases). Let us immediately note, however, that Laplace's formulation is not superfluous. It allows, for the special case of equiprobable events, for calculation: 'favorable cases' and 'all possible cases' can indeed conveniently be calculated by the mathematical branch of combinatorics (a theory of counting, initiated by the fathers of probability theory).

In sum, it appears that the classical interpretation is only applicable to a small subset of all probabilistic events and systems, namely those characterized by discrete and equiprobable basic outcomes. For this subset Laplace's interpretation can be seen as a *formula*, indeed handy for calculations, *rather than an interpretation of probability*. After calculation *the only way to verify the prediction is by testing for frequency stabilization*. It is among others for this reason we believe the latter property is the natural candidate for a basis of an overarching interpretation.

C6. We have defined frequency stabilization and probability by means of the notion of 'convergence' of a certain ratio when n, the number of trials or repetitions of an experiment, grows. It is clear that this phrasing is close to the original definition by von Mises of probability as $P(R_j) = \lim_{n\to\infty} n(R_j)/n$. However, our phrasing "P(Rj) = n(Rj)/n for a number of trials n that is large enough for the desired precision" avoids the notion of infinity; it may therefore avoid problems of mathematical rigor (see discussion in Appendix 1). Note that from a pragmatic point of view, our definition allows to derive, if one would use it to experimentally determine a probability, numbers that are equal to those identified by von Mises' definition to any desired precision. At least operationally there is no difference in the definitions: they lead to the same results. Von Mises' definition may, however, be more satisfactory to the metaphysical aspirations of some: one could say that probability indeed 'is' limiting relative frequency 'at infinity'. An 'ideal' number that one can seldom calculate with infinite precision (except, for instance, if one can use combinatorics), but that one can always measure with a precision that grows when n grows.

[41] The 'basic events' or 'basic outcomes' of coin tossing are: {heads, tails}, of die throwing: {0, 1, …, 6}, etc. The probability of 'non-basic', or rather composed, events (two consecutive heads, etc.) can be calculated by using probability calculus and combinatorics.

These notes conclude the basic description of our model. Needless to say, they are a first introduction to the model; we are well aware that questions will remain unanswered. The only way to validate and strengthen a model is to show that it applies to (all) non-controversial cases, and that it solves problems left open by other models. Concerning the non-controversial cases, we could have shown in more detail how the model applies to other typical systems, e.g. gas molecules (this is one of the model systems von Mises investigates). We believe that the reader will however have no problem in applying the above model also to this case (see also von Mises 1928 p. 20); again, the key for such an application is to recognize that the initializing system here is simply the environment. Since the above model is a direct elaboration of von Mises' interpretation, it should be able to tackle the probabilistic systems that the latter can tackle. However it is much simpler: we do not need the concept of collective, nor its calculus; our calculus is Kolmogorov's.

**4. Applications.**

The aim of the present Section is to show that above model allows to solve paradoxes of probability theory (Fine (1973), von Plato (1994), Gillies (2000)). We will tackle problems R1 – R4.

R1. As is well known, according to frequency interpretations it makes no sense to talk about the probability of an event that cannot be repeated. In condition (i) of our definition of p-randomness, repeatability is an explicit requirement. Therefore, it makes no sense to talk about the 'probability' of Hitler starting a world-war and the like: no experiments can be repeated here, and even less experiments in well-defined conditions. Popper's propensity interpretation of probability was an attempt to accommodate such probabilities of single events – quantum events were his inspiration, like the disintegration of one atom (Gillies (2000) Ch. 6). But again, any quantum property of any single quantum system can only be attributed a probability if the property can be measured on an ensemble of such systems –as in the case of macroscopic systems. Measurement (verification) of a probability is always done on an ensemble of similar systems and events, whether quantum or classical. Therefore the added value of the 'propensity' interpretation seems not obvious to us.

According to the above model, probability is not a property of an object on its own; it is a property of certain systems under certain repeatable human actions; or, more generally, of certain systems in dynamical evolution in well-defined and massively repeatable conditions. The following slogan captures a part of this idea: *probability is a property of certain composed systems*. Similarly, probability is not a property of an event *simpliciter*, but of an event in well-defined conditions, in particular initializing and probing conditions. An even more precise way to summarize these ideas would be, it seems, to attribute probability to *experiments* (van Fraassen 1980, Ch. 6), or *experimental conditions* (Popper 1957). The advantage of the term 'experiment' is that it only applies to 'composed events' for which the experimental conditions are defined in a scientifically satisfying manner – exactly as we believe is the case for the use of (objective / scientific) 'probability'. Note that a scientific experiment is also, by definition, repeatable – Hitler's 'experiment' is not.

R2. A classic paradox of probability theory is Bertrand's paradox. It goes as follows: "A chord is drawn randomly in a circle. What is the probability that it is shorter than the side of the inscribed equilateral triangle ?" Bertrand showed in 1888 that apparently three valid answers can be given. A little reflection shows however that the answer depends on exactly how the initial randomization is conceived: there are many ways to 'randomly draw a chord' (which may not be obvious upon first reading of the problem). One can for instance randomly chose two points (homogeneously distributed) on the circle[42]; a procedure that leads to the probability 1/3, as can be measured and calculated. But other randomization procedures are possible, leading in general to different outcomes. In other words, *Bertrand's problem is simply not well posed*, a conclusion that seems now generally accepted (see e.g. Marinoff (1994)). But that conclusion is obvious within our model: probability is only defined for experiments in well-defined conditions, *among others initializing conditions*. More precisely

[42] for instance by fixing a spinner at the center of the circle; a pointer on the spinner and two independent spins generate two such independent random points.

(see DEF1+2), probability exists only for events which show frequency stabilization in experiments under precise conditions, initial, final and 'environmental'; for a situation that is experimentally ambiguous no unique probabilities can be defined. It will be seen that the literature abounds with 'paradoxes' which stem from formulations that are experimentally ambiguous. But probability theory is a theory of real-world events, of p-random experiments; not only a mathematical theory.

R3. A popular idea is that, somehow, "probability depends on our knowledge, or on the subject". This is the key idea of subjective interpretations of probability, or subjective Bayesianism, associating probability with strength of belief or degree of ignorance. When I throw a regular die, I consider the probability for any particular throw to show a six to be 1/6. But what about my friend who is equipped with a sophisticated camera allowing him to capture an image of the die just before it comes to a halt ? For him the probability seems to be 0 or 1. Or what about following case: imagine Bob, waiting for a bus, only knowing that there is one bus passing per hour. He might think that the probability he will catch a bus in the next five minutes is low (say 5/60). Alice, who sits on a tower having a look-out over the whole city, seems to have a much better idea of the probability in case (she might quickly calculate, based on her observations, that it is close to 1). Are these not patent examples of the wide-spread idea that a same event can be given different probabilities, depending on the knowledge (or degree of belief) of the subject ? And is in that case probability not simply a measure of the strength of belief of the subject who attributes the probability ?

Paradoxical examples as these are unlimited, but if one adheres to our model, it seems they stem from a neglect of the 'boundary conditions' that are part of the definition of probability. Probability is only defined – only exists – for repeatable experiments on well-defined systems, including, among others, well-defined conditions of observation. Doing a normal die throw, and observing the result on a table as is usually done, corresponds to a well-defined p-system, with a specific initiating system, probing system etc. In the example, the second observer does not measure the same probability: he does not measure the probability of finding a six on a table after regular throwing, but of finding a six after measurement of whether a six will land or not on the table. The latter is a very different, and indeed fully deterministic, experiment; *at any rate, the observing subsystem is very different*. A similar remark holds for the bus-case; the measurement system (even if just the human eye) is part of the p-system; one cannot compare observer Alice and observer Bob if their means of observation are widely different. One could also summarize by stating that Alice and Bob do not measure the probabilities of the same event or system (contrary to what is thought): that is why they measure different probabilities. Probabilistic events and systems are composed.

If one can generalize examples as the above, it seems that our model allows to safeguard the objective interpretation of probability by emphasizing that probability belongs to events under well-defined probing conditions. (We believe the just illustrated neglect of the probing subsystem may be at the basis of subjectivist interpretations of one or the other form.) If one includes in the experimental series all boundary conditions, and especially the probing subsystem, the probability of a given outcome *is* an objective measure. True, 'objective' (and 'observer-independent' even more) is a somewhat tricky word here: it means '*identical (and mind-independent) for all observers performing identical experiments*', so *objective in the scientific sense* – even if the observer, or rather the observing subsystem, is in a sense part of the system ! (Remember that the observing subsystem is part of the p-system.) Stated more precisely, the probability of an event is an 'objective' property of the p-system that generates the event, in that independent observers can determine or measure the same probability for the same event.

R4. At this point the step to quantum mechanics is immediate. Indeed, it seems that our investigation has brought in focus a striking similarity between classical and quantum systems. It belongs to the key elements of the standard or Copenhagen interpretation that "the observer belongs to the quantum system", or at least that the measurement detectors belong to it, and influence it. Suffices here to cite Bohr in his famous debate with Einstein on quantum reality (Bohr 1935, see also Gauthier 1983 on the role of the 'observer' in quantum mechanics). The debate concerned the completeness of quantum mechanics, questioned by Einstein, Podolsky and Rosen in the case of two non-commuting observables of "EPR electrons". The key point of Bohr's counterargument is summarized in following quote: "The procedure of measurement has an essential influence on the conditions on which the very definition of the physical quantities in question rests" (Bohr 1935, p. 1025). The latter statement is probably one of the most famous quotes of the Copenhagen interpretation, corresponding to one of its most basic ingredients; however, even experts of the foundations of quantum theory have complained that the quote is incomprehensible (Bell (1981) p. 58). It seems however we are now armed to make Bohr's phrasing transparent. According to Bohr *the definition of quantum properties depends in a fundamental way on the measurement* *conditions*. Now quantum systems are probabilistic systems; and we have argued throughout this article that the numerical value of a probability depends in a fundamental way on the observing subsystem or conditions – in general different probing conditions may lead to a different probability for otherwise identical events. In classical systems one has to look carefully for examples to exhibit this in principle fact (we gave several examples), but in the quantum realm it apparently becomes basic. As an example, the probability an x-polarized photon passes an y-polarizer obviously depends on x and y (x and y are the parameters that describe the initializing and probing conditions). At any rate, we believe Bohr's quote is now well understandable; and that a careful inspection of the concept of probability shows that in *all* probabilistic systems, quantum or classical, the measurement system plays an essential role. We thus come to the surprising conclusion that, in this respect, quantum systems are not as exceptional as generally thought. In this context, see Gauthier (1983) for the role of the observer in quantum mechanics.

## 5. Conclusion.

We have provided in the present article an analysis of the concept of probability that can be seen as a variant of von Mises' frequency interpretation. Whether all essential aspects of what (objective) 'probability' is are captured by our model obviously remains a question; but we believe that the uniqueness of Kolmogorov's axiomatic system, and the massive existence of scientific data to which the calculus applies, are convincing indications of the idea that one overarching objective interpretation is at least a defendable working hypothesis. We hope that our model can contribute to the elaboration of such a unifying theory.

We proposed a model in which it only makes sense to define probability for systems or events (or better, experiments) that exhibit frequency stabilization or p-randomness – the essential characteristic of probabilistic systems. It appeared useful, if not necessary, to partition probabilistic systems into three parts: if natural, in object and environment; if artificial, in object, initializing subsystem, and probing subsystem. That a precise definition of initial and final conditions or subsystems is necessary for defining probability, is often neglected, and leads to countless paradoxes. Most importantly, including the *probing subsystem* into the probabilistic system allows to define probability in an objective, mind-independent manner – and thus to see probability as an objective category, in contrast to subjective interpretations. Only if the probing is defined, the probability is. By the same token we showed that there is an essential parallel between quantum and classical systems: in order to be able to define a probability for an event, *in both cases* one needs to specify the 'observer', or rather the probing subsystem. Including the *initializing subsystem* into the probabilistic system also allows to solve paradoxes, such as Bertrand's paradox.

Acknowledgements. For detailed discussion of the issues presented here I would like to thank Mario Bunge, Yvon Gauthier, Henry E. Fischer, and Jean-Pierre Marquis.

## Appendix 1. Alleged mathematical problems of von Mises' theory.

Let us have a look at some criticisms of von Mises' mathematical theory, which is based on the concept of 'collective'. Some of these criticisms seem superficial ("infinite collectives do not exist because real experiments always end"; see our comments in the main text); others, stating that von Mises' probability as a limit for $n \rightarrow \infty$ is not always well defined (see e.g. Richter (1978)) are not really to the point as soon as one realizes that von Mises' theory is a physical theory, not a strictly mathematical one (see von Mises' clear arguments in his (1928/1981), e.g. p. 85). The most interesting and cogent critiques concern the 'condition of randomness' that von Mises imposes on collectives. According to him, the randomness of a collective can be characterized as 'invariance under place selection' (roughly, the limiting frequencies of a collective should remain invariant in subsequences obtained under 'place selections', certain functions defined on the original collective). But which and how many place selections are required ? – von Mises' critics ask. An important result was obtained by A. Wald, who showed that for any denumerable set of functions performing the subsequence selection, there exist infinitely many collectives à la von Mises: a result that seems amply satisfying for proponents of von Mises' theory (see the reviews in von Plato (1994), Gillies (2000) and Khrennikov (2008) p. 25). Even the famous objection by J. Ville can be shown not to be a real problem (see Khrennikov (2008) p. 27 and also Ville's own favorable conclusion reproduced in von Plato (1994) p. 197).

As a side remark, let us note that von Mises' attempts to mathematically describe randomness led to interesting developments in the mathematics of string complexity, as produced by no-one else than his 'competitor' Kolmogorov, and mathematicians as Martin-Löf. A general result of these developments can broadly be stated as follows: real physical randomness seems to defy full mathematical characterization (Khrennikov (2008) p. 28). In our view, this is not really surprising: if a series of experimental results can be generated by an algorithm, one would think it is not random by definition (it may of course look random). A real series of outcomes of coin tosses cannot be generated (predicted) by an algorithm; but – and this is close to magic - it does show frequency stabilization.

This observation allows to counter a curious critique by Fine (1973), who derives a theorem (p. 93) that is interpreted by the author as showing that frequency stabilization is nothing objective, but "the outcome of our approach to data". However, closer inspection shows that Fine's argument only applies to complex mathematical sequences (of 0's and 1's) that can be generated by computer programs. But if a sequence can be generated by a computer algorithm, it is by definition not random, but deterministic, even if it looks random. The essential point of real series of coin tosses is that they cannot be predicted by numerical algorithms… *and* that they show frequency stabilization, as can easily be shown by experimentation. From this perspective, mathematical randomness or complexity, as discussed in relation to the complexity of number strings, has little or nothing to do with physical randomness[43].

To end our succinct review of critiques of von Mises' theory, the decisive point is the following. It is of course possible to withhold von Mises' interpretation of probability (as a limiting frequency in a collective), even if his calculus would have shortcomings in mathematical strictness - the present article only needs the interpretational part, as often emphasized. On top of that, the sometimes criticized randomness condition appears to be not necessary for the theory, in the sense that the equivalence with Kolmogorov's measure-theoretic approach can be proven without using that condition (see an explicit proof in Gillies (2000) p. 112). This is again, we believe, a clear indication of the fact that the essential feature of probabilistic randomness is frequency stabilization, not invariance under place selections.

[43] Except if the number strings are generated by a real physical number generator, which is, as far as I know, never or almost never the case.

# Chapter 5

# The Interpretation of Quantum Mechanics and of Probability: Identical Role of the 'Observer'

**Abstract**. The aim of the article is to argue that the interpretations of quantum mechanics and of probability theory are much closer than usually thought. Indeed, a detailed analysis of the concept of probability reveals that this notion always refers to an observing system. The enigmatic role of the observer in the Copenhagen interpretation therefore derives from a precise understanding of probability. Besides explaining elements of the Copenhagen interpretation, our model allows to solve paradoxes of probability theory, and to reinterpret recent claims from 'relational quantum mechanics' [1-2] and of the 'subjective approach to quantum probabilities' [3-5]. Finally, we suggest a way to reconcile the objective interpretation of probability and moderately subjective variants.

## 1. Introduction.

A key element of the Copenhagen interpretation of quantum mechanics is the role played by the observer, or rather the observing system. The observing system, or the measurement, makes the wave function collapse. By the same token it causes the 'measurement problem': why is an observing system any different from any 'normal' physical system, e.g. the natural environment - which leaves the wave function of the system in its superposition state ? Bohr and Heisenberg are reputed to be the first to have recognized the role of the observing system, giving an 'instrumentalist' or 'operationalist' flavor to quantum mechanics. To some, reality even seemed to depend on, or be determined by, apparatuses. Since then a further shift in the interpretation of quantum mechanics has been proposed by several authors – a shift towards subjectivism, in which it is now the observer *as a human being*, including his or her mind, who plays the starry role. The degree of subjectivism is of course different for different authors, but some of these interpretations (see Section 3.2.)

explicitly claim that quantum probabilities have a 'subjective element'. Besides the standard Copenhagen interpretation, we will in the following investigate some of the better known new interpretations of quantum mechanics, namely 'relational quantum mechanics' of Rovelli and others [1-2], and the Bayesian or subjective interpretation of quantum probabilities of Bub, Caves, Fuchs, Schack and others [3-5]. As an excellent representative of the orthodox Copenhagen interpretation, we will use Peres [6-8], especially his textbook [6].

The aim of the present article is to show that the role of the 'observer' in quantum mechanics is not new: it is exactly the same as he/she/it plays in classical probability theory. More precisely, we will argue 1) that a precise definition of probability (within the frequency interpretation) always refers to an observing system, 2) that as a consequence the instrumentalist aspects of the Copenhagen interpretation stem from the probabilistic nature of quantum phenomena, and 3) that also essential claims of other interpretations of quantum mechanics [1-5] can be re-interpreted within the standard interpretation of probability. The 'understanding' of quantum mechanics, beyond the formalism, would therefore heavily draw on the interpretation of the concept of probability.

In contrast to philosophers and mathematicians, many a physicist will wonder whether anything new can be learned from the interpretation of probability. Is everything about probability not entirely said with Kolmogorov's simple axioms, dating from 1933 [9] ? Unfortunately not, otherwise probability theory would not be termed the branch of mathematics in which it is easiest to make mistakes (Ref. [10], p. 4). Indeed, in order to apply probability calculus to the real world, one needs to know to *which type of events* exactly to apply it[44]; in other words, one needs an interpretation of the concept of probability, beyond Kolmogorov's axioms. The most widespread interpretation in science is the relative frequency interpretation (in the limit of infinite trial series), which is generally attributed to Richard von Mises [11-12]. However, other interpretations such as the classical interpretation of Laplace, the propensity interpretation of Popper, and the subjective interpretation, associating probability with 'degree of belief', exist (general references are [13-16] and the condensed [17] Ch. 4). It should be noted that at least one (controversial) version of the subjective (or 'Bayesian') interpretation of probability, namely the Bayesianism of E.T. Jaynes [32-33], has proponents in the physics community since decades, also outside quantum mechanics. Jaynes seems an important source of inspiration for proponents of the subjective interpretation of quantum probabilities, since he is cited by both the authors of [2] and [4-5].

[44] Many or most real world events are not probabilistic.

As a matter of fact (and much to our own surprise), it appears that the concept of probability contains a few implicit notions, and that a minimalistic definition as P(R) = probability of result R = $\lim_{n\to\infty}$ {number of occurrences of R / total number (n) of trials}, does *not* guarantee flawless application, as we already argued in Ref. [18]. The main result of [18] is simple: in order to calculate P(R) (with R a result, outcome, or event), that 'R' needs to be precisely specified, including the 'initializing' and 'probing' conditions in which it occurs. In other words, P(R) can be completely different from P(R') if R' is (at first glance) the same event as R but probed or measured in different circumstances (see Section 2). Of all available interpretations of probability, the findings presented here resonate best with von Mises' ideas [11-12]. But the link with von Mises' theory[45] is rather minimalistic. Von Mises interprets any probability as a relative frequency in an (in principle infinite) series of *experimental outcomes*; here we will analyze and highlight the role of the *experimental conditions*, which include in particular an observing subsystem; the link ends here. In view of the recent debate in quantum mechanics, it seems not superfluous to recall that not only von Mises but also Kolmogorov and his pupil Gnedenko, were well aware of, and actively engaged in, the objective – subjective debate. They offered vigorous arguments in favor of an objective interpretation of probability (see e.g. Ref. [11] p. 75ff., 94ff., and Ref. [19] p. 26ff.). We will come back to this ancient and multi-faceted debate in Section 2, Section 3.2. and Appendix 1, and suggest that there is room for convergence: it seems that less radical variants of the subjective school may be made to coincide with an objective interpretation, under the conditions we will specify. What did come as a surprise to us, is that a detailed conceptual study of the notion of probability appears so useful for solving paradoxes.

Finally, let us emphasize that recently another author, A. Khrennikov, has convincingly highlighted the relevance of von Mises' work for the present topic [16]. Ref. [16] offers a detailed mathematical treatment of the foundations of probability and its link with quantum mechanics. As compared to Ref. [16], we will not use von Mises' formal

[45] Besides an *interpretation* of probability (as a relative frequency), von Mises developed a full-blown probability *calculus* (the calculus of 'collectives') [11-12] that leads to the same numerical results as Kolmogorov's measure-theoretic approach. The formal theory fell almost into oblivion because it is somewhat less simple or elegant than Kolmogorov's axioms, but it has as essential advantage that it makes the link with physical experiments much clearer: it is, in von Mises' words, a *physical* theory, not only a mathematical one. The calculus has occasionally been criticized by mathematicians for flaws of mathematical rigor. Note that we are immune to such criticisms: our calculus is Kolmogorov's.

‘calculus of collectives’ but start from an analysis of what a ‘probabilistic system’ is [18]; we will remain strictly within ‘Kolmogorov - von Mises’ probability theory and not extend it to non-standard theories [16]; and, most importantly, we will focus on the role of the ‘observer’ and the objective – subjective debate. As we will indicate below, some of our conclusions coincide with Ref. [16]; we would like to promote the latter work as an essential reference.

Before turning to the interpretation of quantum mechanics, let us in the next Section briefly recall the main results of Ref. [18].

## 2. Results of a detailed interpretation of probability (Ref. [18]).

Typical textbooks state that probability is attributed to ‘random events’. When we want to experimentally determine the probability of the event e = ‘this coin shows result R = heads’, we all know what to do: we toss the coin many times and determine the relative frequency of heads (P(e) = P(R) = ½ for a regular coin). Now if we are looking for the *in-principle conditions on which P(e) depends* (these in-principle conditions will become essential in quantum mechanics), we quickly realize that these at least contain the initial and final conditions of the tosses (rather of the ‘tossing experiment’): the way we start the experiment and probe (observe) the result. Normal tossing involves a vigorous and random momentum to start with, and normal probing takes place, for instance, on a table. We could obtain *any* result for P(e) if we would put glue on the table and cast the coin with a sufficiently refined momentum distribution (with sufficient technological trickery we could moreover make the whole affair look perfectly random and regular if we wanted, but that isn’t important here). *Therefore, in principle P(e) depends on the ‘initializing’ and ‘probing’ or ‘observing’ conditions in which e occurs* [18]. (Equivalently, one could say that P(e) depends on the initializing and observing subsystem that form, together with the die, the probabilistic system, or ‘p-system’. In our example the initializing subsystem is typically a randomizing hand, and the observing subsystem a table. More examples are given further on, and in Ref. [18].) In the above example ‘e’ is a probabilistic event ‘created’ by a human; in such an artificial experiment we have no problem to conceive the role of probing or observing conditions. Now does the above mentioned dependence also hold for *natural* probabilistic phenomena (diffusion, gas collisions, quantum phenomena,…) that happen every instant everywhere in the universe without anybody observing ? In [18] we argued that the role of the initiating and probing conditions is taken here by the *environment* (temperature, force fields etc. determine the probabilities in question in an obvious way). At any rate, when we *verify* or *measure* the probabilities which e.g. quantum theory predicts for an event occurring in a specified environment, we need to do the experiment in that same environment or in a lab reproducing the environment – thus the initiating and probing conditions reappear. Therefore, it is safe to generalize above idea: for calculating or determining a probability, as soon as things get somewhat subtle, one should remember the in-principle dependence on the initiating and probing conditions (or the environment if one prefers). (Ref. [18] proposes a detailed definition of probability that includes these conditions.)

However simple the idea, inspection of the countless paradoxes which the fascinating topic of probability has to offer, left us little doubt that it is this idea which is at the basis of many if not all of these paradoxes. Von Mises exposes many tens of them in his works [11-12]; many textbooks contain at least a few (see e.g. [10] and [19]). Let us look at a few well-known examples. Bertrand's paradox (dating from the end of the 19th century) goes as follows: "A chord is drawn randomly in a circle. What is the probability that it is shorter than the side of the inscribed equilateral triangle ?" Note that this paradox is a topic of research and debate since its birth: all 'fathers of probability theory' have discussed it, and often given different answers [21]. Now, in the light of the above discussion the solution is straightforward: since a chord can be drawn randomly in a variety of ways (as may not be obvious at first reading), the answer is simply undefined [18, 21]. Each method of random drawing (read: each initializing condition of the random experiment, or each initializing subsystem) leads to a specific probability (1/2, 1/3, or 1/4). In other words: to have a defined probability, one needs a defined initializing condition. The infamous 'Monty Hall' problem [22], amply discussed and popularized in the nineties in all media by Marilyn vos Savant, will be seen to become equally straightforward if one remembers that, here to, one looks for a relative frequency of an outcome R of a well-defined experiment[46].

Essential for the following is the role of the probing or observing conditions. Any numerical value of *any* probability depends, in principle, on the observing conditions, as was already suggested by the coin example [18]. In a macroscopic and thoroughly known case as a coin toss, when asking a person to determine 'the probability of heads for this coin', no-one feels compelled to specify these conditions. Everyone assumes that everyone knows what is meant with the event 'e'. But in the quantum world we don't know much anymore; as we will see further the conditions become essential.

[46] R = 'I win a car by switching'; P(R) = 2/3.

However, on further inspection they play a crucial role too in classical systems, namely to safeguard the objective nature of probability. Indeed, a question that almost automatically pops up in discussions on probability, is: "but doesn't probability depend on our knowledge ?" (It is this question that opens the door to subjective interpretations of probability.) Consider for instance a regular die throw. 'For me', the probability that a 6 shows up is 1/6$^{th}$, but 'for Alice', who is equipped with a sophisticated camera and can register the die movement in real-time until it comes to a halt, that same probability seems to be either 0 or 1. Or, 'for me' the chance that I will catch a bus during the next 10 minutes is 10/60 (I know that one bus passes per hour). But someone who sits in a tower and has a lookout over the city traffic (and has therefore a different 'knowledge state') might disagree [18].

It is not difficult to devise paradoxical situations as these, but for an advocate of an objective interpretation they stem from a neglect of the fact that P(R) is only well-defined – only exists – if R is well-defined, including its probing conditions. Doing an experiment in which a die is thrown and probed in the usual way is one thing; throwing and then monitoring it in real-time with a camera is another. These two events are different due to different observing conditions (or observing subsystems); for *that* reason the probabilities are different, not because of two different knowledge states or degrees of belief.

Now, the subjective shift is extremely tempting. Indeed, it is obvious that in the above die example one could introduce *conditional* probabilities (e.g. the probability that the upper surface shows 6 at halt *if it is given that* the surface shows 6 one μsec before halting as observed with a camera). *Any* probability can in principle be considered a conditional probability (see Kolmogorov, Ref. [9], p. 3; Ref. [19] p. 67). And the expression "if it is given that" seems to be equivalent, in this context, to "if it is known that", thus invoking the knowledge state of an observer in a subliminal manner. It seems to us that this idea, however simple, may be at the origin of why subjective interpretations are so popular. We refer to Appendix 1 for a more detailed argument.

If we assume that probability theory is part of science, then it seems it should necessarily concern 'objects' that can be separated from the 'subject' (we): if our minds would play any role, i.e. determine the objects (= probabilities) under study (as the most radical subjectivist interpretations suggest), scientific, i.e. 'objective' comparison of results would be impossible (in the context of quantum mechanics, this idea has been analyzed and stressed by Bohr at several occasions). In [18] we showed that if one wants to see probability as an objective measure, there is a simple way to do so: attribute it to events under defined probing conditions. True, 'objective' (and 'observer-independent' even more) is a somewhat tricky word here: it means '*identical (and mind-independent) for all observers performing identical experiments*', so *objective in the scientific sense* – even if the observer, or rather the observing subsystem, is in a sense part of the system ! (Remember that the observing subsystem is part of the p-system [18].) It is therefore clear that in objective accounts 'observer' must be understood as a physical system without a mind. The subjective shift is equivalent to including a human mind in the 'observer'. As already stated, there are many brands of subjectivism. In Appendix 1 it is shown that certain forms of Bayesianism, in particular Jaynes' [32-33], may be reconciled with our interpretation (actually, calling these variants 'subjective' is a misnomer, if one compares them to radical versions as these exist in economics, sociology, philosophy etc. [14-15]).

A next remark that will prove useful is the following. In [18] we argued, following authors as von Mises and Popper [20], that it is often helpful to remember that probability belongs to 'experiments', rather than just events: probability theory concerns repeatable and real physical tests or experiments. The 'event space' of axiomatic probability theory [9] is in a sense a vague concept[47].

As a last remark, let us make a link with Ref. [16]. Above we highlighted the role of the experimental conditions in determining probabilities. This element can be directly extracted from von Mises' concept of 'collective', the basic notion of von Mises' theory. A collective is an - in principle infinite - series of outcomes in a probabilistic experiment. These outcomes obviously refer to, or depend on, experimental conditions. Ref. [16] uses von Mises' original calculus of collectives; since these collectives refer to experimental conditions, some of our conclusions (highlighting the role of in particular the observing conditions), should be backed-up by formal results of [16]. This is indeed the case, as we will indicate below.

Let us now see how the above findings compare to the interpretation of quantum mechanics.

---

[47] It is of course mathematically well-defined within a system as Kolmogorov's; but it does not say to which events probability theory can be applied. Many or most real-world events are not probabilistic.

## 3. Conceptual link between probability and quantum mechanics.

### 3.1. Objective probability and the Copenhagen interpretation.

In view of what we learned above, it seems that the importance of the 'observer' in quantum mechanics should not surprise us, if quantum phenomena are probabilistic phenomena. One of the key ideas of the Copenhagen interpretation is expressed in the following quote by Bohr, taken from his 1935 reply to Einstein, Podolsky and Rosen in their debate on the completeness of quantum mechanics [24]. "The procedure of measurement has an essential influence on the conditions on which the very definition of the physical quantities in question rests" [24]. It is often complained that this passage is incomprehensible (we did not understand it before our investigation [18]). In Ref. [23] Bell not only identifies the above passage as the essential ingredient of Bohr's reply, but also admits and details his complete incapacity to understand it (p. 58). However, within the objective interpretation of probability of Section 2, we can simply re-interpret what Bohr claims. It appears now quite clearly that Bohr says that quantum properties *are determined by the observing conditions or in other words the observing subsystem*[48]. But we found in the previous Section that this dependence holds – in principle - for all probabilistic systems, quantum or classical. For classical systems one needs some attention to find examples of the influence of the observing system; in quantum mechanics the 'in principle' becomes basic - as correctly emphasized by Bohr. As an example: the probability that an x-polarized photon passes a y-polarizer obviously depends on x and y. *Any* numerical probability value, quantum or classical, is a function of the analyzer / detector parameters.

Seen from this angle, one would think that the 'measurement problem' becomes, in essence, transparent. The collapse of the wave function (the selection of one of the outcomes of the experiment that is encoded in the wave function), strongly reminds us, of course, of classical probabilistic measurements in which the observing subsystem selects one possibility (e. g. 'six up' in a die throw selected by a table). We believe the concept of collapse contains in essence no more mystery than the 'determination' of a classical probabilistic system during the act of observation. (This does not preclude that more refined mechanisms might underlie the simple probabilistic manifestation that emerges from these mechanisms.)

---

[48] This 'determination' can of course only materialize through an interaction between the trial system and the observer system (think of how any classical probability is measured). This idea is elaborated by Bohr in Ref. [24] by invoking the 'quantum of interaction'.

Let us have a look at a few salient statements of Peres in Ref. [6]. "*A state is characterized by the probabilities of the various outcomes of every conceivable test*" ([6] p. 24, where these 'tests' should be understood as pertaining to an experimental set-up). A little further: "Note that the word 'state' does not refer to the photon by itself, but to an entire experimental setup involving macroscopic instruments. This point was emphasized by Bohr […]" ([6], p. 25). We are by now inclined to say that these quotes hold for *any* probabilistic system, quantum or classical. To a die throw one could obviously associate a 'state' consisting of the six discrete results and their probabilities. At any rate, also the probability of an outcome of a die throw depends crucially on the 'entire experimental setup' that is used to perform it, in particular the initial and probing conditions (Section 2). Probabilities, and in particular quantum probabilities, are measures pertaining to repeatable experiments, *not* objects per se [18]. Compare to the following passage of [6] (p. 73): "The notion of density matrix – just as that of state vector – describes a *preparation procedure*; or, if you prefer, it describes an ensemble of quantum systems, whose statistical properties correspond to the given preparation procedure". In Section 2 we termed the 'preparation' of this quote the 'initializing' of the probabilistic system. We can review here only a few of the instrumentalist passages of [6], but it appears they all can be understood by using the same notions as probability theory already uses, explicitly or implicitly.

Note that we of course do not claim that every element of the Copenhagen interpretation has its counterpart in probability theory (à la von Mises). Notably, the latter has no uncertainty relations, neither the corresponding commutation relations. But interestingly enough, *even these are foreshadowed by an adequate interpretation of probability*. The commutation relations stipulate, among other things, which observables A and B can be measured simultaneously. Now, within von Mises' framework, P(A&B), the joint probability of A and B, is defined *only if A and B can be measured by an experimental set-up that allows to measure A, B, and A and B (combined) by using the same observing system*[49]. In symbols: $P_C(A.B) = P_C(A).P_C(B|A)$, with C the *same* experimental conditions in the *three* experiments. One could equally well write P(A.B|C) = P(A|C).P(B|A.C): see Appendix 1. In other words,

[49] This follows from von Mises' exposition in Ref. [12] pp. 26 – 39, even if it is not entirely explicit. The key procedure is the following: start from two collectives (probabilistic experimental series), one in which A is measured, one for B. If A and B are 'combinable', then one can construct a collective for the joint measurement of A and B, and determine their joint probability. On a straightforward interpretation of von Mises, the latter collective must correspond to an experimental series *using the same equipment as used for measuring A and B separately*. Thus a 'joint' measurement is a 'simultaneous' measurement in this sense. Von Mises warns explicitly that not just of any A and B the joint probability can be measured.

within von Mises' probability theory the question of simultaneous measurement of two random quantities is already crucial – just as it is crucial in quantum mechanics (Ref. [16] comes from a different angle to basically the same conclusion, see p. 54). Von Mises warns explicitly that not of all quantities A and B the joint distribution exists; it is not difficult to find conditions C in which A and B cannot be simultaneously measured. We see here again that (via the commutation relations) quantum mechanics fills in, in a stringent manner, the conditions under which to apply classical probability theory in the atomic and subatomic realm.

At this point it is extremely tempting to suggest a link with the notorious quantum non-locality or contextuality revealed by, e.g., the theorems of Bell and Kochen-Specker. Let us succinctly indicate why. We argued that typical quantum probabilities and correlations as P(A) and P(A.B) should be understood as P(A|C) and P(A.B|C) and thus depend *in principle* on C, i.e. on all the parameters of the set-up, for instance polarizer or magnet directions (call this the 'dependence thesis'). Now this idea, taken at face value, seems to offer an elegant solution to Bell's theorem [30]. Indeed, according to the above thesis Bell's well-known correlation functions $P(a,b) = \int d\lambda.\rho(\lambda).A(a,\lambda).B(b,\lambda)$ (same notation as Bell [30]) have to be replaced by expressions as $P(a,b) = \int d\lambda.\rho(\lambda|a,b).A(a,\lambda).B(b,\lambda)$. The probability distribution of the hidden variables $\rho(\lambda)$ may, *in principle*, depend on the whole experimental set-up. It is then obvious that Bell's inequality cannot be derived anymore. However, at this point the caveat '*in principle*' in our dependence thesis becomes essential and needs to be specified: it seems clear that the observing conditions C (the polarizer directions a and b, in general the 'left' and 'right' observing subsystems) can only determine $\rho(\lambda)$ *if both a and b are in causal contact (in the relativistic sense) with the particles* (at the source, or at impingement). Now in the most relevant experimental tests of Bell's theorem there is a spacelike separation between the polarizers and between the polarizers and source: the particles cannot be influenced by both polarizers at the moment of measurement. In other words, we believe that the 'dependence thesis' (dealing with probabilities as P(A|C)) should be restricted to parameters C that are in causal contact with A. This specification, saving Bell's theorem, seems to be in perfect harmony with von Mises' view on how probabilities are determined (see in this context also the first footnote of this Section). For a different position, also based on the idea that the relevant measure in Bell-type correlation experiments is P(A.B|C) rather than P(A.B),

see Ref. [16] pp. 65-84 and Ref. [31]. There it is argued that within a strict von Mises framework Bell's theorem does *not* hold[50]. See also the interesting Ref. [27] in this context.

Let us note, finally, that probability theory is less authoritative than the Copenhagen orthodoxy. On the Copenhagen interpretation, the commutation relations stipulate not only which observables can be measured simultaneously, but also which observables *exist* simultaneously. Note that this is indeed *the second essential ingredient of Bohr's answer to EPR*, besides above passage. (As we read it, Bohr's argument can thus be summarized as follows: measurement brings observables into being through an inevitable interaction with an observing system; if two observables cannot be measured simultaneously, they do not exist simultaneously.) Probability theory is of course silent about the existence of outcomes when they are not measured (i.e. about their being determined by hidden variables, whether measured or not). Laplace and Einstein famously argued that probabilities hide causal mechanisms - an option that is fully left open by probability theory, but prohibited by the Copenhagen interpretation. (Determinism has become much less plausible in the eyes of many physicists since the experimental tests of Bell's theorem; but we believe much remains to be said about the famous theorem [16, 25, 28, 31].)

As we will see in the next Section, a detailed frequency interpretation allows to re-interpret, in a unified manner, also more recent claims dealing with the interpretation of quantum mechanics.

### 3.2. Relational Quantum Mechanics; Subjective Quantum Probability.

An interesting interpretation of quantum mechanics is due to Rovelli and others [1-2], who termed it 'relational quantum mechanics' (RQM). The essential idea of this theory is to consider any state vector as relative to an observer, or rather observing system (for Rovelli anything can be an 'observer', e.g. an atom). A relevant quote is this: "The notion rejected here is the notion of absolute, or observer-independent, state of a system; equivalently, the notion of observer-independent values of physical quantities. The thesis of the present work is that by abandoning such a notion (in favor of the weaker notion of state – and values of physical quantities - *relative* to something), quantum mechanics makes much more sense" [1] (Quote 1). Just as Einstein's rejection of the obsolete notions of absolute time and simultaneity allowed to reinterpret the Lorentz transformations, Rovelli conjectures that the replacement of absolute states by relative states allows to found quantum mechanics on more solid grounds [1].

[50] Note added for the present PhD thesis: I believe now that these criticisms of Bell's theorem should be taken seriously, as explained in detail in Chapter 2 and 3 of the thesis.

Now, according to orthodox quantum theory [6], a quantum state represents outcomes of an experiment and their probabilities (see Section 3.1.). But we have argued in Section 2 that these probabilities necessarily depend on, e.g., the observing subsystem; therefore, also the corresponding outcomes necessarily depend on the observing subsystem[51]. But this seems to offer an explanation of the main claim of relational quantum mechanics (Quote 1), stating that quantum states and the physical quantities they represent are *relative* to an observer. They indeed are: they numerically depend on it, or are determined by it. It thus seems that an objective probability interpretation of quantum mechanics provides a natural basis to understand the relational nature of quantum mechanics (see also [34] in this context).

The subjective interpretation of quantum probability is vigorously defended in recent Refs. as [3-5]. Ref. [5] states: "In the Bayesian approach to quantum mechanics, probabilities – and thus quantum states – represent an agent's degrees of belief, rather than corresponding to objective properties of physical systems." Let us remark from the beginning that we do not intend to scrutinize these articles, containing sometimes surprising statements. In view of the centennial history of the subjective – objective debate in probability theory, it may well be impossible to convert adherents of the subjectivist interpretation to an objectivist position (and v.v.): the premises seem too different; they may ultimately be metaphysical; and identical facts can be interpreted in different ways. We will therefore not venture into a detailed analysis of [3-5], but just relate a few statements of [5] to our arguments of the preceding Sections.

Let us first recall that arguments against the subjective interpretation were already offered by the fathers of probability theory (Section 1). One of the main ideas of von Mises (Ref. [11] pp. 96-97) and Refs. [17, 26] is that it is risky to attribute probability to *propositions* instead of experimental outcomes (except if the link with experiments is unambiguous): there is a danger of running into unscientific interpretations[52]. The main question that remains unanswered within the subjective approach is, it seems: if probabilities are subjective degrees of belief of some observer, why do observers all over the world measure the same (quantum) probabilities when they do the same experiments ? Why do these different observers measure the same electron energies in solids, the same EPR state probabilities, the same radioactive decay rates etc. ? Needless to say, *all* probabilistic predictions of quantum mechanics are measured or verified by determining relative frequencies.

[51] For instance, a certain judiciously constructed die throw experiment, containing as probing system a table covered with glue, may have as unique outcome R = 6, with probability 1. With another probing system (remove the glue) the outcomes are 1,…,6 with probability 1/6. Probabilities *and* outcomes depend, in principle, on the probing.

[52] For instance, within our strict scientific interpretation, talking about the 'probability' of the propositions p = 'Theory X is correct' or p = 'Person X is guilty' makes no sense. Which experiments would have to be performed in order to measure P(p) ? See [18] and Appendix 1 for a more detailed argument.

The only point of [5] we will briefly comment here is the following. In order to justify and substantiate any difference between the subjective interpretation and the usual frequency interpretation, the authors (but this remark holds mutatis mutandis for all subjective approaches) have to invoke a premise such as Lewis' 'principal principle' [5], distinguishing objective chance and (Bayesian) probability. Bayesian (real) probability (Pr) should satisfy, according to this principle: Pr(E|C&D) = q, with E an event, C the proposition "the objective chance of E is q" (q *is* supposed to be the objective chance (!)), and D "some other compatible event, e.g., frequency data" [5]. Even if we make abstraction of the fact that the probability function 'Pr' has as argument an improbable mixture of events, propositions and data, Lewis' principle sounds circular to us. What is the use of introducing Pr if it is determined by q ? Why not stick to the objective measure q, which the authors interpret themselves as von Mises' relative frequency ? The authors of [5] do propose a justification for a 'subjective element' in quantum probabilities; but it strikes us as uneconomic. Let us indicate why.

In fact, we believe that with enough patience it is possible to build a bridge between a subjective interpretation as proposed in [5] and the objective interpretation defended here. In [5] the authors state, when discussing a coin toss: "An advocate of objective chance is forced to say that the chance is a property of the entire 'chance situation', including the initial conditions and any other relevant factors." Exactly this point offers the key to recognize the objective nature of probability, but, as we argued in detail in Section 2 and Ref. [18], only when 'initial conditions' are understood as the *initializing conditions of the repeated experiment* [18], and *not* of every individual coin toss (as the authors of [5] understand it). In their search for a 'subjective element' in quantum probabilities, the authors state: "The subjective Bayesian interpretation of quantum probabilities contends, in contrast, that facts alone never determine a quantum state. […] The prepared quantum state always depends on prior beliefs in the guise of a quantum operation that describes the preparation device" [5]. It is quite clear that the 'subjective element' (the 'prior beliefs') of passages as these correspond in our objective model to the experimental conditions (this time in particular the initializing conditions) that determine any probabilistic outcome. Any quantum state is only defined with respect to a preparation (Section 3.1.); any probability is only defined with respect to an initializing system (Section 2). But this turns quantum probabilities into objective quantities, as explained in Section 2. It seems that the objective model of Section 2 is more economic, since it only refers to experimental conditions of random tests. It explains a wide variety of paradoxes without referring to subjective prior beliefs, which are a very unnatural element in physics. (The above argument is essentially the same as the one we gave for Jaynes' Bayesianism, see Appendix 1, to which the authors of Ref. [5] refer.)

There are other elements in the subjective interpretation that remind us of our objective frequency interpretation. For instance: "This approach [the subjective interpretation] underlines the central role of the agent, or observer, in the very formulation of quantum mechanics" [5]. We have emphasized the role of the observer – *as physical system* – throughout this text. But we have also argued that by attributing probability to experiments, or to composed events including initializing and observing conditions, probability can be defined in a fully objective manner. True, the subjective shift is tempting: in order to make the concept objective, one has to include the observing *subsystem* (*not* the mental state of a human observer) into the probabilistic system.

## 4. Conclusion.

The aim of the present article was to show that the interpretations of probability and of quantum mechanics are more overlapping than usually thought. The bridge between the two is the essential and identical role that the 'observer' plays in both frameworks. Indeed, a detailed conceptual study of probability reveals that any numerical value of a probability (P(R)) depends on, is determined by, the initial and probing conditions in which R occurs (an idea that is compatible with von Mises' view of probability, but also Kolmogorov's). Therefore, any scientific probability is only defined if the observing subsystem is defined; it only exists 'relative' to such an observer system. In a natural environment, when nobody looks, the role of the 'observer' is played by the precise physical conditions imposed by the environment (the latter determine the probabilities in exactly the same manner as the conditions imposed by a human observer in a laboratory test; more precisely, they intervene in an exactly symmetric manner in a precise definition of probability [18]). This simple observation allows to interpret many elements of the Copenhagen interpretation: the 'understandable' (and measurable !) ingredients of the Hilbert space formalism are probabilities – and these should satisfy the above mentioned dependence on the observing system. In classical probabilistic systems this dependence is rarely or never explicitly mentioned (everyone knows how to perform and probe a regular die throw, no need to specify the conditions). But quantum theory has shown that in the quantum realm these conditions are stringent and essential.

Thus we could tackle, within a unified approach, several problems and paradoxes of probability theory and of the interpretation of quantum mechanics. We revisited the 'collapse of the wave function' and some typical elements of the Copenhagen interpretation [6-8, 23-24]. This appeared not a luxury, since even experts as Bell kept on finding some of these elements mysterious – in particular Bohr's response to EPR [24]. Within our model Bohr's response (which is often termed incomprehensible) becomes transparent. Finally, we could derive the main claim of relational quantum mechanics [1-2]; question the central premise of the subjective approach to quantum mechanics [3-5]; and propose a possible bridge between the objective frequency interpretation and Bayesianism à la Jaynes [32-33] (Appendix 1) – a longstanding problem in the interpretation of probability.

Acknowledgements. I would like to acknowledge much valued comments of participants at the 6th Conference on the Foundations of Probability and Physics, Växjö 2011, in particular detailed discussions with Andrei Khrennikov. I am indebted for in-depth discussion to Mario Bunge, Henry E. Fischer, Yvon Gauthier, Jean-Pierre Marquis, and Vesselin Petkov.

## Appendix 1. Objective Probability and the Subjective Shift.

In this Appendix we will have a closer look at the subjective or Bayesian interpretation of probability, due to authors as De Finetti, Ramsey, Jeffreys, Jaynes and others (see the excellent and neutral reviews [14-15]). The subjective interpretation comes in as many variants as there are authors; they range from full-fledged subjective versions such as De Finetti's variant, interpreting probability as a measure of strictly individual belief, to variants which seem easier to adopt in scientific contexts, such as Jaynes' theory. Neutral assessments often come to the conclusion that the hard-boiled subjectivist versions 'become empty' (Ref. [15] p. 84) after detailed scrutiny. In the case of Bayesianism à la Jaynes [32-33], the situation is less clear. It should be noted that there exists a small but fervent fan group among physicists of the latter theory (see the contributors to [33]). The model seems to gain popularity in quantum mechanics [3-5]. We will start by opposing the full-fledged subjectivist interpretation to the frequency interpretation; this will allow us to propose a hypothesis of why the 'subjective shift' is so tempting. In the last part we will turn to Jaynes' Bayesianism and suggest that a bridge with the frequency interpretation is possible.

Let us have a more detailed look at the example of the die throw in Section 2. One may consider the probability of event $e_1$ = "the outcome R = 6 in a regular die throw (probed as usual on a table)". One could also consider the probability of $e_2$ = "R = 6 after a camera has registered that R = 6 one microsecond before the die came to a halt". The probability of $e_1$ = $P(e_1) = 1/6$, while $P(e_2) = 1$ (or suppose so). These probabilities are *not* different because of different subjective knowledge states. First note that also for $e_2$ one can define the probability without referring to 'subjective knowledge' or 'information'; $P(e_2)$ could be measured by an automat[53]. So $P(e_1) \neq P(e_2)$, not because someone has a different strength of belief, but because $e_1 \neq e_2$: both probabilities concern different events, different experiments, and in particular different observing conditions and systems. This is a simple consequence of the basic idea of von Mises' theory: every probability can be seen as the result of a (series of) automated experimental tests.

Now since nothing is simple in the foundations of probability, it is at the same time true, as is obvious for any practitioner of probability calculus, that $P(e_2)$ can also be expressed in a manner that invokes, implicitly or explicitly, the status of knowledge of some observer – whence a potential confusion. Indeed, $P(e_2)$ can be considered a conditional probability, namely the probability that R = 6 *if it is given* that R = 6 one μsec before halting; in symbols $P(e_2) = P(e_1 \mid R = 6$ one μsec before halting). *Now it indeed seems that the phrasing "if it is given that" is equivalent in this context to "if it is known that"*; and therefore one is tempted to interpret conditional probability as depending on the knowledge of some observer. According to our point of view, this is fine as long as one realizes that this interpretation is a shortcut of thought, opening the door to subjective interpretations – it is in any case by no means the only interpretation. As we just illustrated, and as is explicitly proven by von Mises ([12], pp. 22-24), *all* conditional probabilities can very well be seen as corresponding to series of automated experiments in which no intervention nor belief state of a human agent is necessary (whether these probabilities are related to chance games as in the above case, or to natural phenomena). But is this not a happy argument for the homogeneity of probabilistic phenomena ? Under the usual interpretation all *natural* stochastic phenomena occur according to probabilistic laws also without a human pondering about them; therefore the same should hold for *artificial* chance phenomena such as die throws[54]. It is worth to emphasize that conditional probabilities can be interpreted in an objective way, since they play an essential role in probability theory. As already implicit in Kolmogorov ([9], p. 3), and explicitly stated by Gnedenko ([19], p. 67), even so-called unconditional probabilities can be regarded as conditional *on the circumstances of realization*. Note that we stressed throughout Ref. [18] that it is useful to remember that these circumstances or conditions are composed. Therefore the natural generalization in our model is to replace, if helpful, P(R) by P(R|C) where C contains all relevant parameters that describe the initial, final, and 'environmental' experimental conditions.

[53] Such an automated experiment is this: let a robot launch a die, let it select by camera vision those trials that show R = 6 one μsec before the die comes to a halt, and let it measure on that ensemble R again at full stop of the die (all this could be done by a machine). The relative frequency of these results will converge to 1, as our robot could determine.

Let us now have a look at Bayesianism as taught by Jaynes [32-33]. In Jaynes' words ([32] p. 88): "We want probability theory to indicate which of a given set of hypotheses $\{H_1,H_2,\ldots\}$ is most likely to be true in the light of the data and any other evidence at hand". As usual in Bayesianism the key role is played by Bayes' formula which Jaynes writes as follows ([32] p. 89):

$$P(H|DX) = P(H|X).P(D|HX) / P(D|X),$$

with 'X = prior information, H = some hypothesis to be tested, D = the data'. The formula allows to derive the 'posterior' probability of H in the light of new data or information, P(H|DX), as a function of its 'prior' probability, P(H|X). "Equation (4.3) [the above formula] is then the fundamental principle underlying a wide class of scientific inferences in which we try to draw conclusions from data" ([32] p. 89). Thus, Bayesianism understands probabilities as attached to hypotheses of which the likelihood can be calculated in view of '(background) information, data or evidence'.

[54] Gnedenko gives an enlightening analysis of the subjective shift ([19], p. 26ff.).

Now, if one investigates the concrete physical cases to which the above formalism is applied, there seems little doubt these can be recast in the vocabulary of the frequency interpretation. The translation would be immediate if in all cases 'H' and 'D' appearing in the Baysian probabilies P(H|DX) or P(D|HX) would correspond to *outcomes of a random experiment*, and if X would describe the precise *initial, final and 'environmental' conditions* ('C' in our notation). In all physical examples we examined, this appears indeed to be the case. Let us show it in detail for one example, the one that Jaynes treats first and in most detail ([32] p. 93 – 96). In this case we have:

X = 11 machines fabricate widgets, which pour out of each machine into a box (1 box per machine); 10 of these 11 machines produce 1 defective widget out of 6 (these are 'good' machines); one machine (the 'bad' one) produces 1 defective ('bad') widget out of 3. We randomly choose one of the 11 boxes without identifying from which machine, and randomly pick a widget from it. Take then:

D = The picked widget is bad.

H1 = We chose a bad box.

H2 = We chose a good box.

Jaynes then calculates probabilities as P(H1|D), the probability that 'hypothesis H1 is true if data D is known'. It is however clear that all the calculated probabilities (p. 93 – 96) can be understood as probabilities of outcomes of an automated experiment, in which a robot randomly picks first a box, then a widget from that box, and finally identifies whether box and widget are good or bad. In the objective phrasing P(H1|D) is the probability of *outcome* H1 (a bad box is picked) given *outcome* D (a bad widget is picked), i.e. the probability that a bad box is picked if it is given that the picked widget is bad, all in the experimental conditions described by X (the result is P(H1|D) = 1/6). This is of course a most classical application of probability theory. Clearly, Jaynes prefers a subjective vocabulary; but in all concrete applications of Bayesianism we could find ([10, 14-15, 32-33]) 'hypotheses' and 'data' or 'evidence' directly correspond to outcomes of experiments; their probabilities can univocally be given a numerical value as a relative frequency in a large set of experimental data. If this link with experiments cannot be made, Bayesianism cannot invent new results (e.g. concerning the probability of 'hypotheses'; see Ref. [26] and Ref. [10] p. 252 for misuses of Bayesianism). Occasionally, this seems the conclusion to which Jaynes himself tends. When investigating whether Bayesianism can attribute a probability to 'the hypothesis that Newton's

theory is right', he concludes: "[…] there is little hope of applying Bayes' theorem to give quantitative results about the relative status of theories" ([32] p. 139). Von Mises, or any objectivist, would have relinquished hope from the start: the hypothesis 'Newton's theory is right' is not a possible outcome of a random experiment.

In conclusion, we believe that Bayesianism à la Jaynes, if applied to physical experiments occurring in univocal conditions, can be translated into the frequency interpretation: the (background or prior) information or knowledge of the subjective phrasing corresponds to the precise experimental conditions of the objective interpretation. We therefore believe that this coherent form of Bayesianism, to which recent authors in quantum mechanics refer (e.g. Ref. [5]), should *not* be termed subjective. In comparison to truly subjective interpretations of probability, as they exist for instance in economy, philosophy, etc. ([14-15, 26]), it is in fact an objective model, even if phrased in subjective terms. This is again almost explicitly stated by Jaynes himself ([32] p. 45).

# Chapter 6

# Epilogue

In the preceding two chapters a model for the interpretation of probability was presented. Since these texts are by their nature rather condensed, I am well aware that many questions on this subtle topic remain unanswered in these chapters. The proposed model can doubtlessly be made more precise and be applied to many more case studies, which would increase its cogency. On the other hand our goal was to provide a minimal model that allows to solve paradoxes and that is helpful for reassessing the debate investigated in this thesis. Let us also repeat that von Mises has extensively answered to critics of the frequency interpretation in the introductory chapters of his works (von Mises 1928/1981, 1964). In defense of von Mises I should add that these rebuttals seem to be unknown to some of the present-day critics, since some questions about the frequency interpretation always reappear. But our model belongs to philosophy and one may disagree with our interpretation; of course other interpretations exist.

Before formulating a general conclusion of the present thesis, let us see, then, whether the model of Chapters 4-5 can teach us something about the debate on determinism. As elaborated in Chapters 2-3, a detailed investigation of the solutions to Bell's theorem shows that determinism remains a possible hypothesis, and that choosing between determinism and indeterminism is (at best) a philosophical choice, depending on one's metaphysical desiderata, intuitions, hypotheses.

In the following I will argue that probability theory, under the interpretation described in Ch. 4-5, offers arguments in favor of the hypothesis of determinism, and therefore of a deterministic worldview. I will succinctly list my philosophical arguments (Ph1 – Ph4) below. Most of them could be elaborated in much greater detail.

**Ph1. A homogeneity argument**. First of all, many physical systems have probabilistic properties and are yet known to be governed by deterministic laws. For instance, random flow, eddies in rivers, diffusing substances in fluids, turbulences, gases etc. all exhibit properties that are probabilistic, i.e. that show frequency stabilization; yet they are governed by the fully deterministic laws of fluid mechanics. Every individual particle follows a deterministic trajectory (even if in practice unpredictable and possibly chaotic); but the time-averaged or ensemble-averaged trajectories are probabilistic in the sense of Chapter 4. One other rather general example is the following. In numerical statistical physics systems are simulated by computer programs of which the outcomes are probabilistic properties, such as distributions of velocities, of positions, pressure, energy etc. In some cases these probabilistic properties are calculated by giving to the individual constituents of the system a fully deterministic trajectory, in agreement with what we just said. In other cases some 'input' parameters are given a certain probability distribution (in which case it seems less surprising that the output is also probabilistic). However even in this second case, the input distribution is attributed via a numerical random number generator that randomly and equiprobably chooses a number, typically in the real number interval [0,1]. Now it is well-known that these 'random' number generators are actually pseudo-random: they function with a deterministic but *random-looking* algorithm (i.e. a mathematical function). Therefore, on closer inspection such simulations – and these are extremely widespread in physics – simulate probabilistic behaviour by a process that is fully deterministic underneath the surface.

Our question has been recurrent in this thesis: if some probabilities can be retraced to deterministic processes, why not all ? If determinism would hold for all events, one would have the simplest, most homogeneous worldview, needing one ontic category, that of deterministic events (or deterministic laws). I have to admit that this 'simplicity' or 'homogeneity' argument is appealing to me.

**Ph2. A semantic argument**. Looking back at the definitions DEF1 and DEF2 of frequency stabilisation and probability (Ch. 4), one notices that they explicitly contain the terms 'to apply conditions or actions' and 'events generated in experiments'. Innocent as it may seem, these are causal notions; to the very least they invoke causal notions. For instance, 'generation' is one of the primary concepts of determination (Bunge 2009). Also, an 'initializing' system is a 'causing' system. If one may trust these definitions and this semantic hint, even without taking the precise content of the definition into account, then it seems that something *causes* frequency stabilization and therefore probability. It thus seems that the concept of probability needs causal concepts, concepts of determination. I will leave it to the reader's intuition whether she/he will judge this innocent or not.

**Ph3**. **Kolmogorov's question**. Kolmogorov writes on p. 9 of his seminal work on probability (Kolmogorov 1933/1956, our italics): "In consequence, one of the most important problems in the philosophy of the natural sciences is – in addition to the well-known one regarding the essence of the concept of probability itself – to make precise *the premises which would make it possible to regard any given events as independent*. This question, however, is beyond the scope of this book."

In probability theory events A and B are defined as independent if and only if their joint probability P(A,B) factorizes as P(A).P(B), in other words if and only if the conditional probability P(A|B) is given by P(A) (which is equivalent to P(B|A) = P(B)). Now, when one applies probability calculus to concrete problems involving the joint probability of two events (A and B), one typically needs to start from an assumption about the independence or dependence of A and B – one needs to guess whether they are independent or not. (For instance, a school book example is the throwing of two dice. It is rarely explicitly stated that these throws are supposed to be independent: one intuitively 'feels' that the results of two separate and normal die throws are independent. The interesting question is: why is this intuitive ?) Starting from this assumption one then calculates what the consequences are (for instance, supposing independence between the throws, what is the probability of having two sixes ?); consequences that if necessary can be compared to experiment[55].

What Kolmogorov asks, then, is whether it is possible to know in advance if two events are dependent or not. Are there premises that would make hypothesizing – guessing – unnecessary ? Intriguingly, Kolmogorov considers this as one of the most important problems in the philosophy of natural science. That indeed is a thesis for which I have a large sympathy.

Now it seems that determinism, the 'causal' interpretation of probability, allows to identify such premises in a natural manner: two probabilistic events ($y_1$ and $y_2$) are independent *if they share no common causes*. This idea can be formalized, loosely along following lines. In physics probabilistic 'events' correspond to probabilistic properties or variables. Under the hypothesis of determinism a stochastic property or parameter $y_1$ is considered as being caused by, *i.e. a function*[56] *of*, other variables ($x_1,\ldots,x_n$); similarly for $y_2$. So $y_1 = f(x_1,\ldots,x_n)$ and $y_2 = g(x_{n+1},\ldots x_r)$ for some mathematical functions f and g. *In practice we of course do not know these functions* – else the properties would be deterministic. Under a natural assumption, $y_1$ and $y_2$ are independent if $\{x_1,\ldots,x_n\} \cap \{x_{n+1},\ldots x_r\} = \phi$, the empty set.

[55] Von Mises found it important to emphasize that when one applies probability theory to a concrete problem, one always starts from assuming a certain probability distribution. One of his best known maxims is "Probability in, probability out". E.g., when one calculates the chances in urn pulling by combinatorics, one typically (and almost always implicitly) starts from assuming that the chance that a given ball is picked is equal for all balls. Then one proceeds to the problem posed. In the mentioned problem of throwing two dice, one typically assumes that the six outcomes of each of the two dice are equiprobable.

In conclusion, the hypothesis of determinism allows to give an answer to Kolmogorov's question: two events are independent if they don't share common causes – an analysis that is closely related to Hans Reichenbach's 'common cause principle' (Reichenbach 1956). Note it seems indeed that this is the intuition that one uses when guessing whether given probabilistic outcomes (e.g. two outcomes on separate dice) might be dependent or not. One tries to intuit whether they are causally connected or not.

Indeterminism, it seems, is silent with respect to Kolmogorov's question.

**Ph4**. **The Central Limit Theorem**. As is well-known, there is one ubiquitous probability density in the physical world, namely the 'normal' or 'Gaussian' distribution (the bell-curve $\exp(-x^2)$). It characterizes random phenomena such as human length distribution, weight, diffusion processes, temperature distributions, accident rates, distributions of fabrication errors etc. This distribution is so predominant in the real world, that one might say that 'most' phenomena are normally distributed. Now there is a fascinating theorem, namely the Central Limit Theorem, which at least partly explains the predominance of the Gaussian probability density in our world. The Central Limit Theorem (CLT), due to giants of mathematics as de Moivre, Laplace, Liapunov, Chebychev, Pòlya, states and proves that any variable that is the average of many other independent variables, will always be normally distributed, independently of the distribution of the contributing variables (see e.g. Tijms 2004, Gnedenko 1967; the CLT is treated in any good book on probability theory[57]). Thus, one typically reads in textbooks on probability calculus that since physical quantities are often the average of many unobserved random events, the theorem explains the ubiquity of the normal probability distribution.

[56] Notice that identifying causes with arguments in a mathematical function is precisely how Bell defines causes. It remains an essential and fascinating philosophical question to relate this definition to existing definitions of 'cause' in philosophy. Preliminary results are given in Vervoort (2013).

[57] Kolmogorov, Gnedenko and several other members of the great Russian school made important contributions to elaborations and new variants of the CLT.

It is straightforward, but in the present context essential, to observe that the above is most naturally interpreted in causal terms. Any variable that is the *effect* of many hidden (and independent) *causes*, or in other words that is *a function* of many hidden and independent causes, is normally distributed, independently of the distribution of the causes (the latter even may have extremely simple, fully non-random-looking distributions[58]). And indeed, this causal terminology is often used to illustrate the theorem. In other words, '(normal) probability' is interpreted in terms of the occurrence of causes, more precisely, in terms of the massive averaging effect of a large number of causes. One is inclined to say: anything that has many causes will look probabilistic.

If the majority of probabilistic properties of the world, namely those described by the normal distribution, can be understood as emerging from deterministic individual processes, it is very tempting to conjecture that they all can. (Along these lines, and for what it is worth, any continuous distribution can be approximated to any desired precision by a sum of Gaussians or by a shorter sum of ideal and distorted Gaussians.) Also, note that even typical chance games can be included in the discussion. The outcomes R of chance games typically have a discrete and flat (equiprobable) distribution (a die throw has 6 outcomes $R_i$ each with probability 1/6). First, such a discrete distribution can be seen as a special case of a continuous distribution: it can formally be represented by the distribution $\rho(R) = \sum_{i=1}^{J} \frac{1}{J}\delta(R-R_i)$ with δ the Dirac-δ (in the case of a die throw J = 6). As is well-known, such a Dirac-comb can mathematically be approximated by a sum of (infinitely) sharp Gaussians. Another way to see things is to consider a flat distribution as approximated by a Gaussian with an (infinitely) large dispersion – again a Gaussian. The typical deterministic interpretation of a die throw is that there is a quasi-infinite amount of causes determining its outcome; and that dies are constructed by craftsmen in such a way that none of the outcomes is privileged (Chapter 4).

This concludes our arguments; arguments that seem to us to favor the hypothesis of determinism over indeterminism. On our account, a key point is that the hypothesis of a deterministic reality underneath probabilistic behaviour gives an answer to certain questions,

[58] As an example, consider five variables $x_1,\ldots,x_5$. Give to each variable 5 values with a probability 1/5; attribute these values in a fully regular, deterministic way, e.g. $x_1$ will assume first the value $x_{11}$, then $x_{12},\ldots,x_{15}$; then come $x_{21}$, $x_{22}$, until $x_{55}$; then the same regular sequence is repeated a few times. The variable $y = \Sigma x_i$ will to a very good approximation be described by a Gaussian. The input values can be described as a deterministic function of for instance a time-index; the output is probabilistic and Gaussian.

whereas indeterminism remains, as far as we know, silent. Clearly, these arguments are not a proof within a generally accepted physics or mathematics theory. It may be possible to construct counterarguments against (Ph1 – Ph4); but we are not aware of them.

Now, if (in)determinism is a hypothesis of philosophy rather than of physics, then it will remain impossible to proof or disproof it. The best one can hope for is embedding it in a philosophical theory. Here we made a first attempt in this direction by linking it to a model for the interpretation of probability.

Let us now formulate a general conclusion of this thesis. The subject of these chapters was the debate 'determinism versus indeterminism' in the light of quantum mechanics and probability theory; we used both the physical and conceptual / philosophical parts of these theories. According to the dominant position our world, in any case the quantum world, is indeterministic. The general aim of the thesis was to reassess this claim and to propose a more equilibrated treatment of the debate and of the principle / hypothesis of determinism.

One of the most precise tools to investigate determinism is Bell's theorem. The experimental violation of the Bell inequality is often considered as a (near) *proof* that local determinism is impossible; this is certainly the case in the quantum physics community but also in the a priori more neutral quantum philosophy community (see e.g. van Fraassen 1991, Dickson 2007; but any work on the philosophical interpretation of quantum mechanics will do). In Chapter 2 we recalled that one cannot speak of a 'proof': indeterminism is one of several possible metaphysical hypotheses. Two well-known deterministic solutions to Bell's theorem are nonlocal determinism (but that seems to be in contradiction with relativity theory) and total or superdeterminism. In Chapters 2-3 the possibility of other deterministic solutions, gathered under the term supercorrelation, was investigated. Superdeterminism is usually considered implausible because it would be incompatible with reasonable definitions of free will and/or it would be 'conspiratorial'. In Chapter 2 it was argued that this depends on the philosophical standpoint one takes; within a theory as Spinoza's superdeterminism is the evident interpretation of Bell's theorem. Spinozists and many other philosophers would argue that total determinism is only in apparent contradiction with free will; and that conspiracy is in the eye of the beholder.

In this hugely complex and vast debate, I believe it is helpful to see things from different perspectives, physical, philosophical, historical, and even sociological. As far as I know, in the quantum physics and philosophy communities determinism as a solution to Bell's theorem is rarely or never discussed with reference to the most relevant philosophical theories on free will, conspiracy, and the history of determinism. In this collection of articles I only suggested a few entries into this vast program. But realizing that the debate is also philosophical seems essential in itself – once more, one usually believes it is physical (empirical) and terminated. Superdeterminism is not violating any physical law, and it has, arguably, the pleasing feature it corresponds to the most homogeneous, simple worldview (Ch. 2). From this point of view the indeterminism of the Copenhagen interpretation has, in a sense, a somewhat dogmatic taste to it: it boils down to the idea that the probabilistic outcomes / values of quantum experiments / properties cannot be completed *by any future theory*. 'Completion' has to be understood here in the sense (2) or (4) of Chapter 1. On the other hand, it *is* after all possible that quantum mechanics is the final theory for the Bell experiment and beyond. It is possible that the famous EPR-Bell correlations cannot be further explained by other variables – let alone controllable variables – however much they cry out for an explanation to some of us. It seems we are in a Kantian situation, and that it is wise to keep a good dose of agnosticism.

Superdeterminism is a 'local' solution to Bell's theorem because it offers an explanation of why MI, measurement independence, can possibly be violated by a local mechanism. In Chapters 2-3 we investigated existing physical systems, spin lattices, in which MI (and the BI) can be violated *without superdeterminism*. These systems are highly correlated, or supercorrelated in the sense defined in Chapters 2-3. Since such systems are arguably local, this result incited us to investigate whether a similar correlation could be at work in real Bell experiments. Surprisingly, I believe this is the case. It seems the assumption that the analyzers interact with a (background) field and that the latter is consequently modified and interacts in turn with the particles, is physically allowed; it is then quite obvious that this fully local mechanism may lead to violation of MI. Since MI is a premise of the BI, Bell's restriction on background-based theories is not operational. This is explained in detail in Chapter 3. The spin lattices actually can be seen as a simple static variant of such a background-based system.

This seems a surprise, because the quantum community has essentially discarded the possibility that a bona fide local and deterministic theory may underlie quantum mechanics. Therefore this solution will need corroboration by further theoretical or experimental work to gain a broad recognition; I have indicated in Chapter 3 a few ways to do so. We also attempted an explanation of why the solution may have escaped from discovery until now (Ch. 1 and 3). Other authors have conjectured that such a solution exists, e.g. based on an investigation of probability theory (Khrennikov 2008, Kupczynski 1986, Nieuwenhuizen 2009; cf. Ch. 3 for other references). Of course, even if our analysis is correct, there is no guarantee that *in practice* a realistic HVT for quantum mechanics can be constructed. But it is hoped that supercorrelation as a solution to Bell's theorem will motivate physicists to engage more actively in this research program. We have also emphasized throughout Chapters 2 and 3 that there is an interesting – possibly revealing – link to be made with early attempts to construct HVTs for certain quantum phenomena, such as double-slit interference (Grössing et al. 2012); *and* with macroscopic experiments mimicking quantum behaviour (Couder et al. 2005, 2006). These theories and experiments are based on a stochastic background field interacting with particles *and* with the environment (or 'context'), thus determining the particle trajectories.

Within this thesis in philosophy, let us give the last word to philosophy. The fact, then, that superdeterminism is a legitimate solution to Bell's theorem, saves determinism *at least as a philosophical theory*. Supercorrelation, as a potential solution, would furthermore open the door to new physical investigations. It could thus make determinism more attractive to some; but for others this is not needed.

If local determinism is not an impossible worldview, as is so often said, it is in a sense on a par with indeterminism. Of course, a pragmatic philosopher might argue that "indeterminism reigns as long as no deterministic theory is at hand". But isn't the history of science the history of finding determinism behind apparent randomness ? And especially: is it a good idea to cast a spell – for speculative reasons – on a whole research program that simply aims at explaining strange correlations and strange quantum behaviour in general ? It is the almost univocal predominance of the adoption of indeterminism that seems unbalanced and unwarranted to us. It may have hindered interesting science, and interesting philosophy.

As repeatedly noted, many of the here addressed philosophical topics can be elaborated e.g. by applying the usual tools of analytical philosophy. But I believe also within larger philosophy. The reader will have noticed I fancy Spinoza's philosophy. I would welcome research efforts that deepen the here only succinctly suggested link between modern science and Spinoza. On a very personal note, philosophies as Spinoza's, exhibiting a link of necessity between all things and beings and showing at the same time how pro-active ethical action is possible, seem to me immensely instrumental to deal with the modern world. More so than theories that are not based on this link. But agreed, that is another story.

In the last part of the thesis we gave arguments stemming from the interpretation of probability that seem to favor determinism. One of my preferred arguments is the following (essentially Ph1 above). There is a literally infinite number of probabilistic systems, from such diverse areas as fluid mechanics, diffusion, ballistics, error theory, population dynamics, population statistics, chance games, quantum mechanics, information processing, all fields of engineering etc. All these profoundly different systems show the *same* frequency stabilization – the same necessity to converge towards well-defined ratios. They all obey the same simple laws of probability theory. The only possibility I can imagine to explain this 'necessity' shared by all these systems, is that they share the necessity of laws governing the evolution of their individual constituents, i.e. the necessity of determinism. Essentially we are back at the 'homogeneity' argument (Ph1) above.

Thus this thesis started and ends with philosophy. It started with the intuition that something might be wrong with the physicists' claim that a philosophical principle is 'decidable' by experiment. It ended with a conceptual study of the notion of probability. It is hoped that on the way some examples were given showing that philosophy and science are not only solidly intertwined, but that both fields can fruitfully interact to lead to new results in both domains. Unfortunately science education neglects or even combats this idea. Biased as my physics education had left my (vague) opinion, I have to admit I was marvelled by how surprisingly rich the interaction can be.